\documentclass[12pt,twocolumn]{scrartcl}
\usepackage{amsmath, amsfonts, amsthm, amssymb,epsfig}
\usepackage{times}

\usepackage{color, colortbl}
\definecolor{Gray}{gray}{0.9}
\usepackage{fancyhdr}
\usepackage{morefloats}
\usepackage{pdflscape}
\usepackage{graphicx}
\usepackage{changepage}
\usepackage{url,hyperref}
\usepackage{natbib}
\usepackage{graphicx}

\usepackage[labelfont={color=blue,bf}]{caption}

\usepackage{natbib}
\bibpunct{(}{)}{;}{a}{}{,} % to follow the A&A style

\begin{document}

\title{Probing Accretion Disk Winds in AGN \vspace{2mm}\\
I. Asymmetric broad Balmer emission lines}

%\subtitle{I. Asymmetric broad Balmer emission lines}

\author{Cosmos DUMBA\\
 Department of Physics, Faculty of Science \\
 Mbarara University of Science \& Technology\\
 P. O. Box 1410, Mbarara, Uganda}
  %\and Wolfram Kollatschny\inst{1}
  %\thanks{\emph{Present address:}
    %Institut f\"ur Astrophysik, Universit\"at G\"ottingen, Friedrich-Hund Platz 1, D-37077 G\"ottingen, Germany}
     %\and Matthias Zetzl\inst{1}
     %\thanks{\emph{Present address:}
    %Institut f\"ur Astrophysik, Universit\"at G\"ottingen, Friedrich-Hund Platz 1, D-37077 G\"ottingen, Germany}}

%\offprints{C. Dumba, \email{cosmos.dumba@must.ac.ug or cosmos.dumba@yahoo.com}}

%\institute{Institut f\"ur Astrophysik, Universit\"at G\"ottingen, Friedrich-Hund Platz 1, D-37077 G\"ottingen, Germany
%  \and Department of Physics, Faculty of Science, Mbarara University of Science \& Technology,
%  P. O. Box 1410, Mbarara, Uganda}

%\date{Received:~\today~/~Accepted:~\today}

\twocolumn[
  \begin{@twocolumnfalse}
    \maketitle
    \begin{abstract}
     The Broad Line Region of Active Galactic Nuclei is characterized by broad Balmer emission lines in their optical spectra. 
The broad Balmer emission lines are found to be asymmetric, some blue sided and others red sided in their asymmetry. One of the components
behind the asymmetry is thought to be an accretion disk wind. We probe the accretion disk wind using the broad balmer emission
line profiles. \vspace{6mm} \\
This asymmetry of the broad balma emission line profiles is measured in velocity space after a measurement of the line
shift at percentiles from 0, in increaments of 10, up to 90. In addition, the Kurtosis Index is obtained at appropriate points 
of the emission lines' profiles.
This study is based on many hundreds of SDSS spectra, starting with low redshift high signal to noise ratio spectra. We also 
consider a definite number in each bin of their FWHM, in bins of 1000 km/s (atleast 40 per bin), starting from 1000km/s to the
very broad emission lines.  \vspace{6mm} \\
We present how strong the asymmetry (by plotting Asymmetry Index as a function of percentile) of the broad/narrow lines 
(in percent) is, what the Kurtosis $(R20, 80)$ is.
We also present what the Asymmetry Index as a function of line width (FWHM), luminosity (V-band), core-radio flux and Ionization Degree.\vspace{10mm}\\
\textit{AGN: Accretion Disk, Accretion Disk Wind -- Line: Asymmetry, Balmer, Emission, Profiles  }
    \end{abstract}
 
  \end{@twocolumnfalse}
]

%\keywords{AGN: Accretion Disk, Accretion Disk Wind -- Line: Asymmetry, Balmer, Emission, Profiles  }
%\maketitle

\section{Introduction}
Active Galaxies have been widely studied by many authors revealing many fascinating properties which among all include; compact nuclear 
emission \citep{Clavel}, non-thermal continuum emission 
\begin{equation}
        F_\nu \sim \nu^{-\alpha}
\end{equation} \citep{Bregman}. \\
In addition, authors notice that; the continuum stretches from the radio to X-ray/Gamma rays \citep{Mehdipour}, the luminosity of the nucleus 
exceeds that of the host \citep{Osterbrock0}, they have strong emission lines \citep{DeBreuck0}, are highly variable \citep{Ulrich0}, and also 
have X-ray emission \citep{Turner}. \\
The energy source in the AGN is believed to be accretion \citep{Blandford}. This can be demonstrated through the relation	: \\
\begin{equation}
        E = \eta mc^2
       \end{equation} where $\eta$ is the efficiency. \\ The luminosity can then be re-written as:
       \begin{equation}
        L = \frac{dE}{dt} = \eta \frac{dM}{dt} c^2 = \eta \dot M c^2
       \end{equation} where $\dot M = \frac{dM}{dt}$ is the accretion rate. \\
For a typical AGN;
       \begin{equation}
        \dot M = \frac{L}{\eta c^2} \approx 1.8 \times 10^{-3} \frac{L_{37}}{\eta} M_{sun} yr^{-1}
       \end{equation} where $L_{37}$ is the luminosity in units of $10^{37}$ W. \\
This energy, at one point, through the principle of hydrostatic equilibrium, reaches a point at which gravitational forces (causing the accretion)
balance with radiative forces (from the nucleus). This limit defines a characteristic luminosity, the Eddington luminosity ($L_{Edd}$) through 
the relations;
\begin{equation}
        F_{grav} = F_{rad}
       \end{equation}
 \begin{equation}
        \frac{GM(m_p + m_e)}{r^2} \approx \frac{GMm_p}{r^2} = \frac{\sigma_TL_{Edd}}{4\pi cr^2}
       \end{equation}
 \begin{equation}
        L_{Edd} = \frac{4\pi Gm_pc}{\sigma_T}M
       \end{equation}
 \begin{equation}
        L_{Edd} = 1.26 \times 10^{31} \frac{M}{M_{sun}}[W]
       \end{equation}
The important parameter in AGN is $\frac{L}{L_{Edd}}$. \\
One of the consequences of high $\frac{L}{L_{Edd}}$ are winds, Accretion Disk Winds.
Just like in Solar Winds....once a particle exceeds the escape velocity, we see this as a wind from the accretion disk.
Accretion disk winds have been confirmed by some authors by studying BAL Quasars \citep{Hamann}. \cite{Hamann} observed  strong absorption 
trouphs in the rest-frame UV spectrum and estimated outflows with 10,000 km/s.\\
\begin{figure}[!htbp]
\begin{center}$
\begin{array}{c}
\includegraphics[width=0.45\textwidth]{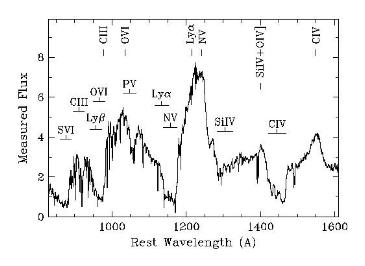}
\end{array}$
\end{center}
\caption[A spectrum of BALs ]{The rest frame of the UV spectrum of a BAL Quasar showing a significant outflow in the absorption lines of up to 
10,000 km/s}\cite{Hamann}
\label{fig:BAL}
\end{figure}
In this work, we study the accretion disk winds by analysing broad Balmer line profile asymmetry and steepness values. \\ We define the two 
parameters in line shape analysis as:
\begin{equation}
A.I = \frac{S}{IPV}
\end{equation}
\begin{equation}
K.I = \frac{IPV_U}{IPV_L}
\end{equation}
where S is the line shift, IPV is the interpercentile velocity, $IPV_U$ and $IPV_L$ are the interpercentile velocities for the upper and lower 
parts of the emission line profile.
The following sections will outline what we did in more detail and explain the most important steps we carried out during the study.
\section{Data}
\subsection{Introduction to Data Analysis}
In order to study any sample of galaxies and quasars, it is important not to forget a few useful conditions that make the sample
produce reliable information:
\begin{itemize}
 \item the size of the sample and
 \item the quality of the spectra, both in terms of resolution and Signal to Noise ratio.\citep{Whittle}
\end{itemize}
Our data meets both criteria thanks to dedicated surveys like the SDSS that have made such homogeneous data sets available to 
the public.\\
We have obtained, in each sample, the top 600 high signal to noise galaxies and quasars from the seventh data release (DR7) \citep{York0}.
The samples have the highest spectroscopic quality in the release \citep{Abazajian0}, and are uniform in terms of calibration let alone being 
complete.\\
We have two samples for the same reasons of obtaining quality spectra in all ranges of FWHM from 1000 km/s to above 10,000 Km/s. This is 
so because as we are selecting the spectra with the highest signal to noise ratio, there is a tendency to obtain only spectra that are both
from sources near (low redshift), thus neglecting those with higher redshift, and also biasing the data primarily on signal to noise ratio.
This is avoided when we split the sample in two samples, one for those between 1000 km/s FWHM and 3000 km/s FWHM, and another for those 
having FWHM greater than 3000 km/s. We end up having spectra with very high signal to noise ratio across the whole spectrum of broad Balmer
emission lines. Remember broad Balmer emission lines are those with FWHM greater than 1000 km/s. \\
In this section, we explain how we download the data, how we treat the data, describing the software we use and all the tasks used. We also 
explain all the steps we carry out, giving details of the output in each step and why it is carried out. In addition we show samples of the 
values extracted from the plots (the whole list being found in the appendices). Lastly we describe the secondary treatment of the extracted 
data from the plots, explaining how and why we carried out the treatment in such ways and show samples of the results in tables (the actual
results being shown in the next chapter). But then, it is necessary to first briefly describe the database (SDSS) and thereafter the telescope
and the surveys it has carried out so that one gets a feeling of the whole process from observing, collection of data, treatment of data and 
analysis.

\subsubsection{SDSS Database}

The Sloan Digital Sky Survey consists of three major surveys that together provide scientists with immense volumes of data
obtained from a dedicated 2.5m telescope located at Apache Point Observatory in Southern New Mexico. This survey has been
collecting data since the year 2000 \citep{Abazajian0, York0}. The three surveys are;
\begin{itemize}
 \item Legacy
 \item SEGUE
 \item Supernova
\end{itemize}

\subsubsection{SDSS Legacy Survey}
The SDSS Legacy Survey provided a uniform, well-calibrated map in ugriz of more than 7,500 square degrees of the North Galactic Cap, 
and three stripes in the South Galactic Cap totaling 740 square degrees. The central stripe in the South Galactic Gap, Stripe 82, was 
scanned multiple times to enable a deep co-addition of the data and to enable discovery of variable objects. Legacy data supported 
studies ranging from asteroids and nearby stars to the large-scale structure of the universe. Almost all of these data were obtained 
in SDSS-I, but a small part of the footprint was finished in SDSS-II. 
\subsubsection{SEGUE - Sloan Extension for Galactic Understanding and Exploration}
SEGUE was designed to explore the structure; formation history; kinematics; dynamical evolution; chemical evolution; dark matter 
distribution of the Milky Way. The images and spectra obtained by SEGUE allowed astronomers to map the positions and velocities of 
hundreds of thousands of stars, from faint, relatively near-by (within about 100 parsec or roughly 300 light-years) ancient stellar embers
known as white dwarfs to bright stellar giants located in the outer reaches of the stellar halo, more than 100,000 light-years away. 
Encoded within the spectral data are the composition and temperature of these stars, vital clues for determining the age and origin 
of different populations of stars within the Galaxy \citep{Yann}.
\subsubsection{The SDSS Supernova Survey}
The SDSS Supernova Survey was one of three components (along with the Legacy and SEGUE surveys) of SDSS-II, a 3-year extension of 
the original SDSS that operated from July 2005 to July 2008. The Supernova Survey was a time-domain survey, involving repeat imaging
of the same region of sky every other night, weather permitting. The primary scientific motivation was to detect and measure light 
curves for several hundred supernovae through repeat scans of the SDSS Southern equatorial stripe 82 
(about $2.5\,^{\circ}$ wide by ~$120\,^{\circ}$ long) \citep{Frieman}.
\begin{figure*}[!htbp]
\begin{center}$
\begin{array}{ccc}
\includegraphics[width=2.0in]{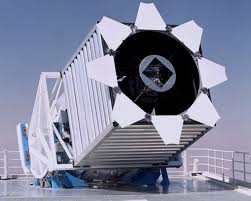} &
\includegraphics[width=2.0in]{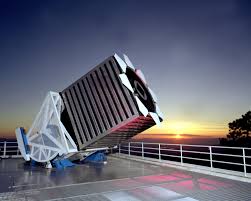} &
\includegraphics[width=2.0in]{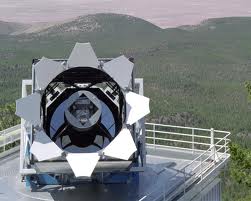} 
\end{array}$
\end{center}
\caption[The 2.5m Telescope at Apache Point Observatory, Southern New Mexico]{The 2.5m telescope located at Apache 
	Point Observatory in Southern New Mexico}\cite{sloan}
	\label{fig:The 2.5m Telescope at Apache Point Observatory, Southern New Mexico}
\end{figure*}
The above three surveys have provided scientists with a catalog derived from the images obtained by the 2.5m telescope.
These images include more than 350 million celestial objects, and spectra of 930,000 galaxies, 120,000 quasars, and 460,000 
stars. This data is not only fully calibrated and reduced, carefully checked for quality, and publicly
accessible through efficient databases, but has also been been publicly released in a series of annual data releases. It is through
this effort that we are able to carry out this study on the asymmetry of the broad emission lines of active galactic nuclei.\\
In this Data release, this study exploits the immense volume of spectra of galaxies and quasars \citep{Abazajian0, York0}. 
The sample obtained from the release
includes around 600 spectra of both galaxies and quasars in the redshift range of 0 and 1, with broad H$\beta$ lines from ~1500km/s enabling 
us contain all groups of broad line emission objects.\cite{sdssdr7}

\subsubsection{Obtaining the Data}
In order to obtain data from the SDSS database with your own constraints on the sample, one needs to write an SQL Query that generates
a list of the sources that meets your requests.

In this study, we focus on the parameters of the $H\alpha$ $\&$ H$\beta$ emission line, looking for asymmetry in these lines.

\subsubsection{Data Properties and Constraints Invoked}
The data in our samples combined consists of around 300 objects, galaxies and quasars, restricted to a redshift range between 0 and 1. The 
samples consist of reduced spectra of these objects in fits files that we renamed in ascending order from those with the highest signal
to noise ratio. This helps us analyze those with the highest signal to noise ratio first. This is important because the results 
obtained from high signal to noise ratio sources are more reliable for the derivation empirical relations. The redshift limit was chosen
so that we have a good coverage of the H$\beta$ line and its adjacent spectral regions. Figure presents the general properties 
of the sample in terms of redshift, signal to noise ratio and the broadening of the H$\beta$line. 

In addition, this data consists of around 283 spectra with measured $H\beta$ profiles around 165 measured $H\alpha$ profiles.
This is because not all spectra could have both the  $H\beta$ emission line and $H\alpha$ emission line due to red-shifting at both ends of 
the optical window.
However, most of the spectra with $H\beta$ lines also have $H\alpha$ lines but not vice versa. Of course spectra that are at the extreme end 
of our redshift
selection are the ones affected with not having the $H\alpha$ line visible, but this is not an issue for us to worry about since we were more 
interested
in the $H\beta$ line profiles.

\begin{figure*}[!htbp]
\begin{flushleft}$
\begin{array}{cc}
\includegraphics[width=3.0in]{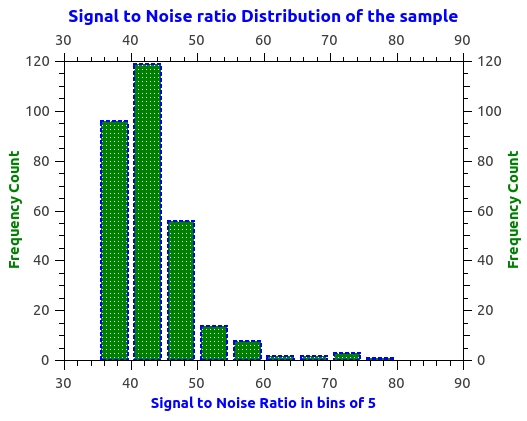} &
\includegraphics[width=3.0in]{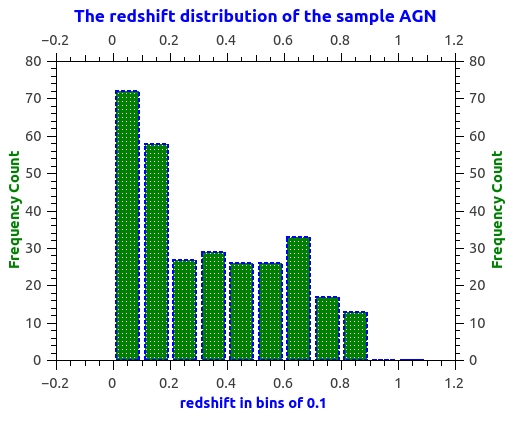} \\
\end{array}$
\end{flushleft}
\caption[The General Properties of the Data Sample - Signal to Noise ratio \& Redshift]{Signal to Noise ratio \& Redshift distribution of Sample}
\label{fig:sn_redshift}
\end{figure*}

Figures ~\ref{fig:sn_redshift} show the distribution of the signal to noise ratio of our sample and the redshift. It is seen here in
the signal to noise distribution that we have a fairly good signal to noise ratio, the minimum being 37 and maximum 79. The distribution
is binned in 5 starting from 35. The figure therefore clearly shows that most of our sample spectra have a signal to noise ratio between 
40 and 45, to be specific around 120 out of 300 (40\%), with a few having more than that. This is not a problem since, given the nature
of our study, a signal to noise ratio of even 30 would be sufficient for good measurements \citep{Thorne0}.

In the same way, the redshift distribution is shown. Because we chose spectra with the highest signal to noise ratio, it seemed obvious 
that most of the sample spectra will be in the near end of the redshift window we chose with those below redshift 0.2 dominating.
However, since the distribution does not fall rapidly as we move to higher redshift, the data will provide a sufficient study of AGN within
this redshift bin chosen.
In a later study, it would be good to also study other redshift ranges, possibly using Ultraviolet emission lines which can be obtained in the 
optical spectra after they have been red shifted.

\begin{figure*}[!htbp]
\begin{flushleft}$
\begin{array}{cc}
\includegraphics[width=3.0in]{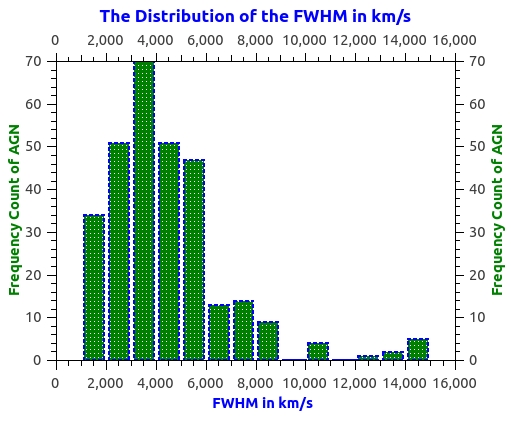} &
\includegraphics[width=3.0in]{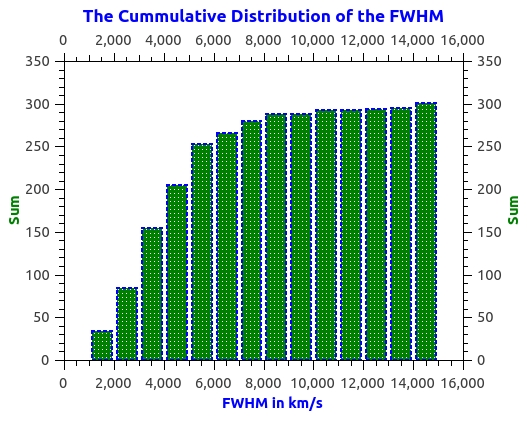} \\
\end{array}$
\end{flushleft}
\caption[The General Properties of the Data Sample - FWHM(km/s)]{Full Width at Half Maximum distribution of Sample}
\label{fig:fwhm}
\end{figure*}

Figures ~\ref{fig:fwhm} show the distribution of the FWHM of all the $H\beta$ emission lines measured. The distribution shows that
there is a peak between 3000 km/s and 4000 km/s. But still it is not a big range from the bins at both ends which are at 50 each.
Our intention is to have a good distribution across the whole range from 1000 km/s to over 10,000 km/s, and since we obtain enough
counts for the first five bins, then the sample will provide a statistically sufficient analysis for our findings.

The other figure here is a cumulative distribution, which shows that from FWHM of 9000 km/s, there isn't any increase in counts
as there are very few AGN with such extremely broad lines. Our study will obtain information about the broad lines basing almost 
entirely on the first six or seven bins with sufficient numbers.

The distribution of the FWHM in our sample is in agreement with the AGN statistics obtained by \citet{Hao0} when they plotted the distribution
of the FWHM values of the $H\alpha$ emission line for over 40,000 emission line galaxies. Their distribution was bimodal since they included
all AGN, narrow and broad line AGN. Their boarder line of broad line AGN was naturally placed at those with a FWHM of over 1200 km/s.
To compare with our distribution, we only consider those over 1000 km/s, and it gives us the same distribution. In their study, defining 
broad line AGN as objects with a FWHM greater than 1200 km/s, they obtained 1,317 objects out of 42,435 emission line galaxies. This makes
our sample of 300 objects not bad since it is a quarter of this value, let alone having been restricted to a redshift between 0 and 1.

Following this simple look at the general properties of the downloaded data, it is now necessary to explain the treatment of the data
in order to obtain measurements that are needed for our later analysis. The detailed spectral properties of the sample will follow later
in the forthcoming chapters.
%----------------------------------------------------------------------------------------

\subsection{Image Plotting and Analysis with IRAF}
IRAF (an acronym for Image Reduction and Analysis Facility) is a collection of software written at the National Optical 
Astronomy Observatory (NOAO) geared towards the reduction of astronomical images in pixel array form.
In this thesis, i used IRAF to plot spectra from our sample AGN and analyzed a few spectral lines needed to derive more information
for analysis later \citep{iraf0, iraf2}.
To be specific, we plotted the Balmer lines, $H\alpha$ and $H\beta$ emission lines, although not all spectra contained both emission
lines. Most of them had the $H\beta$ emission line, a good number had both $H\beta$ and $H\alpha$ emission line, while very few($\sim$2\%) had 
only the $H\alpha$ emission line.

In the following subsections, i will explain briefly the steps and tasks i used to extract the information i needed from the broad Balmer lines 
in my sample. This was principally the part of the thesis that occupied me most since it involved careful visual analysis of the spectrum first 
and accurate execution of the required commands for which a mistake with one of them means redoing the whole process from the beginning. 
Explained below are the steps i took during my data extraction.
\subsubsection{Spectrum Visual Analysis}
After downloading the data, (the fits files), and having renamed them, starting from the one having the highest signal to noise ratio, i plotted
each spectrum and analyzed the Balmer emission lines. The reason why i needed to do this is because each spectrum is unique and the tasks to 
perform on each varied from one spectrum to another. Some of the things i looked out for were;
\begin{itemize}
 \item The availability of both $H\beta$ and $H\alpha$ emission lines.
 \item Iron emission around the $H\beta$ emission line, the Fe Emission. \\ This was important to note because a lot of Fe emission could provide 
 a poor estimate for the continuum measurements. If the emission was only on either side of the Balmer line, i would use the end with less 
 emission as my proxy for the continuum. In cases where there was too much Fe emission, i excluded such a spectrum from my sample since it meant 
 an extra process of using a template to subtract the Fe emission first.
 \item Availability of neighboring narrow lines ([OIII] for $H\beta$ \& [SII] for $H\alpha$). \\This was for use in estimating an equivalent 
 FWHM of the narrow component to subtract from the Balmer line. However, for the case of the $H\alpha$ emission line, the [SII] emission doublet 
 was not accurate enough because the two lines were blended together in most cases. It meant using the [NII] lines that were within the broadened
 $H\alpha$ emission. This meant a visual estimation once both [SII] \& [NII] lines were unavailable. On a positive note, there were very few cases
 where i could not use any of the two.
 \item The function to use while subtracting the continuum. \\This was important because the uniqueness of each spectrum, and the quest to 
 apply a good estimate to a flat spectrum, entails using different functions for each spectrum in order to obtain a flat spectrum from the power 
 law spectra downloaded.
\end{itemize}
All the above mentioned reasons vary from spectrum to spectrum and each of them is important in order for me to obtain the best results from the 
spectrum, and for minimizing errors as much as i can. \\
In our data, which has two samples, sample 1 being that of AGN with FWHM between 1000 km/s and 3000 km/s, and sample 2 being that of AGN with FWHM
from 3000 km/s and above, this is what we observed from the visual analysis;
\begin{itemize}
 \item The AGN in sample 1 have both $H\alpha$ and $H\beta$, meaning they are found in the lower part of the redshift range from 0 to 1, 
 preferably less than z = 0.6. There are less than 10 out of 81 that had only the $H\beta$ emission line. 
 \item Sample 2 has at least 40\% of the AGN having both $H\alpha$ and $H\beta$. This means 60\% of them are found in the higher end of the 
 redshift bin, preferably above z = 0.7. Recall that my redshift range is from z = 0 to z = 1.
\end{itemize}

\subsubsection{Continuum Subtraction}
\begin{figure*}[!htbp]
\begin{center}$
\begin{array}{ll}
\includegraphics[width=3.0in]{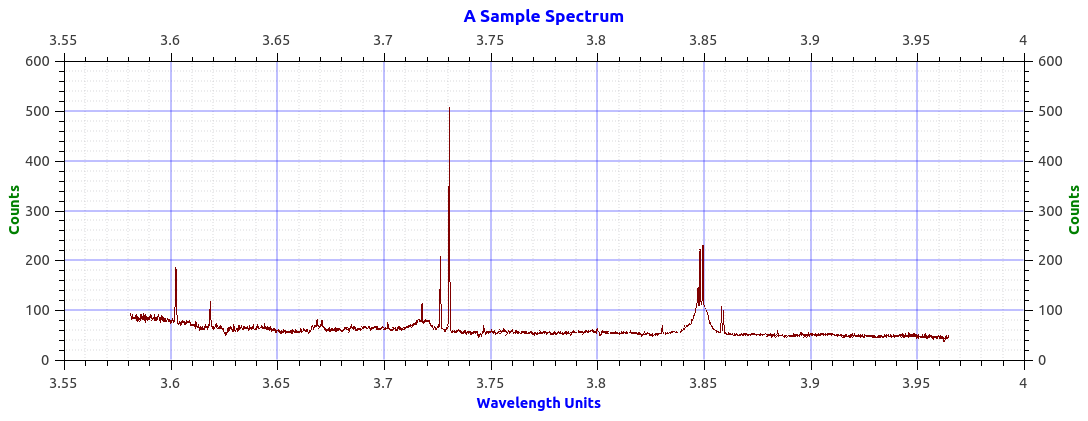} &
\includegraphics[width=3.0in]{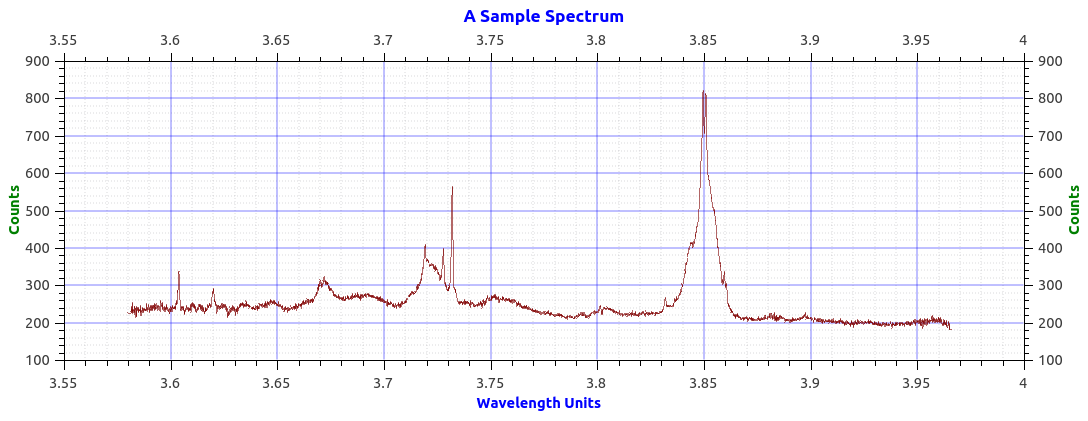} 
\end{array}$
\end{center}
\caption[Sample images of the original spectrum during visual analysis]{Seen above are two samples of the spectra during visual analysis. It should
be noted that not all spectra contain the $H\alpha$ and $H\beta$ emission lines as seen here, reason being that those red-shifted significantly
will eliminate the $H\alpha$ since it will be Doppler shifted to wavelengths outside the optical range. In addition, not all spectra are relatively
flat like the ones shown due to the varying amounts of continuum emission at different wavelengths for many AGN}
\label{fig:samplespectra}
\end{figure*}

\begin{figure*}[!htbp]
\begin{center}$
\begin{array}{ccc}
\includegraphics[width=3.0in]{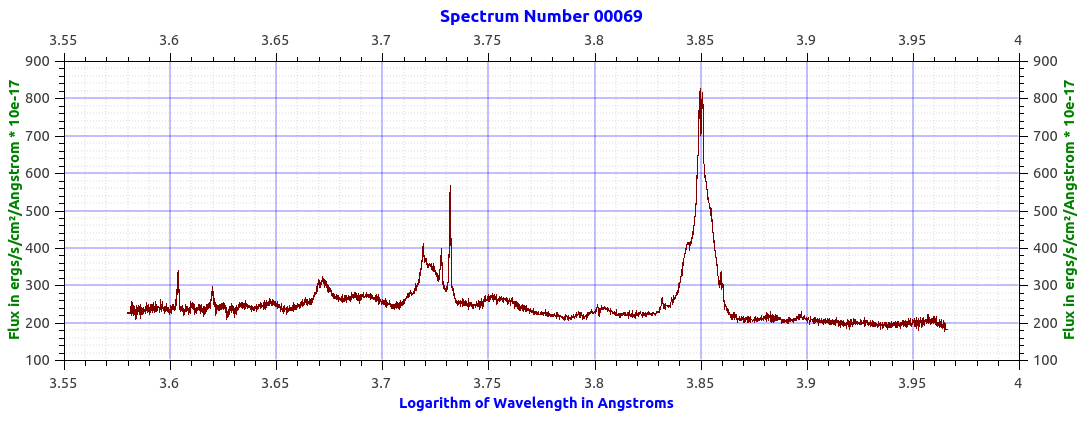} &
\includegraphics[width=1.5in]{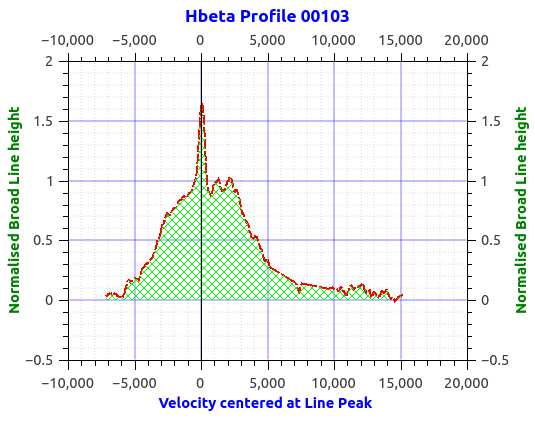} &
\includegraphics[width=1.5in]{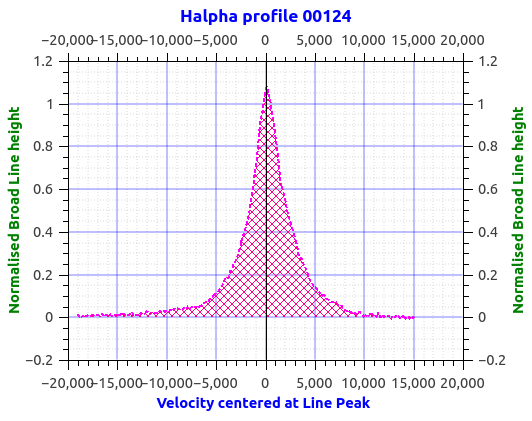} \\
\includegraphics[width=3.0in]{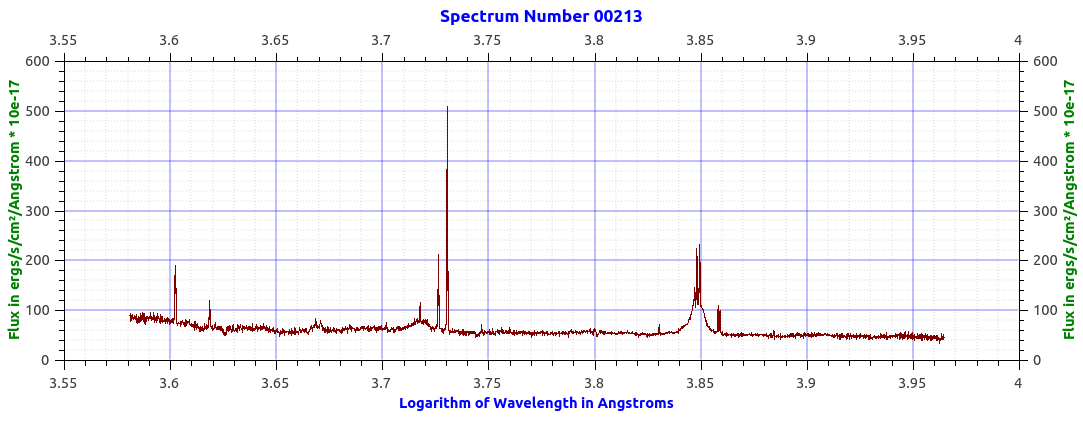} &
\includegraphics[width=1.5in]{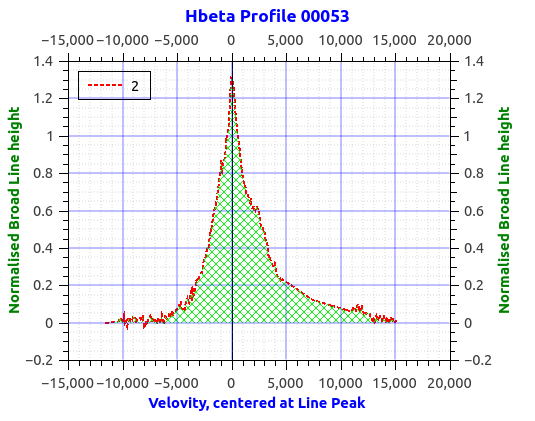} &
\includegraphics[width=1.5in]{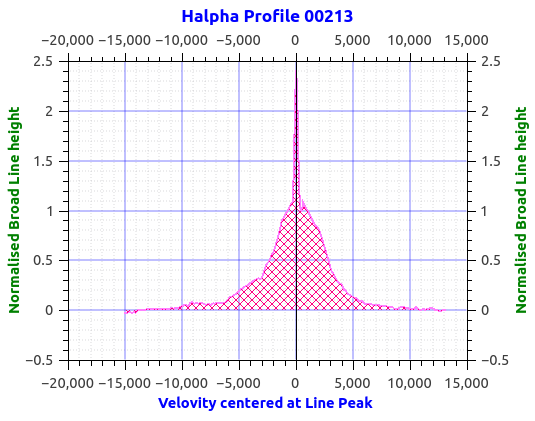} 
\end{array}$
\end{center}
\caption[Extracted Balmer emission lines from sample spectra]{The images show an example of extracted Balmer emission lines after continuum subtraction
on each of them. One should note here that all the extracted emission lines have their continuum at zero, making it the starting point of the percentile
ranges we need examine later.}
\label{fig:continuumsubtraction}
\end{figure*}

This is the task done after visual analysis of the spectrum. Spectra with both the $H\alpha$ and $H\beta$ emission lines look similar to the 
ones in figure~\ref{fig:samplespectra} although some with significant redshift will only have the $H\beta$ emission line still available.
The subtraction of the continuum is done in splot, with the keys "t" followed by a "-", followed with the appropriate function necessary for
the selected region. The continuum is subtracted for each emission line separately and separate images are saved out of the original image
having both emission lines. This can be seen in the sample plots in figure~\ref{fig:continuumsubtraction} on page~\pageref{fig:continuumsubtraction}.
Continuum subtraction is the second step to spectrum analysis for the Balmer emission line features and it sets a vantage point for line
normalization, as we shall discuss in the next subsection.

\subsubsection{Line Normalization}
To normalize the emission line simply means to have its base at a zero point and its peak at unity. It should be noted that we are normalizing
the broad Balmer lines. This means that we either subtract the narrow components or simply leave them visible, but ignore their additional height
as we simply place a unity value at the peak of the broad line.

\begin{figure*}[!htbp]
\begin{center}$
\begin{array}{cccc}
\includegraphics[width=1.55in]{00103} &
\includegraphics[width=1.55in]{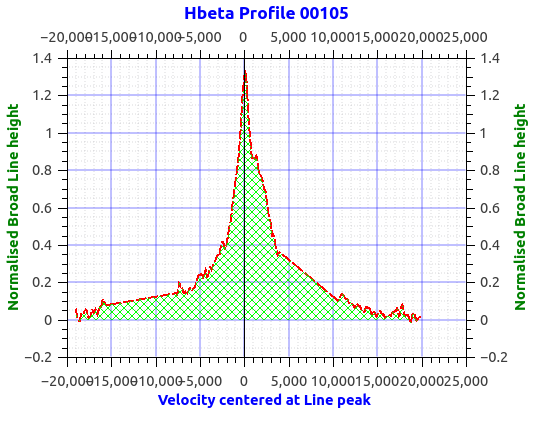} &
\includegraphics[width=1.55in]{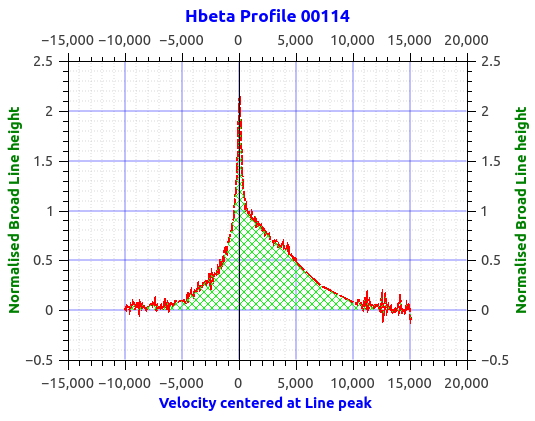} &
\includegraphics[width=1.55in]{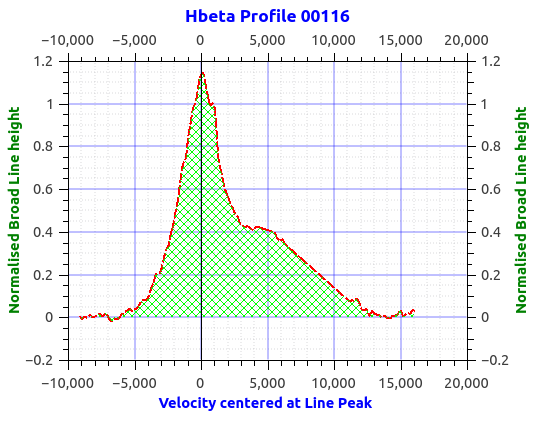} \\
\includegraphics[width=1.55in]{00124} &
\includegraphics[width=1.55in]{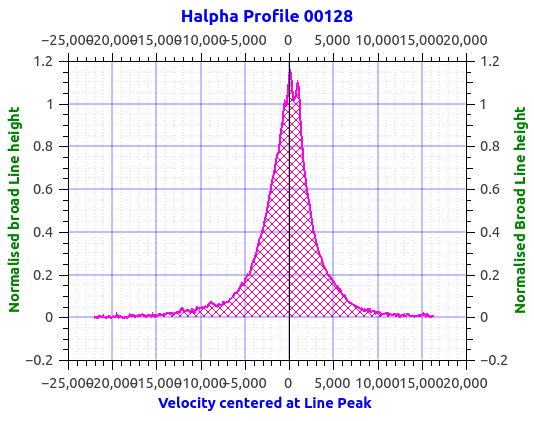} &
\includegraphics[width=1.55in]{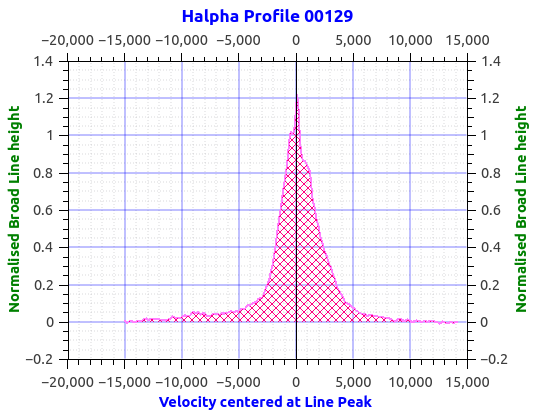} &
\includegraphics[width=1.55in]{00213} 
\end{array}$
\end{center}
\caption[Normalized broad Balmer emission line samples]{The images show an example of the normalized broad Balmer emission lines after 
continuum subtraction on each of them. It is clear from these examples, the diversity of broad and narrow components in AGN permitted 
emission lines}
\label{fig:linenormalisation}
\end{figure*}

In the figure~\ref{fig:linenormalisation}, examples on the diversity of broad and narrow components in our sample with some not having narrow
components at all and others dominated by the narrow component are shown.

To perform the normalization of the broad Balmer emission lines, we use the arithmetic task, "imarith", which can be used anytime in IRAF.
However, after continuum subtraction, the counts at the peak of the broad Balmer line are recorded in a table. Also in the same table is 
recorded the velocity at which the emission line is centered. This velocity will be of use while transforming to velocity space, centering
the emission line peak at that velocity.

Table~\ref{tab:initialvaluessampletable} shows part of the table. The whole table can be viewed in the appendices section. It shows the 
Spectrum Number, the Balmer line heights used to normalize the profile to unity and the corresponding wavelengths at which they are identified.
These wavelengths are the wavelengths at which we center the corresponding lines while transforming to velocity space.

\begin{table*}[!htbp]
\caption[Sample table showing the Initial Values Extracted From Each Spectrum Plot for use in Line Normalization and Velocity Transformation]
{A table showing the initial values extracted from each spectrum plot for Line Normalization and Velocity Transformation.
} \label{tab:initialvaluessampletable}
\begin{center}
\begin{tabular}{ccccc}
\hline
Spectrum Number & H$\alpha$ height & Centering $\lambda$(\AA) & H$\beta$ height & Centering $\lambda$ \\ \hline \hline

00001(ha \& hb.fit) & & & 88.01 & 7353.861 \\ 
00002(ha \& hb.fit) & & & 58.36 & 8599.299 \\ 
00003(ha \& hb.fit) & & & 39.25 & 7235.383 \\ 
00004(ha \& hb.fit) & 354.4 & 7365.362 & 87.10 & 5455.679 \\ 
00005(ha \& hb.fit) & 258.2 & 8927.240 & 88.46 & 6612.635 \\ 
00006(ha \& hb.fit) & 414.5 & 6785.756 & 140.0 & 5026.721 \\ 
00007(ha \& hb.fit) & & & 22.12 & 6819.172 \\ 
00008(ha \& hb.fit) & 173.7 & 6837.628 & 44.66 & 5064.256 \\ 
00009(ha \& hb.fit) & & & 15.91 & 7756.598 \\ 
00010(ha \& hb.fit) & & & 28.17 & 6809.348 \\ \hline

\end{tabular}
\end{center}
\end{table*}

\subsubsection{Line Correction : Filtering out other lines and Smoothening out the noise }
In any spectrum, there is always a probability that the emission line you are interested in is surrounded by other emission or absorption lines.
For some emission lines, the neighboring emission lines or absorption lines are not a threat to any measurements pertaining the emission line 
under study, for example, for the $H\alpha$ emission line, the neighboring Sulphur II emission lines do not pose a threat to any measurements
needed from the $H\alpha$ emission line. This can be seen in the examples below.\\

However, for the $H\beta$ emission line, there is an undeniable problem. The neighboring emission lines are significant enough to pose a threat 
to measurements taken from it. These emission lines are; the Oxygen III lines and the Iron emission at both sides of the $H\beta$ emission line.
The Iron emission will increase on the error in the continuum subtraction since it creates a pseudo continuum and thus making it difficult to 
estimate a continuum level. To reduce on the error in this measurement, one has to subtract the iron emission, preferably using some already 
developed templates before the actual continuum level can be estimated.\\
In our case, since it was a small fraction of spectra that had significant iron emission on both sides, we neglected this since it affects less 
than $5 \%$ of the data.
The Oxygen III emission lines also increase on the uncertainty of the FWHM values measured since they are embedded in the broad Balmer $H\beta$ 
emission lines. The most reliable way to deal with this uncertainty is to subtract them before any values are measured. However, since they are
also broadened at their bases, their subtraction, using a Gaussian model, increases absorption features in the residue $H\beta$ emission line 
left. Thus, since the emission lines are already broad, we decided to simply cut them out in order to leave a smoother profile for the residue 
for better measurements without introducing a smoothing factor. It is noticed that subtracting the OIII lines introduces absorption features
which in turn adds an uncertainty to the measurements, so simply cutting them off and preventing this uncertainty seemed better because the 
uncertainty we deal with by not doing the subtraction itself is much less.\\
When all other neighboring emission lines are removed, the spectrum is saved, and it is this spectrum that is used for all the other needed
measurements. The images of the spectra that will be displayed will in most cases be those that have been cleaned of all their neighboring 
emission line components.

\subsubsection{Display Transformation to Velocity Space : Centering velocity at Line center}

The spectra downloaded from SDSS displays counts on the y-axis against wavelength in angstroms on the x-axis. The wavelength is logarithmic in
scale. In order for us to measure the asymmetry in a clearer manner, we preferred to transform the wavelength scale to a velocity scale.
This is easily done in IRAF using the command ``disptrans'' which is accessed in splot as well. However, since we need to measure the asymmetry, 
we also center the velocity at the emission line peak. This procedure displays the profile in velocity space but centered at the line peak making it 
possible to measure the asymmetry from the line center.

\subsubsection{Spectrum Conversion to Text File : For further analysis of lines}

To measure the asymmetry, there several ways one can use, some measure by hand, others prefer to use a program. In our case, we preferred to use 
IDL to measure asymmetry \citep{IDL0}. This meant that we have to convert the spectrum image to a text file for the program to be able to read values and 
calculate accordingly the asymmetry. For IDL to perform this, we had to write a simple script with the necessary conditions and the desired output
we needed. The output from this program are our results of the asymmetry of all the line profiles we plotted. Another advantage of converting 
the spectrum image to text was that we could easily plot the very image using any other plotting software like GNUplot, Python and QtiPlot
\citep{QtiPlot0, Python0, Python1, Python2, Python3}, making it possible to plot in any way we preferred other than the default plot of splot 
in Iraf.

The values or output data we obtained from the IDL program is by definition our results. By results i mean the Asymmetry Index, because from this,
we can easily obtain the Kurtosis Index as well. It is the Asymmetry Index and Kurtosis Index from out data that was of prime importance from which
we shall relate to other kinematic properties of the host galaxies or Ionizing regions of the AGN.

\subsection{Data Treatment and Analysis}
The data for our study is obtained from a series of treatment processes. The previous sections were explaining how the individual emission lines
are treated in each spectrum to the moment when a text file is extracted from the fits image. This text file is convenient for further treatment
and analysis since we can re-plot the emission line with any other plotting software like GnuPlot, QtiPlot and Python \citep{QtiPlot0, Python0, Python1, Python2, Python3}.

Further treatment on these extracted text files is done with IDL, a program we use to extract values of the velocities at the chosen percentiles
and dump them out as the useful values for further analysis. This is done for each emission line, after which a text file having all the
necessary percentile velocities for all the files is obtained. It is in this file that we shall start our analysis of the asymmetry of our emission
lines. In other words, this file will contain all the information we need for each of our 447 broad Balmer emission lines in order to study the 
asymmetry.

Before we describe the process of extracting percentile velocities for all the emission lines, it is also necessary to briefly describe the observations
made on the broad Balmer emission lines during the initial processes of obtaining the extracted individual plots. 

\subsubsection{Description of Obtained Data}
\begin{figure*}[!htbp]
\begin{flushleft}$
\begin{array}{cc}
\includegraphics[width=3.0in]{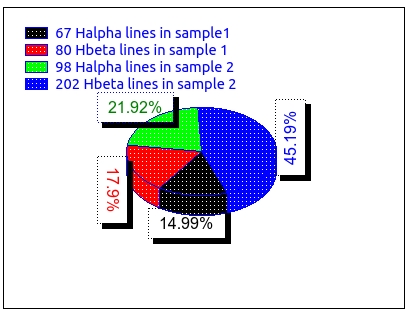} &
\includegraphics[width=3.0in]{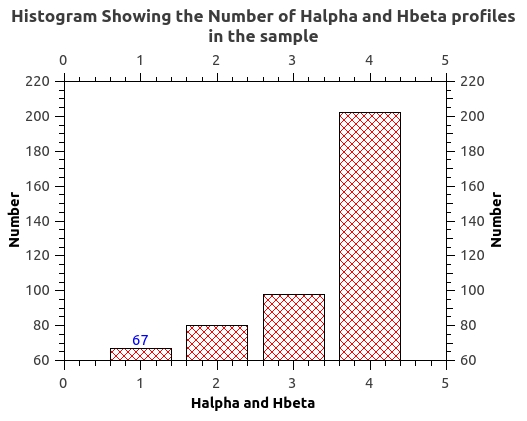} \\
\end{array}$
\end{flushleft}
\caption[The General distribution of the two samples of the data]{A pie chart and bar graph showing the distribution of both $H\alpha$ and $H\beta$ emission lines in the sample}
\label{fig:samplepie}
\end{figure*}
A first analysis of the data samples showed a clear distinction between AGN with broad Balmer lines with FWHM between 1000 km/s and 3000 km/s, and 
those with their FWHM greater than 3000 km/s. It was observed that the former are predominantly in a lower redshift region, while the later are
predominantly high red-shifted. This is evident from the fact that out of 80 spectra in the first sample, 67 contained both broad Balmer emission
lines, while for sample 2, out of 200 spectra, 98 contained both emission lines, the others missing the $H\alpha$ emission line due to being red-shifted.

\begin{figure*}[!htbp]
\begin{flushleft}$
\begin{array}{cc}
\includegraphics[width=3.0in]{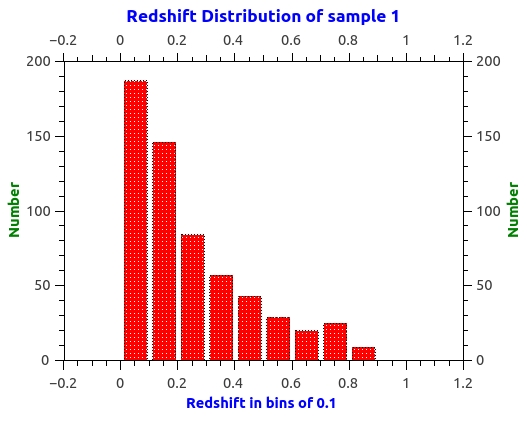} &
\includegraphics[width=3.0in]{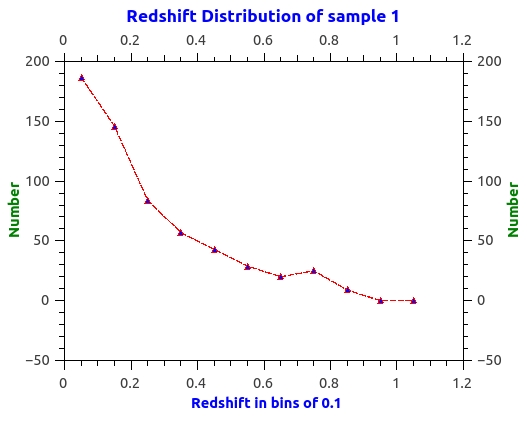} \\
\end{array}$
\end{flushleft}
\caption[The Redshift distribution of sample 1]{Redshift distribution of sample 1}
\label{fig:sample1z}
\end{figure*}

\begin{figure*}[!htbp]
\begin{flushleft}$
\begin{array}{cc}
\includegraphics[width=3.0in]{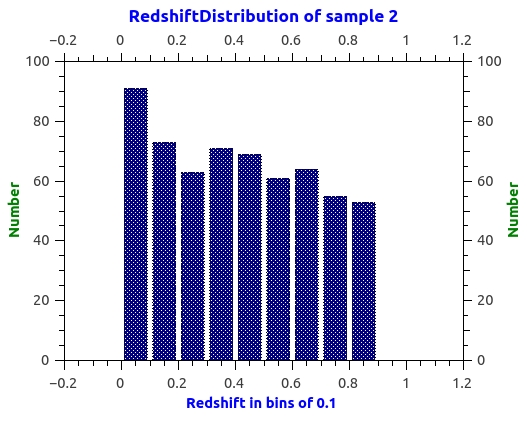} &
\includegraphics[width=3.0in]{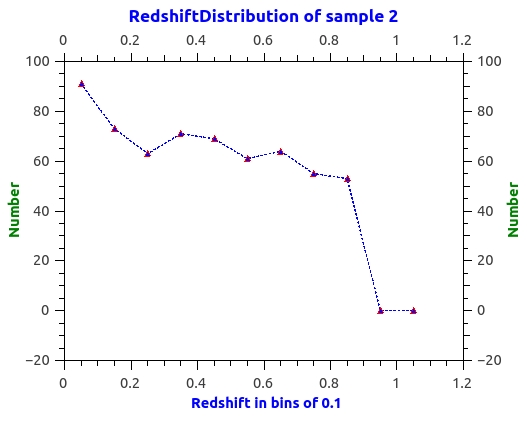} \\
\end{array}$
\end{flushleft}
\caption[The Redshift distribution of sample 2]{Redshift distribution of sample 2}
\label{fig:sample2z}
\end{figure*}

From figures~\ref{fig:sample1z} and~\ref{fig:sample2z}, we can notice that sample 2 is clearly unbiased in its redshift distribution as sample 1 is.

\subsubsection{Other levels of the data}

The data obtained initially is in the form of spectroscopic data, spectra downloaded from $SDSSDR7$ from which the asymmetry is measured using the IDL
script we developed \citep{IDL0}. The asymmetry values are then used to obtain the steepness of the profiles.

In order to relate the asymmetry and kurtosis to other kinematic properties, we needed to obtain more data pertaining the objects whose values of
asymmetry and kurtosis we have. This other data includes:

\begin{itemize}
 \item Flux values of [OIII], [OII] \& [OI] for obtaining the ionization degrees through their flux ratios.
 \item V-Band magnitudes for calculating the Luminosity.
 \item Radio Flux measurements for those sources whose radio fluxes have been databased already.
 \item Line width values of $H\alpha$ and $H\beta$ emission lines.
\end{itemize}

Values of the flux of the oxygen lines were obtained from the downloaded spectra. This was not an easy task as quite a number of sources were red-shifted
in such a manner that either [OI] was missing or [OII] was missing. However, it was still possible to have objects in which we could obtain values
from both emission lines.

The V-Band magnitudes were obtained using an sql script, which we wrote, to download spectroscopic values of $ugriz$ photometry data, from
which i transformed the values to the $UBVRcIc$ system. The V-Band magnitude was then calculated as:\\
\begin{equation}
 V = g - 0.58*(g-r) - 0.01
\end{equation}
It then made it possible to transform the magnitude to Luminosity.

Radio flux values were obtained from the NASA/IPAC EXTRAGALACTIC DATABASE, in which i inputted the coordinates of all my sources. The query returned
all the sources whose radio flux was databased. The radio flux was in milliJansky. Also to note is that since we are interested in the core radio flux,
the radio flux measurements queried were for regions within a $5.0$ arc-sec cone.

All the different sets of data were synchronized to match the asymmetry and kurtosis values for analysis. To note too is that the data was separated
in two sets; that of $H\alpha$ profiles and that of $H\beta$ profiles. In the analysis to follow, it will be noted that plots of $H\alpha$ and $H\beta$ are
placed besides each other, with the $H\alpha$ to the left and $H\beta$ to the right, where possible in different colors as well.

\subsubsection{Plotting}

Plotting in this work is done using quite a number of programs, from IDL \citep{IDL0}, QtiPlot \citep{QtiPlot0} and Python \citep{Python0, Python1,
Python2, Python3}. However, most of the plots in the Data section are plotted using Python, where made a script for each dataset so that it 
minimizes the size of the text file containing the data. It is beyond the scope of this section to give more details of the scripts. However, 
a sample will be attached in the Appendix section for more clarification.
The reasons for using various programs to plot ranged from which type of plot was needed and which data i was analyzing as i always used the most
efficient one for each type of plot i needed.

\section{Results}
In this section, we present the results obtained from the data analysis and a discussion about them. By results here,
we focus on the Asymmetry Index and Kurtosis Index of the Balmer emission lines measured. We analyze the distribution
of the Asymmetry Index across the whole profile from FWZI to its peak. In the later half of the chapter, we relate 
the Asymmetry Index to other kinematic properties such is ionization degree \citep{DeRobertis0, Rafanelli0}, radio flux \citep{Brotherton1}, 
luminosity and the line width (FWHM) \citep{Yu0}.

\subsection{Distribution of results}
On obtaining the Asymmetry Index and Kurtosis Index, it was necessary to briefly show how they are distributed. We chose to plot 
a frequency distribution of the Asymmetry Index and Kurtosis index.\\
The Asymmetry Index varies from -1 to +1, a value of 0 meaning the profile is symmetric while a deviation to either side of the zero
implies asymmetry. The degree of asymmetry in each case will be probed by relating to other kinematic properties. It is to our interest
to find out whether this positive asymmetry and kurtosis is related to either non-thermal or thermal radiation and/or even to the relative
strength and kinematics of the BLR. If the degree of asymmetry is a measure of the radial flow of material from or to the BLR, then it will be
possible to understand the structure of the accretion disk winds using asymmetry index.\\
In our distributions to follow, we separate the $H\alpha$ emission line profile asymmetry from that of the $H\beta$ emission lines.
The Kurtosis Index follows the same pattern.

\subsection{Asymmetry Index}

The values obtained from the IDL script yielded values of the Asymmetric Index for both the $H\alpha$ emission lines and $H\beta$
emission lines. The bar graphs that follow reproduce the statistical distribution of the Asymmetry Index for both Balmer lines, with
the $H\alpha$ on the left of each double image figure (in blue) and $H\beta$ to the right (in green).

It is also important to review the meaning of a value of asymmetry index as seen on the plots. This is seen in \ref{tab:ai0} below

\begin{table*}[!htbp]
\caption[Asymmetry Index]{Interpretation of Asymmetry Index}
\label{tab:ai0}
\begin{center}
\begin{tabular}{|c|l|l|}
\hline
  Asymmetry Index $(x)$ & Measure & Interpretation \\ \hline
  $0 \textless x \leq 0.08$ & Weak & The distribution is relatively symmetrical. \\ \hline
  $0.08 \textless x \leq 0.15$ & Moderate & The distribution is relatively asymmetrical.  \\ \hline
  $x \textgreater 0.15$ & Strong & The distribution is asymmetrical. \\ \hline
\end{tabular}
\end{center}
\end{table*}

\begin{figure*}[!htbp]
\begin{center}$
\begin{array}{cc}
\includegraphics[width=3.0in]{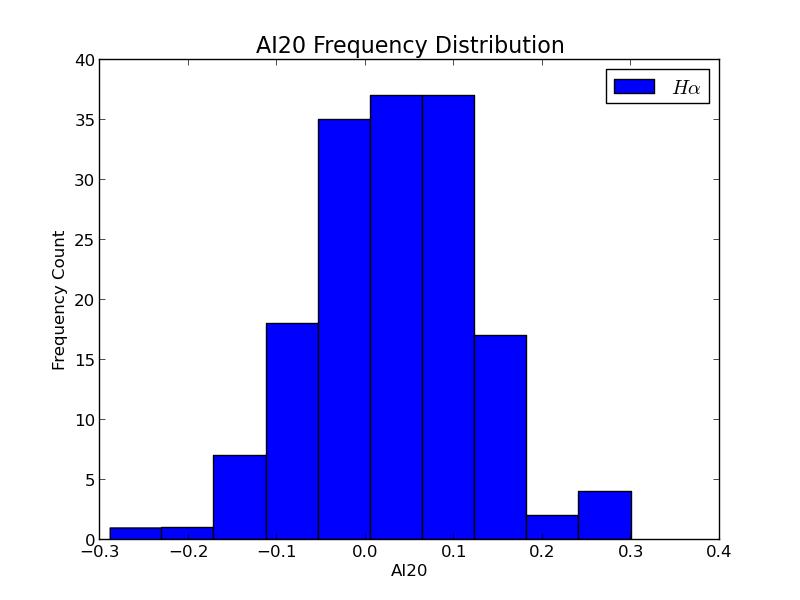} &
\includegraphics[width=3.0in]{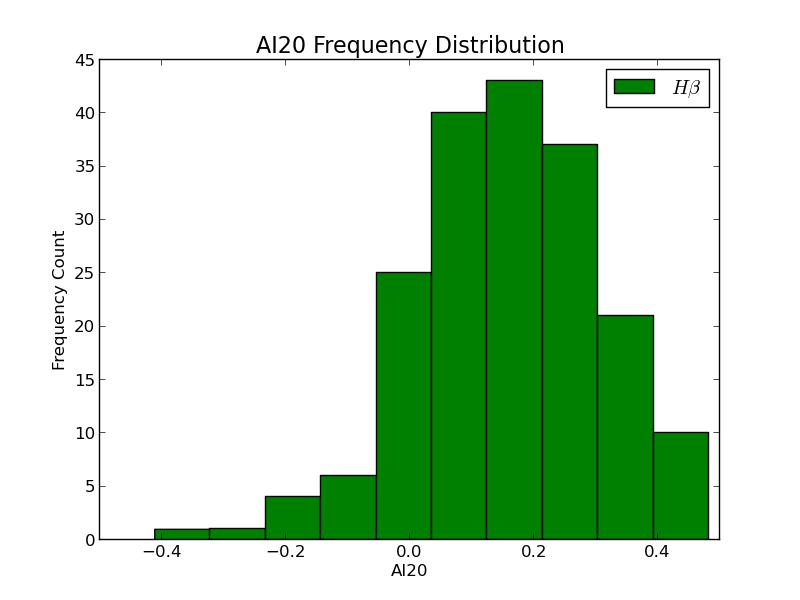} \\
\end{array}$
\end{center}
\caption[The Frequency Distribution of the AI20 of $H\alpha$ and $H\beta$ emission line profiles]
{Frequency Distribution of the AI20 of $H\alpha$ and $H\beta$ emission line profiles}
\label{fig:AI20}
\end{figure*}

Fig \ref{fig:AI20} shows the Asymmetry Index distribution of $H\alpha$ emission lines, on the left hand, and $H\beta$ emission lines,
in the right side at FW20\%I. It is noticed that the AI is almost symmetric and peaking in the red end for $H\alpha$ but still red-shifted for $H\beta$.

\begin{figure*}[!htbp]
\begin{center}$
\begin{array}{cc}
\includegraphics[width=3.0in]{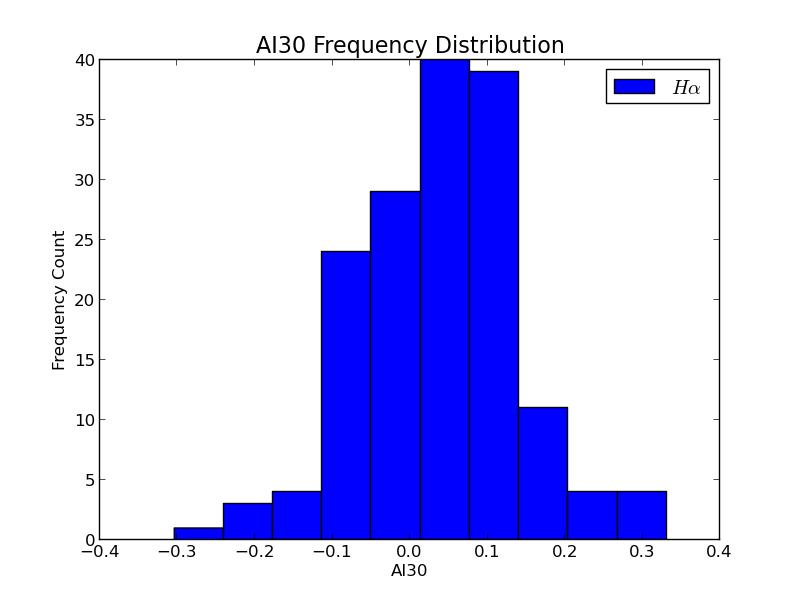} &
\includegraphics[width=3.0in]{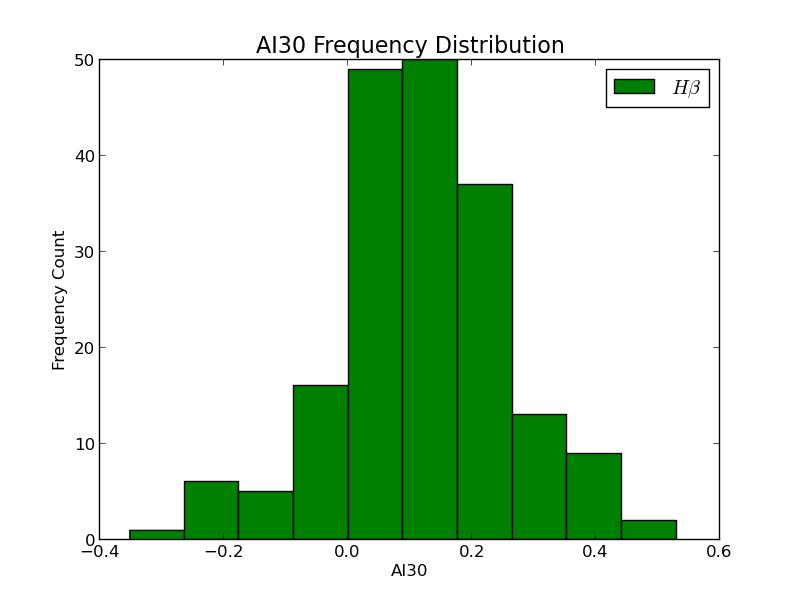} \\
\end{array}$
\end{center}
\caption[The Frequency Distribution of the AI30 of $H\alpha$ and $H\beta$ emission line profiles]
{Frequency Distribution of the AI30 of $H\alpha$ and $H\beta$ emission line profiles}
\label{fig:AI30}
\end{figure*}

Fig \ref{fig:AI30} shows the Asymmetry Index distribution of $H\alpha$ emission lines, on the left hand, and $H\beta$ emission lines,
in the right side at FW30\%I. It is noticed that the AI is almost symmetric for $H\alpha$ but still red-shifted for $H\beta$.

\begin{figure*}[!htbp]
\begin{center}$
\begin{array}{cc}
\includegraphics[width=3.0in]{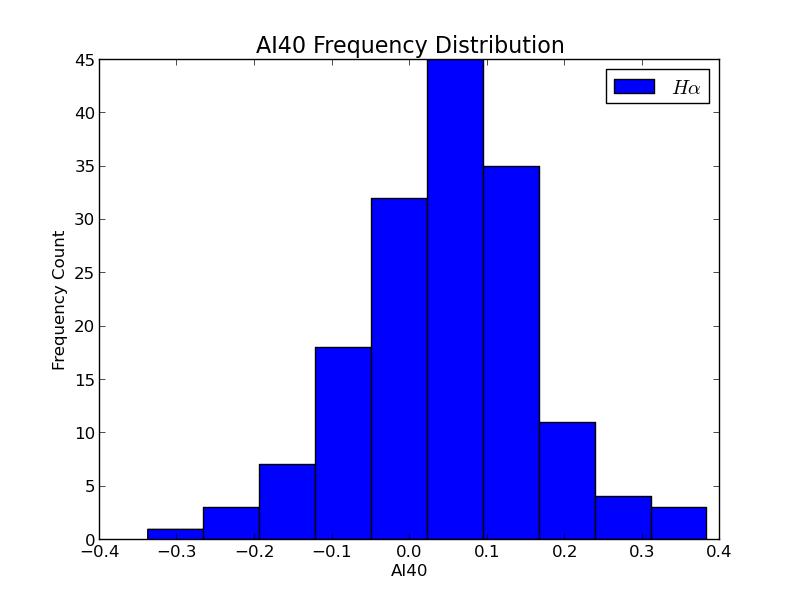} &
\includegraphics[width=3.0in]{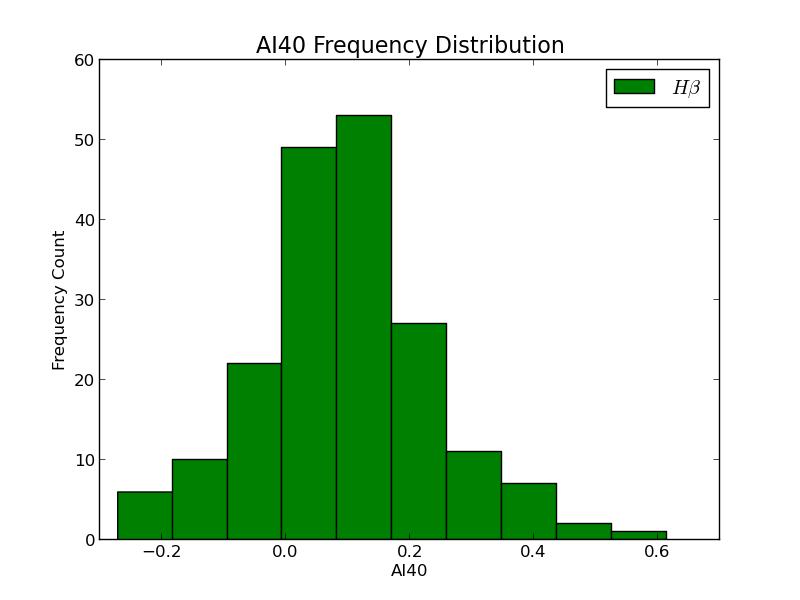} \\
\end{array}$
\end{center}
\caption[The Frequency Distribution of the AI40 of $H\alpha$ and $H\beta$ emission line profiles]
{Frequency Distribution of the AI40 of $H\alpha$ and $H\beta$ emission line profiles}
\label{fig:AI40}
\end{figure*}

Fig \ref{fig:AI40} shows the Asymmetry Index distribution of $H\alpha$ emission lines, on the left hand, and $H\beta$ emission lines,
in the right side at FW40\%I. It is noticed that the AI is almost symmetric for $H\alpha$ but still red-shifted for $H\beta$.

\begin{figure*}[!htbp]
\begin{center}$
\begin{array}{cc}
\includegraphics[width=3.0in]{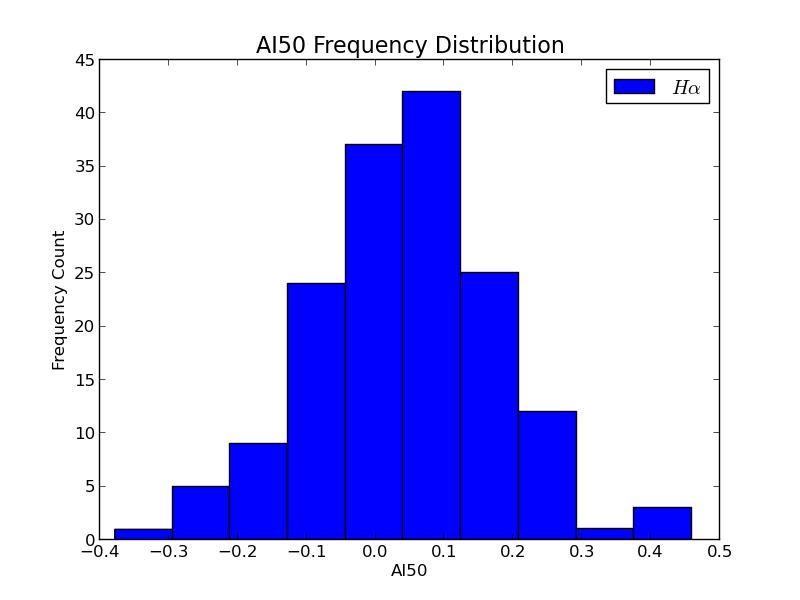} &
\includegraphics[width=3.0in]{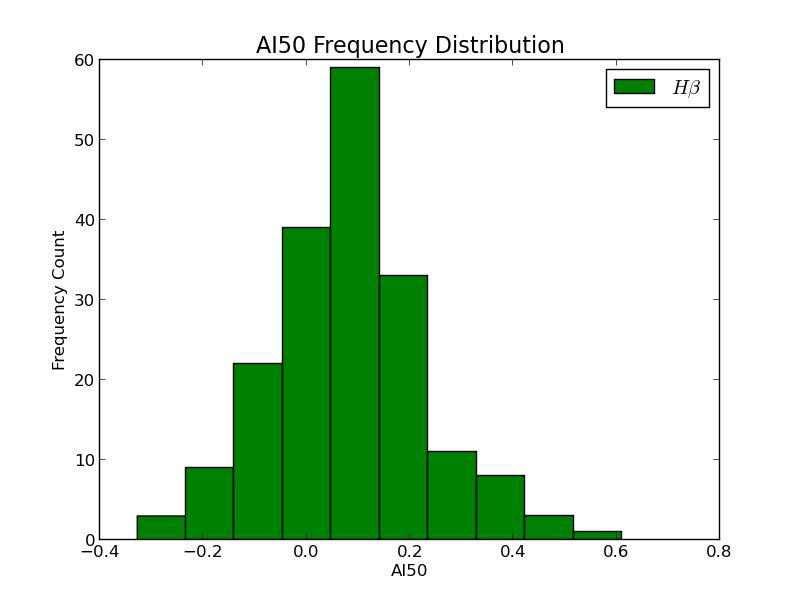} \\
\end{array}$
\end{center}
\caption[The Frequency Distribution of the AI50 of $H\alpha$ and $H\beta$ emission line profiles]
{Frequency Distribution of the AI50 of $H\alpha$ and $H\beta$ emission line profiles}
\label{fig:AI50}
\end{figure*}

Fig \ref{fig:AI50} shows the Asymmetry Index distribution of $H\alpha$ emission lines, on the left hand, and $H\beta$ emission lines,
in the right side at FW50\%I. It is noticed that the AI is almost symmetric for $H\alpha$ but still red-shifted for $H\beta$.

\begin{figure*}[!htbp]
\begin{center}$
\begin{array}{cc}
\includegraphics[width=3.0in]{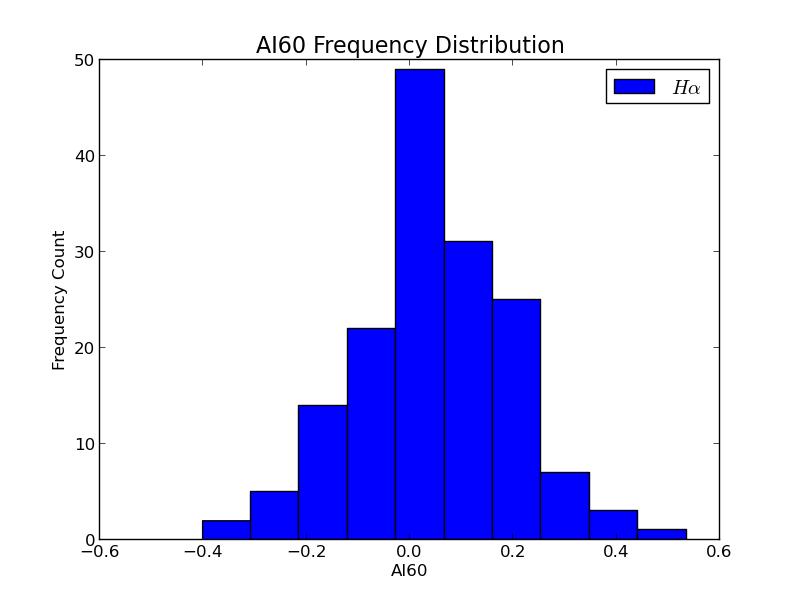} &
\includegraphics[width=3.0in]{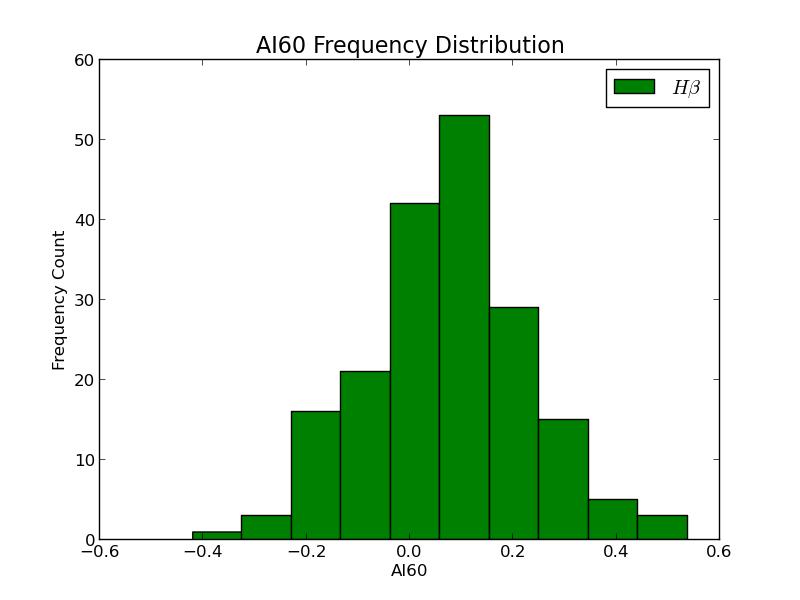} \\
\end{array}$
\end{center}
\caption[The Frequency Distribution of the AI60 of $H\alpha$ and $H\beta$ emission line profiles]
{Frequency Distribution of the AI60 of $H\alpha$ and $H\beta$ emission line profiles}
\label{fig:AI60}
\end{figure*}

Fig \ref{fig:AI60} shows the Asymmetry Index distribution of $H\alpha$ emission lines, on the left hand, and $H\beta$ emission lines,
in the right side at FW60\%I. It is noticed that the AI is almost symmetric for $H\alpha$ but still red-shifted for $H\beta$.

\begin{figure*}[!htbp]
\begin{center}$
\begin{array}{cc}
\includegraphics[width=3.0in]{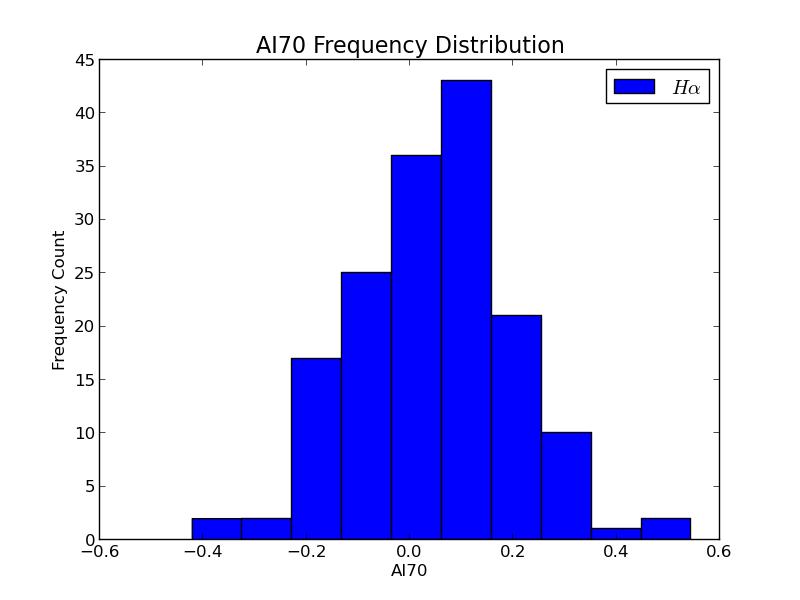} &
\includegraphics[width=3.0in]{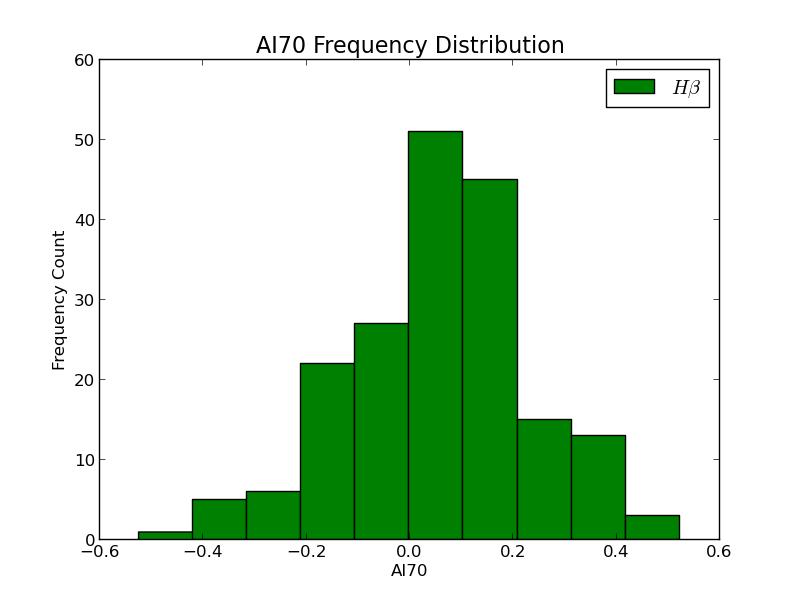} \\
\end{array}$
\end{center}
\caption[The Frequency Distribution of the AI70 of $H\alpha$ and $H\beta$ emission line profiles]
{Frequency Distribution of the AI70 of $H\alpha$ and $H\beta$ emission line profiles}
\label{fig:AI70}
\end{figure*}

Fig \ref{fig:AI70} shows the Asymmetry Index distribution of $H\alpha$ emission lines, on the left hand, and $H\beta$ emission lines,
in the right side at FW70\%I. It is noticed that the AI is almost symmetric for $H\alpha$ but still red-shifted for $H\beta$.

A few key features are noticed here, $H\alpha$ profiles are almost symmetric, they tend to peak in the red end but the degree of asymmetry
is negligible. For the $H\beta$ profiles, it is noticed that most of them are clearly asymmetric, peaking predominantly in the positive side.
This is maintained all through the profile from the base, FWZI, to the core of the profile at the higher percentiles.

\begin{figure*}[!htbp]
\begin{center}$
\begin{array}{ccc}
\includegraphics[width=0.32\textwidth]{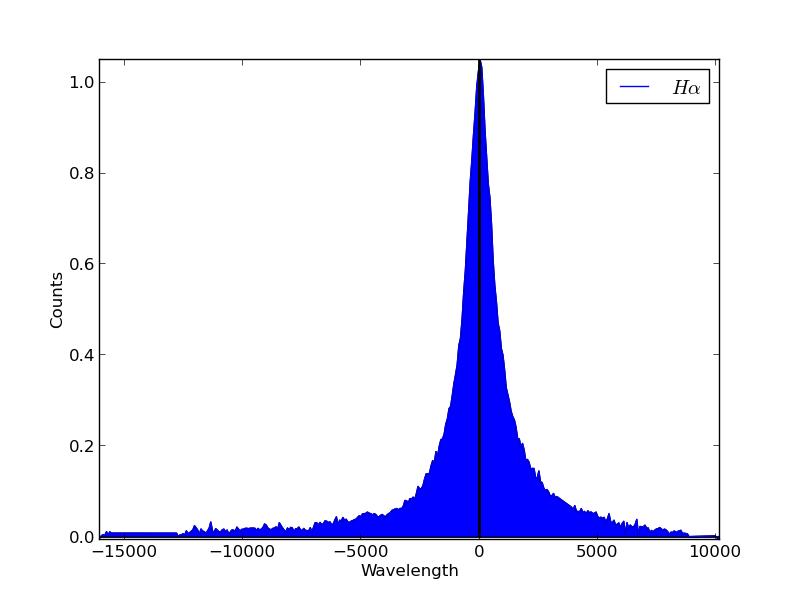} &
\includegraphics[width=0.32\textwidth]{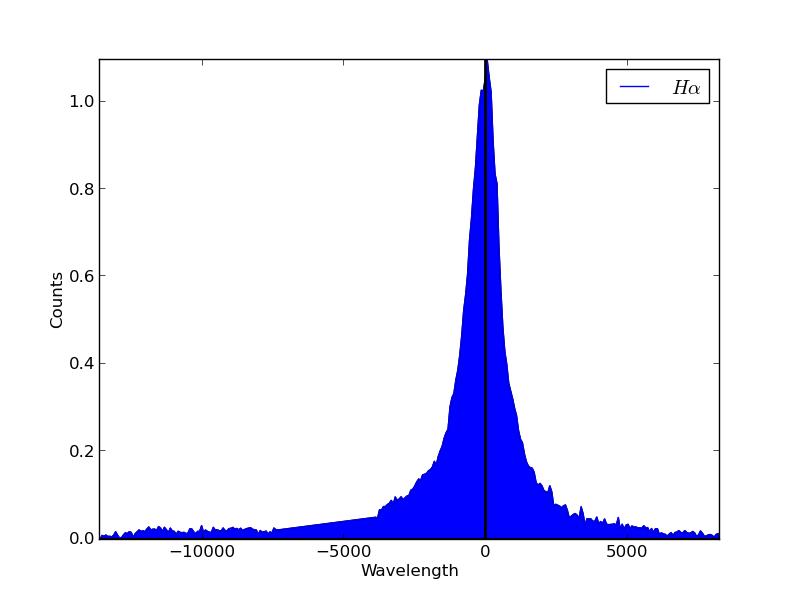} &
\includegraphics[width=0.32\textwidth]{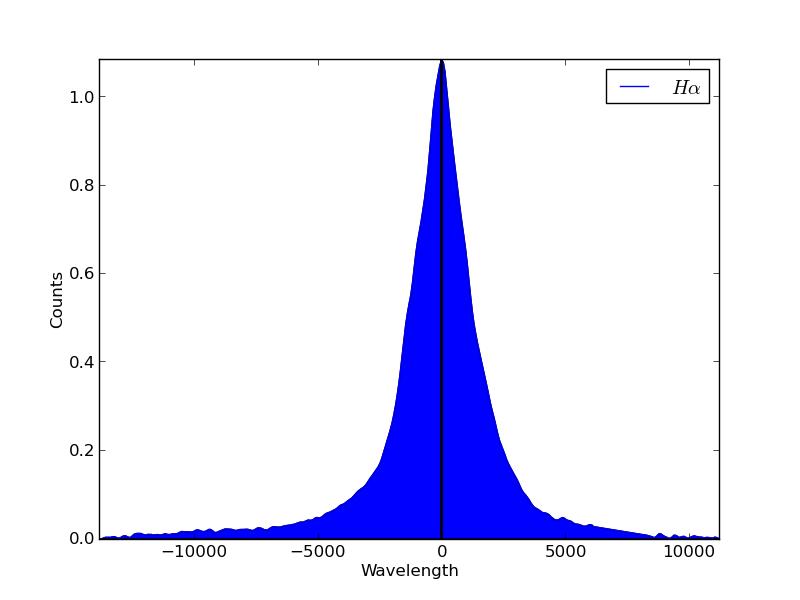} \\
\includegraphics[width=0.32\textwidth]{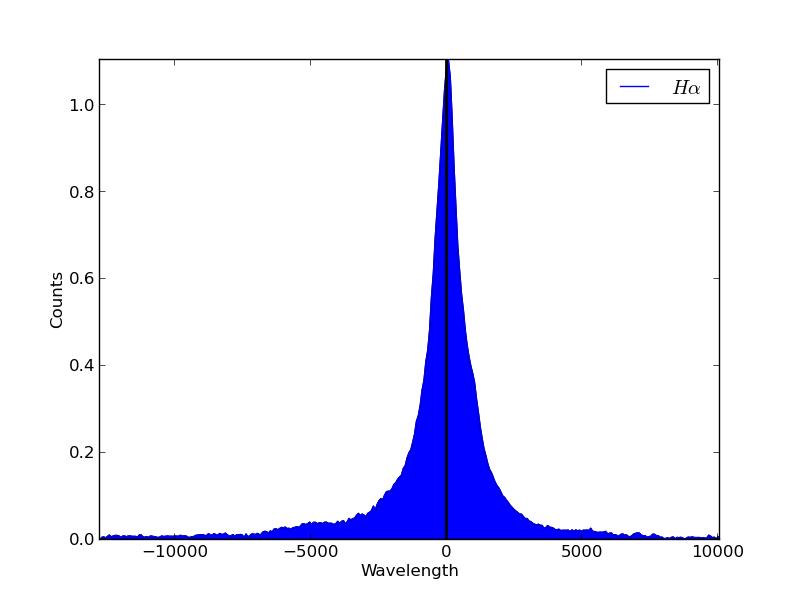} &
\includegraphics[width=0.32\textwidth]{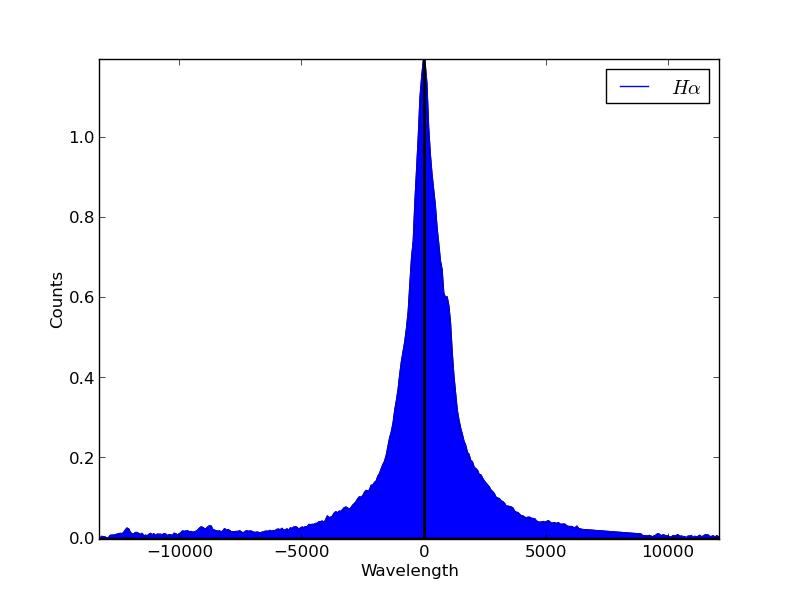} &
\includegraphics[width=0.32\textwidth]{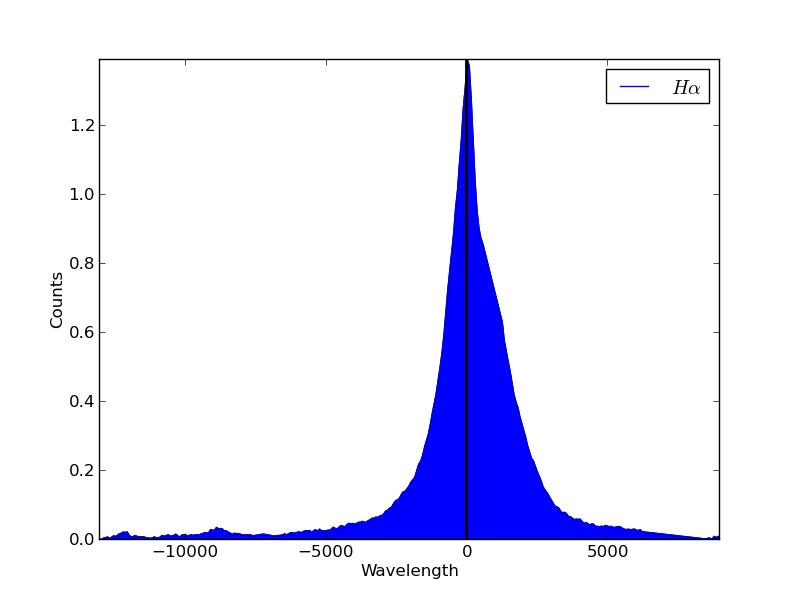} \\
\end{array}$
\end{center}
\caption[Some examples of symmetric $H\alpha$ line profiles]
{Some examples of symmetric $H\alpha$ line profiles}
\label{fig:ha_slp}
\end{figure*}

\begin{figure*}[!htbp]
\begin{center}$
\begin{array}{ccc}
\includegraphics[width=0.32\textwidth]{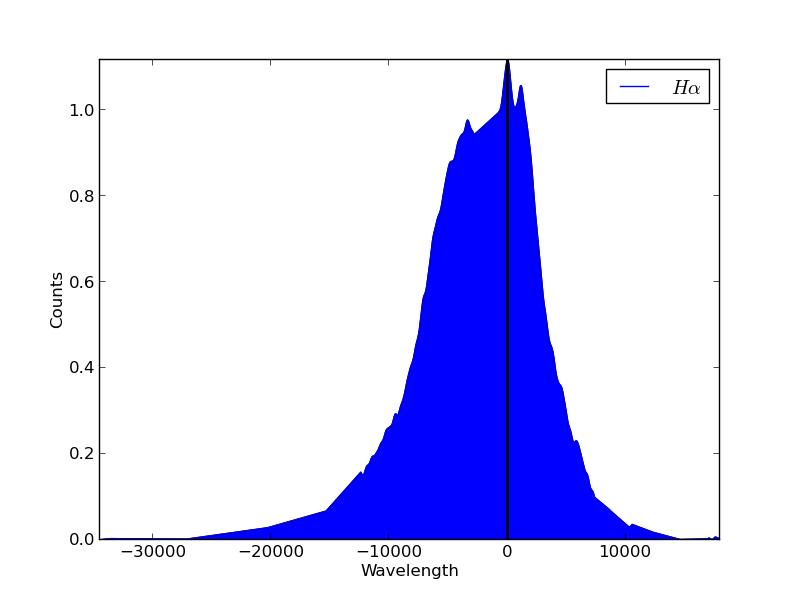} &
\includegraphics[width=0.32\textwidth]{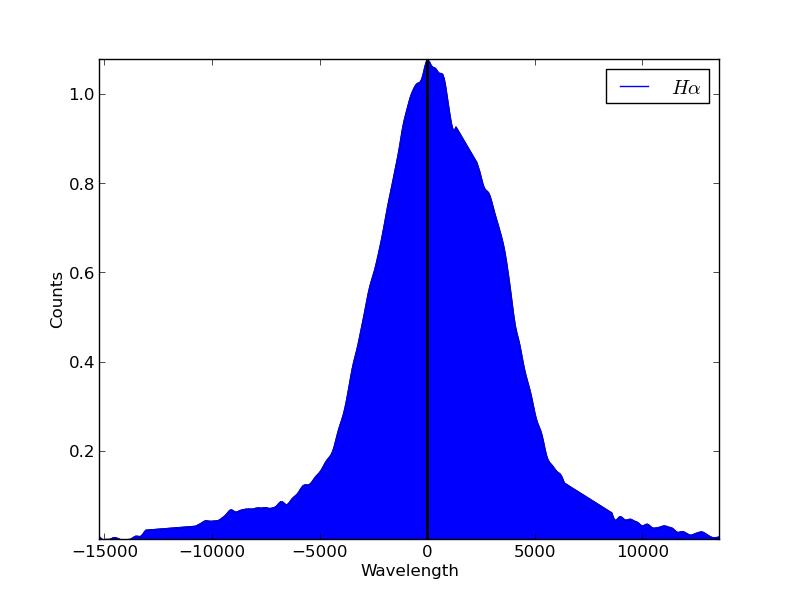} &
\includegraphics[width=0.32\textwidth]{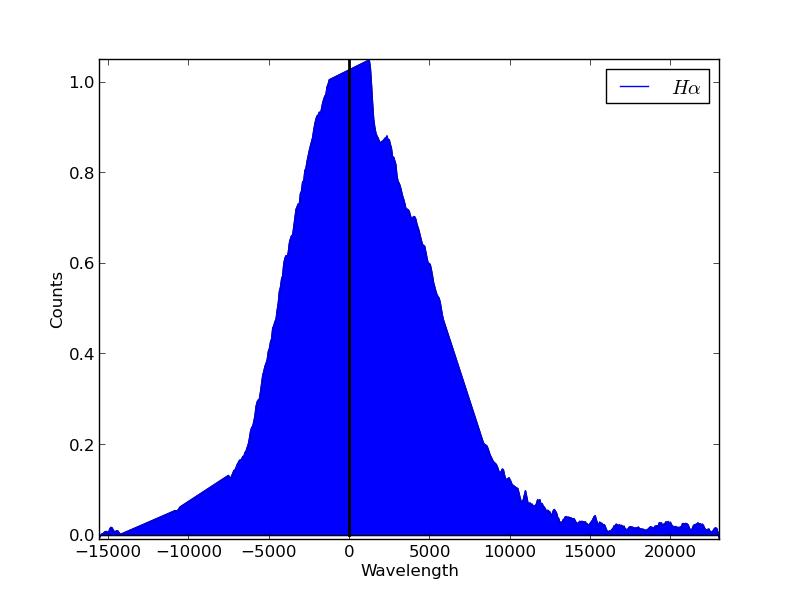} \\
\includegraphics[width=0.32\textwidth]{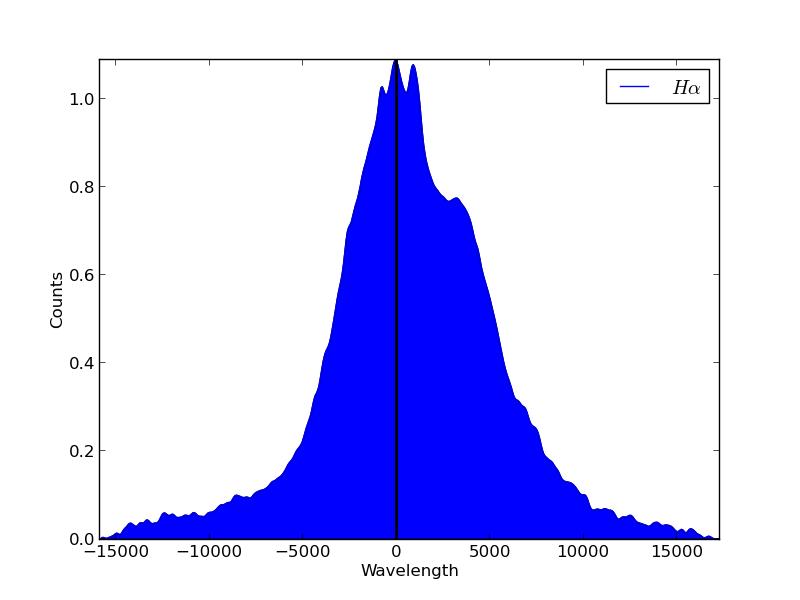} &
\includegraphics[width=0.32\textwidth]{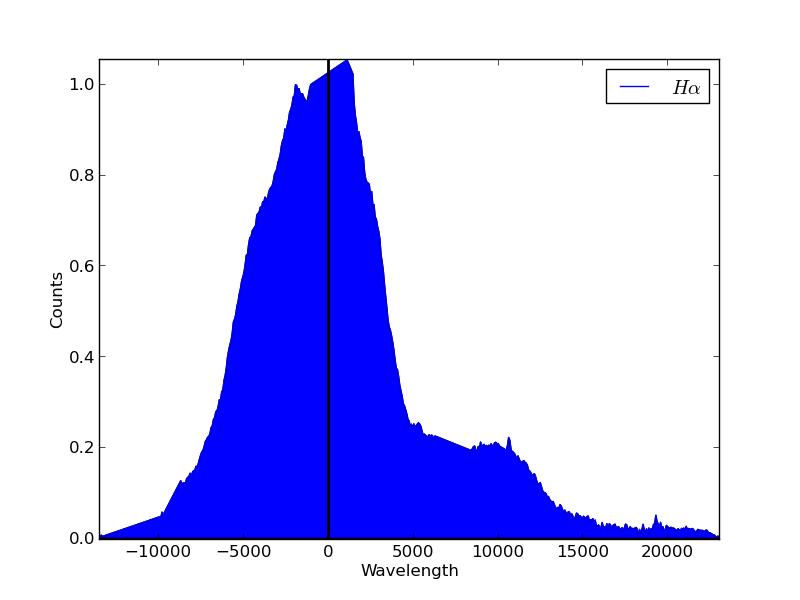} &
\includegraphics[width=0.32\textwidth]{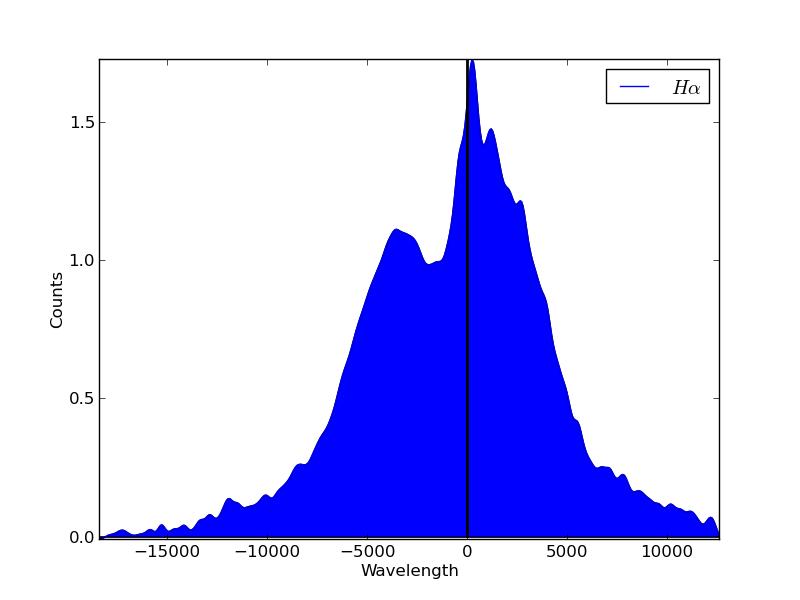} \\
\end{array}$
\end{center}
\caption[Some examples of asymmetric $H\alpha$ line profiles]
{Some examples of asymmetric $H\alpha$ line profiles}
\label{fig:ha_aslp}
\end{figure*}

\begin{figure*}[!htbp]
\begin{center}$
\begin{array}{ccc}
\includegraphics[width=0.32\textwidth]{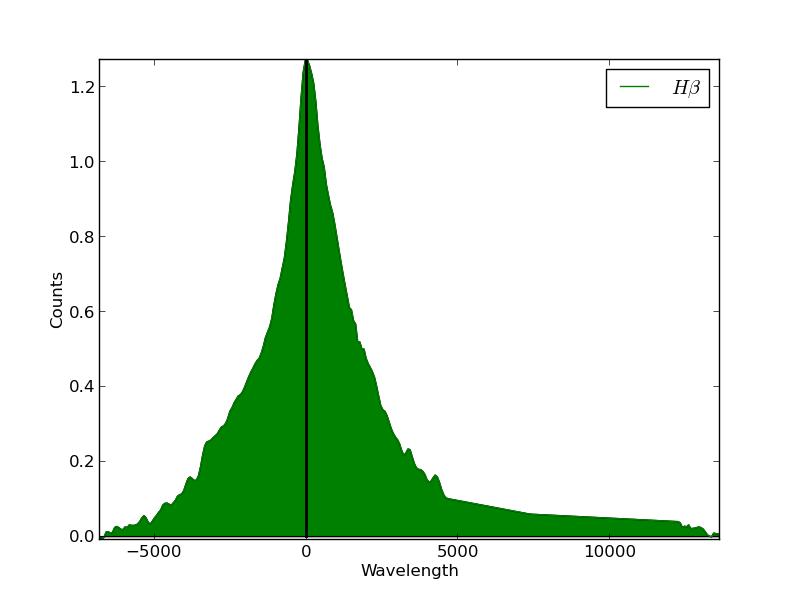} &
\includegraphics[width=0.32\textwidth]{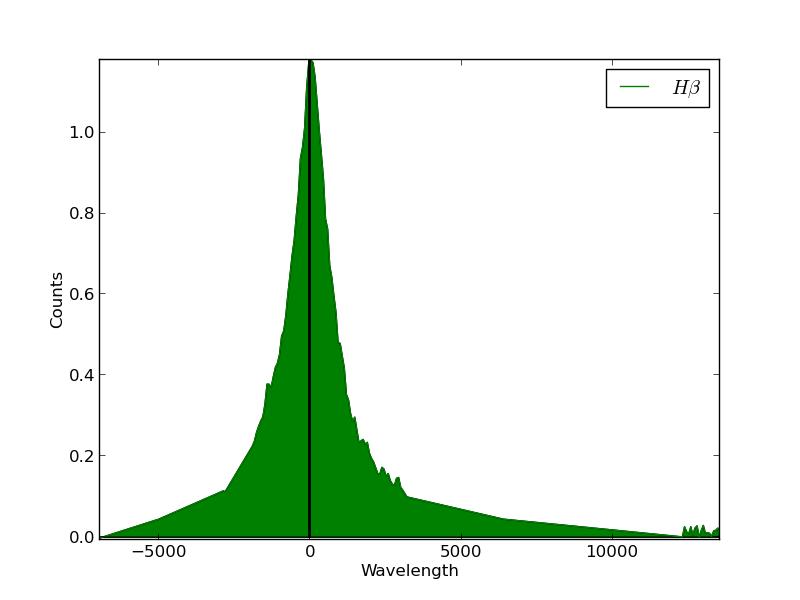} &
\includegraphics[width=0.32\textwidth]{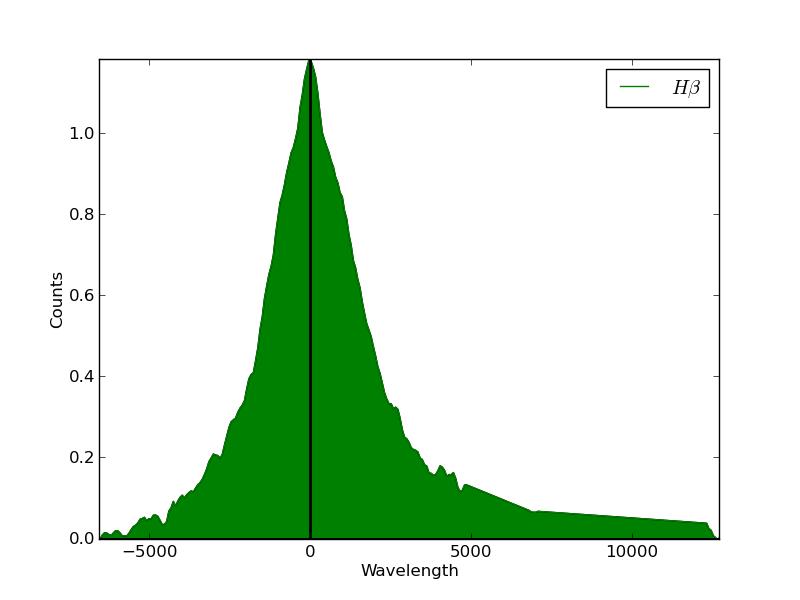} \\
\includegraphics[width=0.32\textwidth]{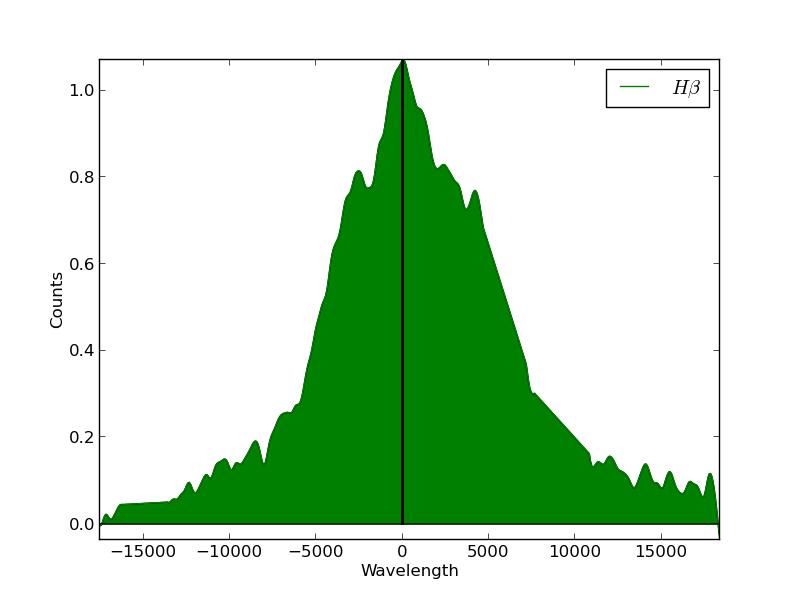} &
\includegraphics[width=0.32\textwidth]{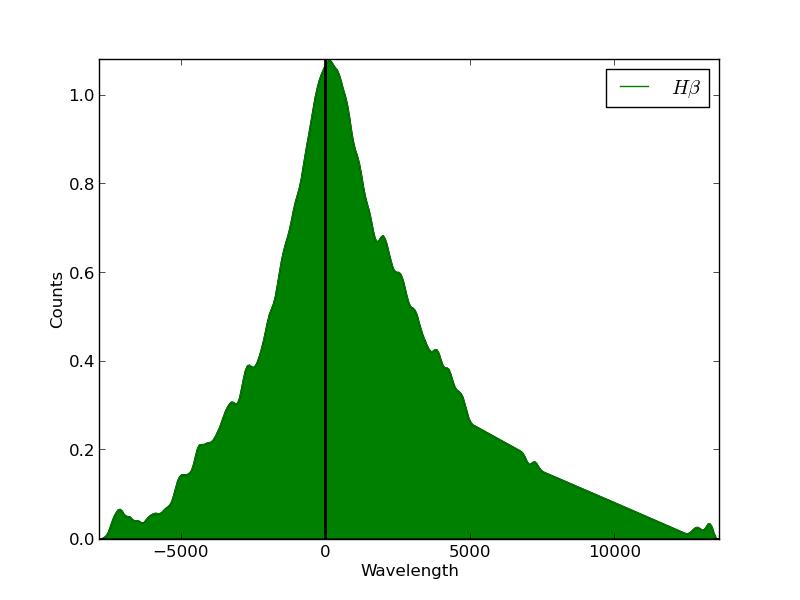} &
\includegraphics[width=0.32\textwidth]{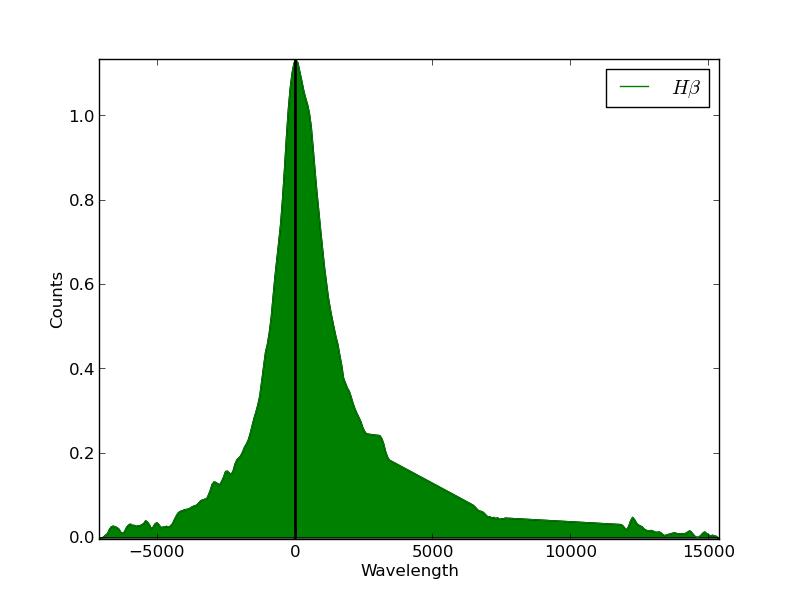} \\
\end{array}$
\end{center}
\caption[Some examples of symmetric $H\beta$ line profiles]
{Some examples of symmetric $H\beta$ line profiles}
\label{fig:hb_slp}
\end{figure*}

A sample of symmetric profiles can be seen in figures \ref{fig:ha_slp} for $H\alpha$ profiles and \ref{fig:hb_slp} for the $H\beta$ profiles.
Asymmetric profiles can be seen in figures \ref{fig:ha_aslp} and \ref{fig:ha_aslp} for the  $H\alpha$ and $H\beta$ emission lines respectively.
It is evident that the $H\beta$ profiles display the highest degree of asymmetry.

%\subsubsection*{$H\alpha$ and $H\beta$ line profiles}

\begin{figure*}[!htbp]
\begin{center}$
\begin{array}{ccc}
\includegraphics[width=0.32\textwidth]{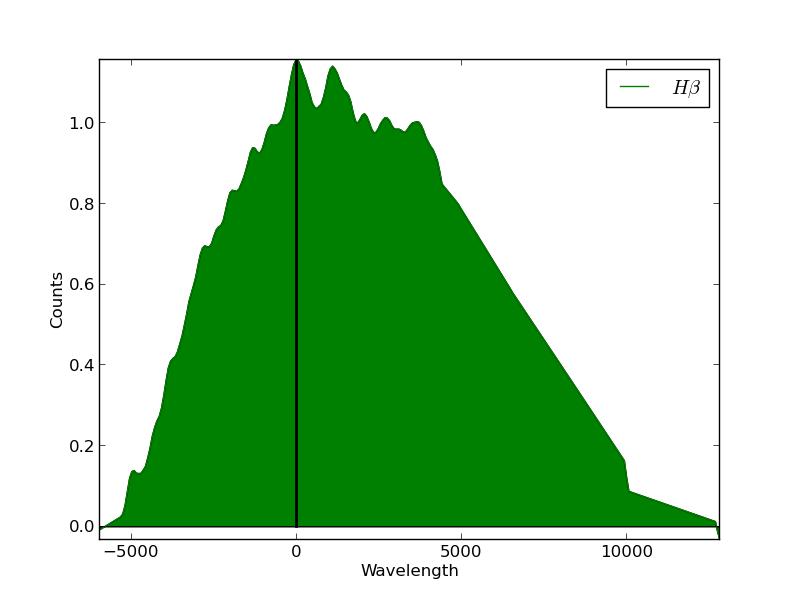} &
\includegraphics[width=0.32\textwidth]{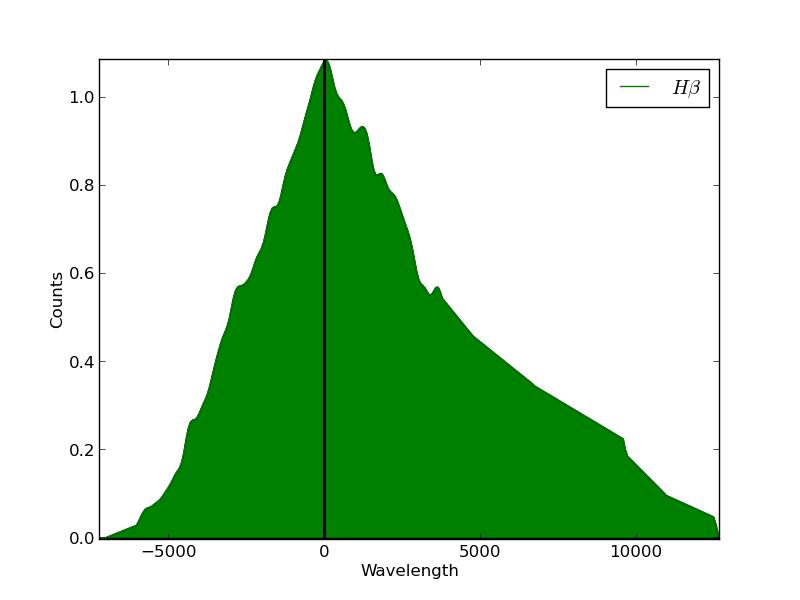} &
\includegraphics[width=0.32\textwidth]{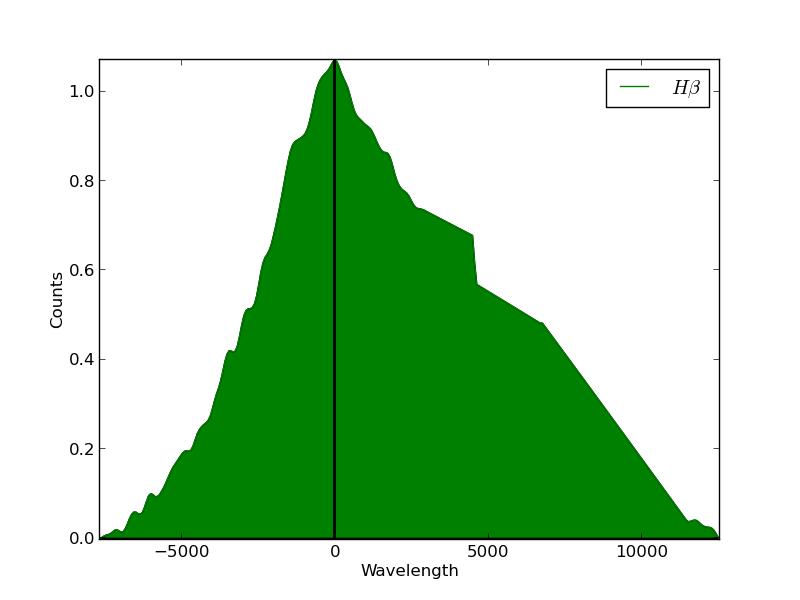} \\
\includegraphics[width=0.32\textwidth]{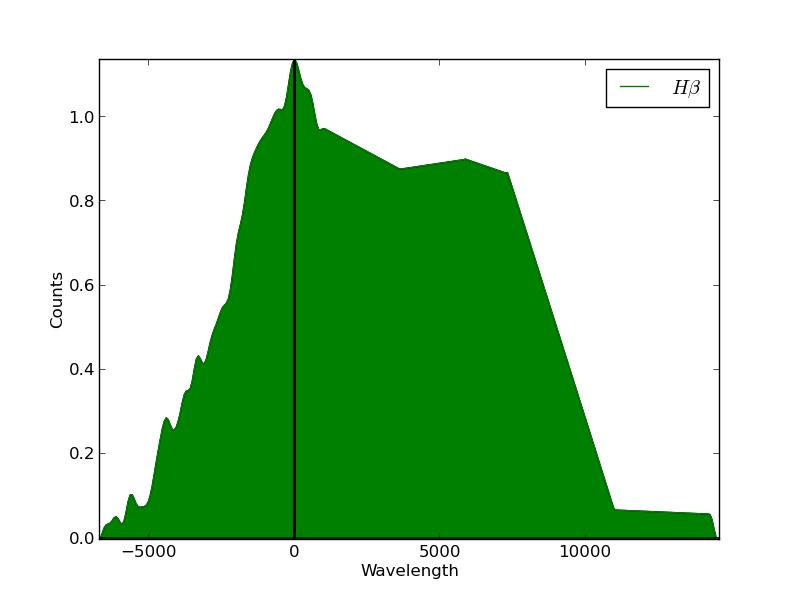} &
\includegraphics[width=0.32\textwidth]{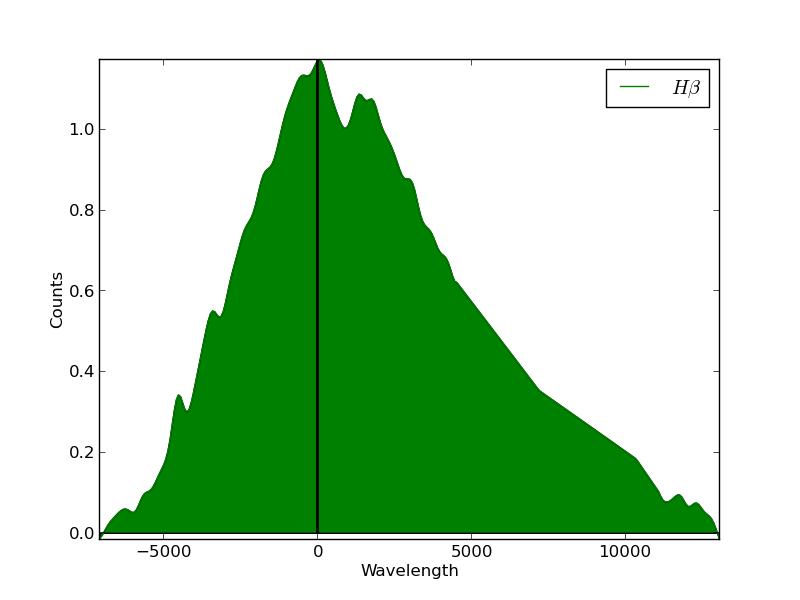} &
\includegraphics[width=0.32\textwidth]{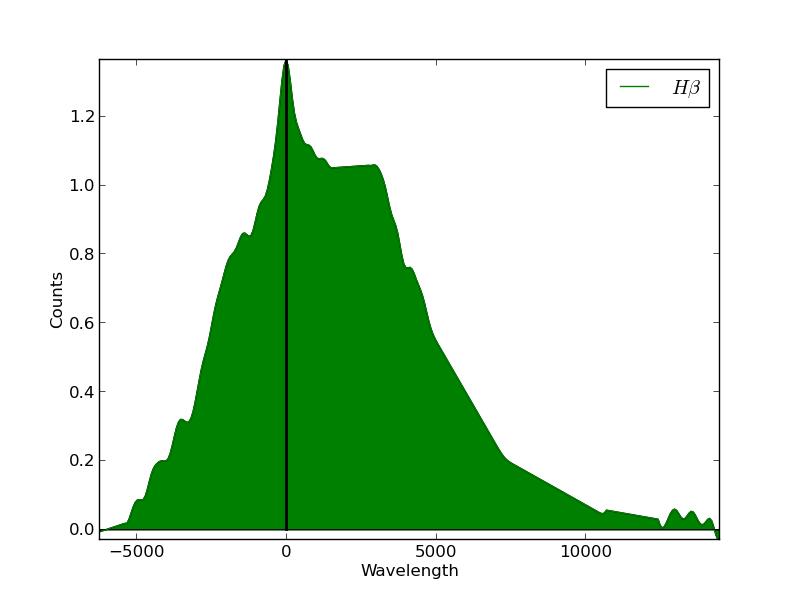} \\
\end{array}$
\end{center}
\caption[Some examples of asymmetric $H\beta$ line profiles]
{Some examples of asymmetric $H\beta$ line profiles}
\label{fig:hb_aslp}
\end{figure*}

\begin{figure*}[!htbp]
\begin{center}$
\begin{array}{ccc}
\includegraphics[width=0.32\textwidth]{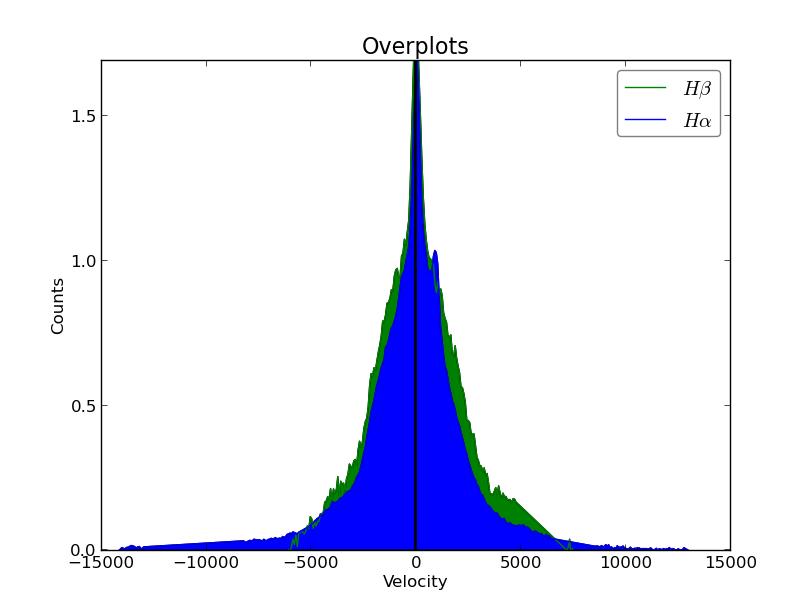} &
\includegraphics[width=0.32\textwidth]{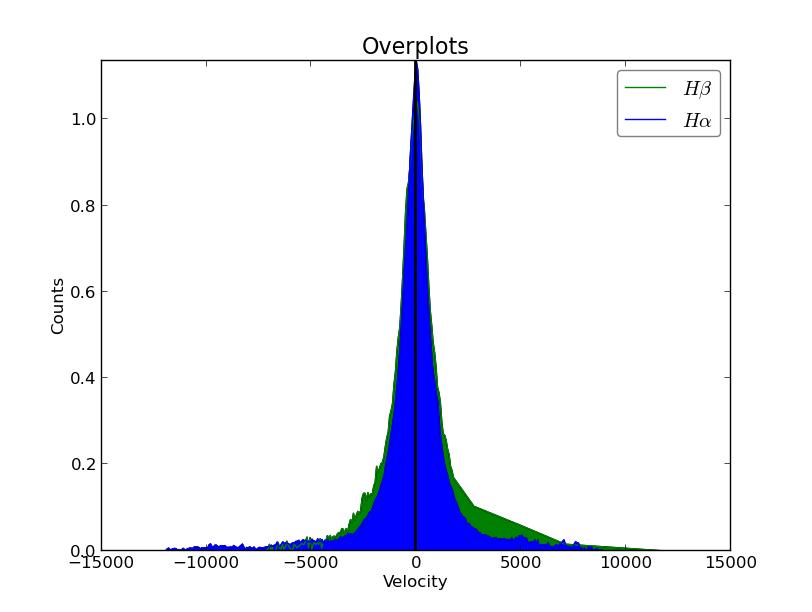} &
\includegraphics[width=0.32\textwidth]{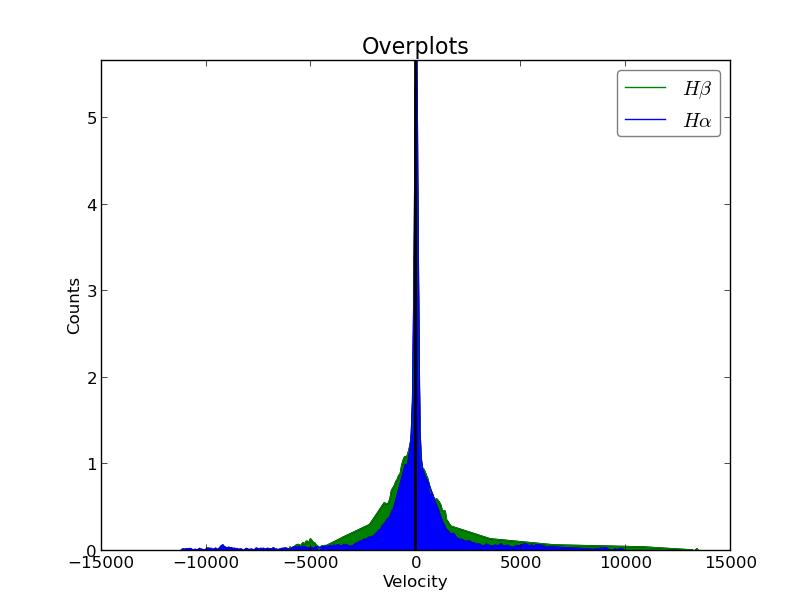} \\
\includegraphics[width=0.32\textwidth]{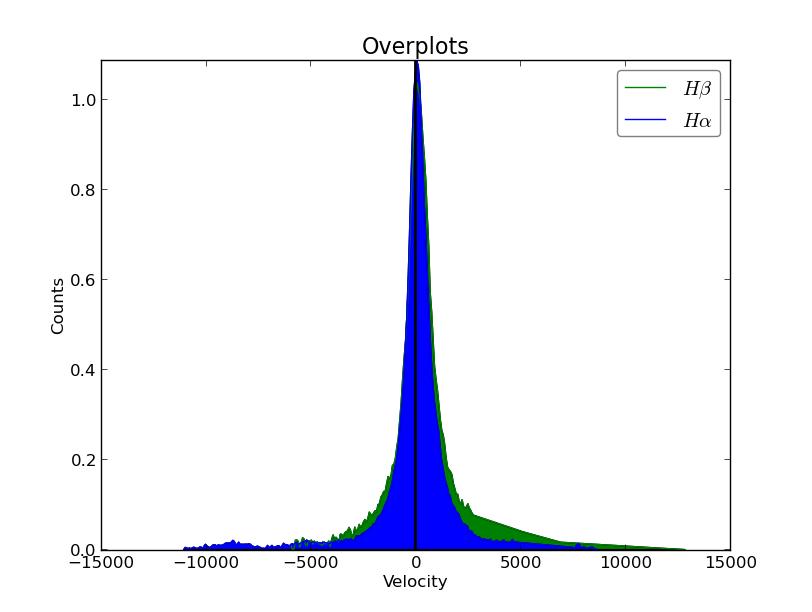} &
\includegraphics[width=0.32\textwidth]{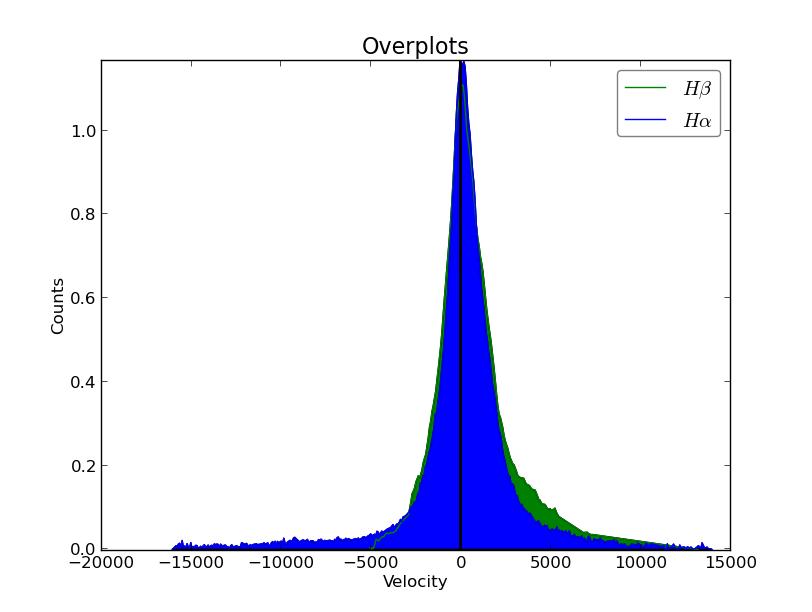} &
\includegraphics[width=0.32\textwidth]{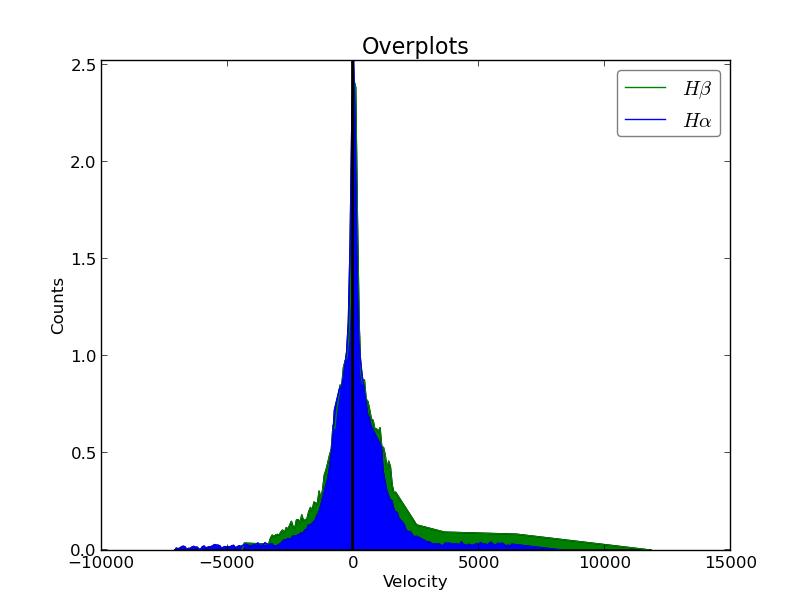} \\
\end{array}$
\end{center}
\caption[Some examples of symmetric $H\alpha$ and $H\beta$ line profiles overploted]
{Some examples of symmetric $H\alpha$ and $H\beta$ line profiles overploted}
\label{fig:ha_soverplots}
\end{figure*}

\begin{figure*}[!htbp]
\begin{center}$
\begin{array}{ccc}
\includegraphics[width=0.32\textwidth]{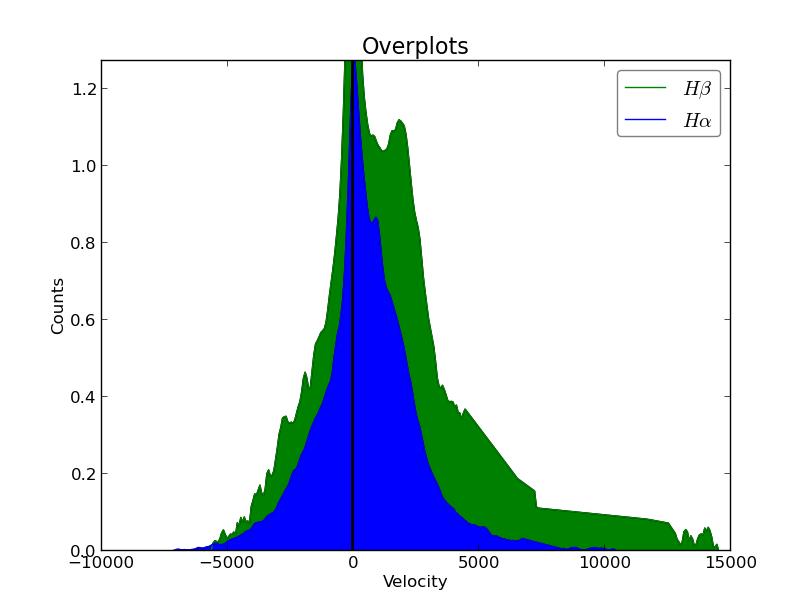} &
\includegraphics[width=0.32\textwidth]{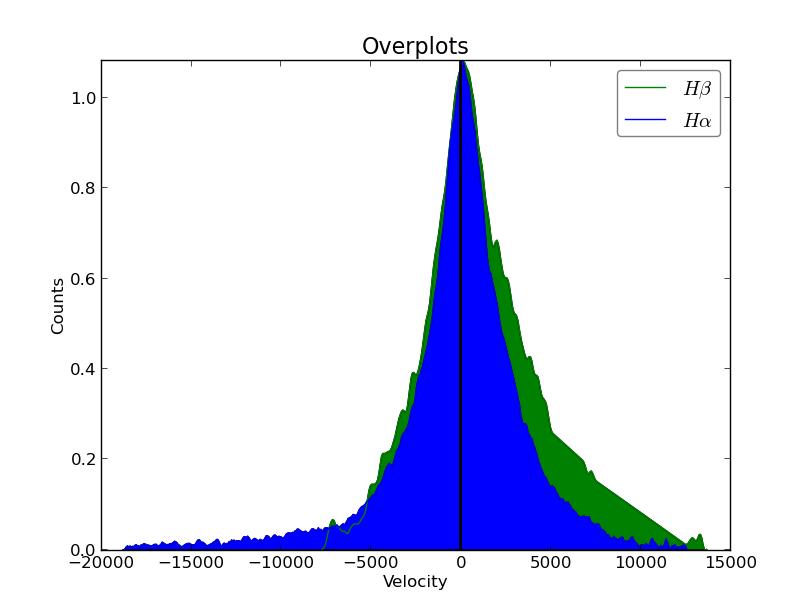} &
\includegraphics[width=0.32\textwidth]{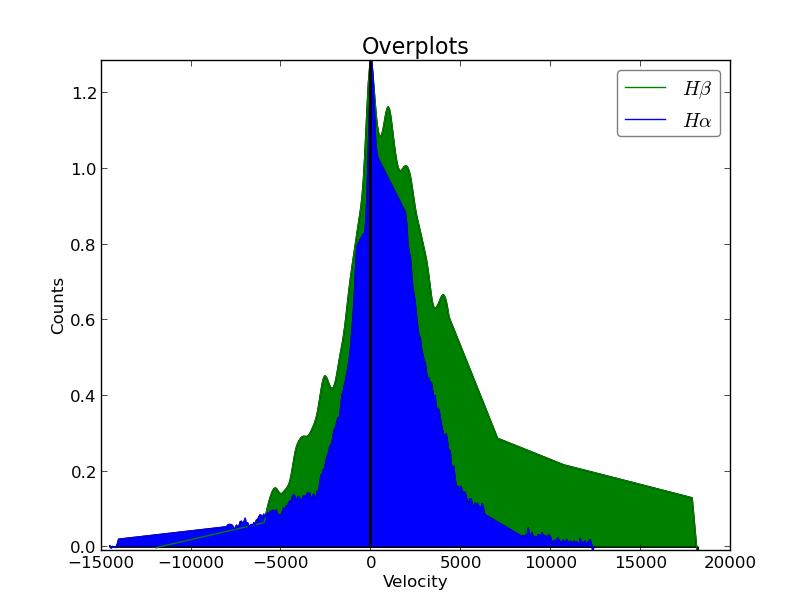} \\
\includegraphics[width=0.32\textwidth]{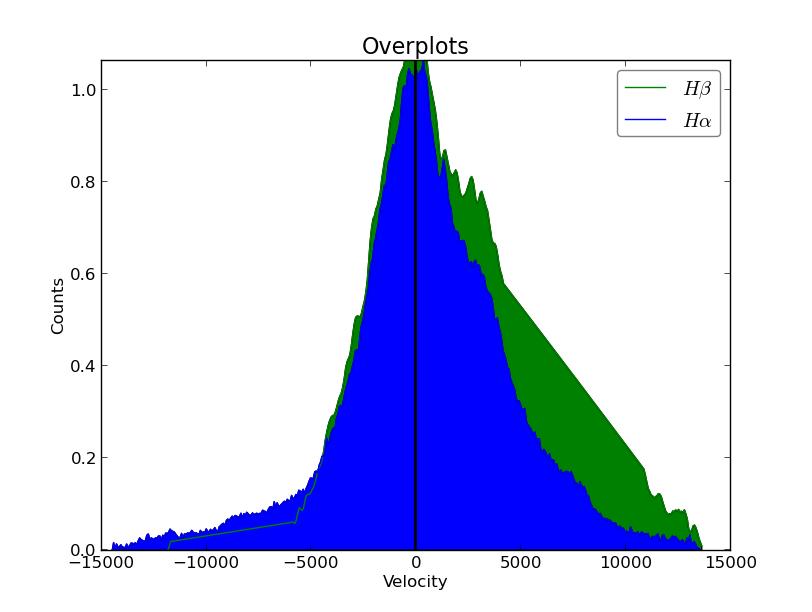} &
\includegraphics[width=0.32\textwidth]{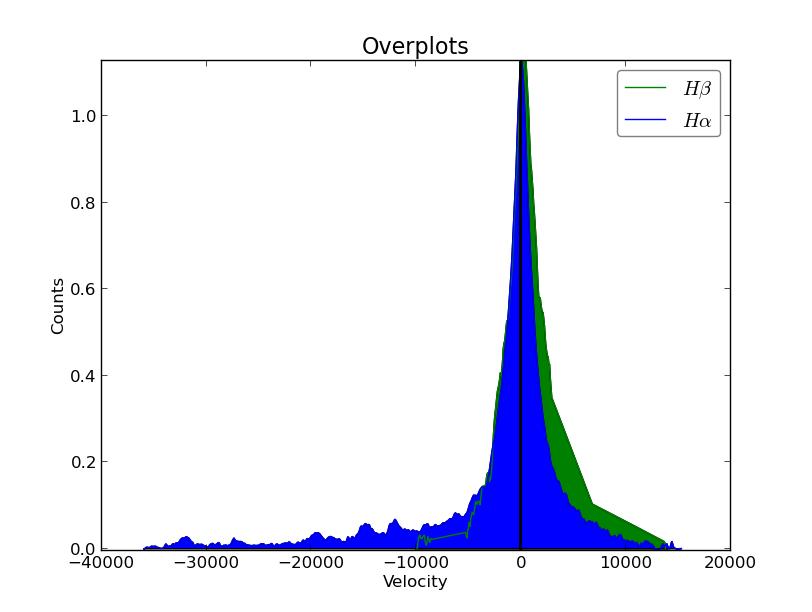} &
\includegraphics[width=0.32\textwidth]{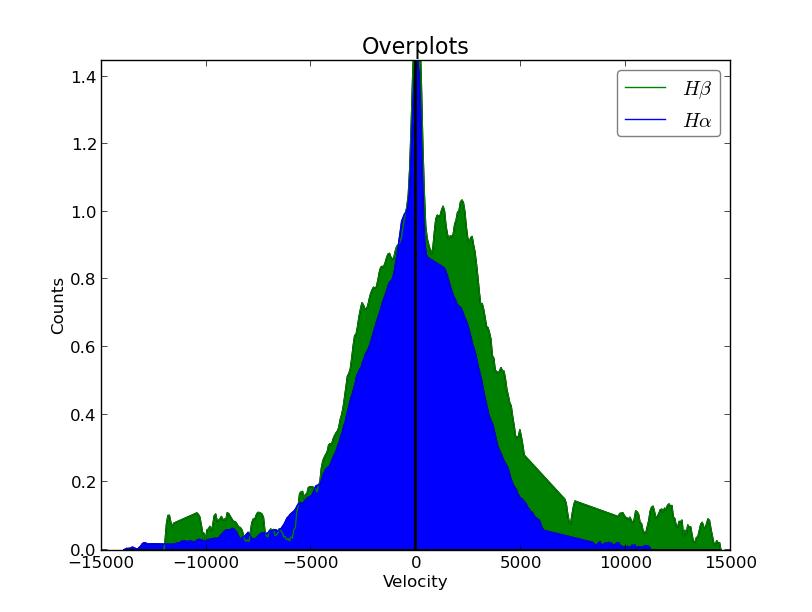} \\
\end{array}$
\end{center}
\caption[Some examples of asymmetric $H\alpha$ and $H\beta$ line profiles overploted]
{Some examples of asymmetric $H\alpha$ and $H\beta$ line profiles overploted}
\label{fig:ha_asoverplots}
\end{figure*}

\subsection{Kurtosis Index}
The Kurtosis Index is a measure of how steep the profile is. The Kurtosis parameters are all less than unity, with a decrease in value
indicating an increase in steepness. The parameters will range from 0.0 to 1.0, but of course with very few profiles having values
close to 1.0. \citep{Whittle}\\
Table \ref{tab:ki0} shows an overview of the Kurtosis parameters and their meaning.

\begin{table*}[!htbp]
\caption[Kurtosis Index]{Interpretation of Kurtosis Index}
\label{tab:ki0}
\begin{center}
\begin{tabular}{|c|l|l|}
\hline
  Kurtosis Index $(x)$ & Measure & Interpretation \\ \hline
  $0 \textless x \leq 0.25$ & Strong & The change in shape is very large. \\ \hline
  $0.25 \textless x \leq 0.65$ & Moderate & The change in shape is normal.  \\ \hline
  $x \textgreater 0.65$ & Weak & The change in shape in insignificant. \\ \hline
\end{tabular}
\end{center}
\end{table*}

For the Kurtosis parameters, it is vital to note that this is a measure of how the profile shape changes from the top to the bottom. A value 
of close to unity signifies almost no change in the steepness. A value close to zero on the
other hand will mean a very large change in the shape as one moves towards the base of the profile. This means the profile shape rapidly changes
from narrow to broad. 
Some profiles may be broad but with low values of Kurtosis, that is if they are consistently broad from the upper part of the emission line to below
the half maximum. Profiles that show this shape are mostly those in which the narrow component is not available or obscured.
There are other broad emission line profiles that will display high values of Kurtosis, that is, having significantly extended wings.
It is also noted that high values of asymmetry in a profile suffice low values in Kurtosis.\\

In dealing with the Kurtosis measure, three regions of the profile are chosen, the top measured with $KI_1$, the middle, measured with $KI_2$, 
and the bottom, measured with $KI_3$. Thus each of these values shows a change in profile in the mentioned regions on the overall profile. To capture
the whole profile, another parameter is defined, $KI_{Gen}$, which measures the overall change in shape of the profile from the top to the bottom.

\begin{figure*}[!htbp]
\begin{center}$
\begin{array}{cc}
\includegraphics[width=3.0in]{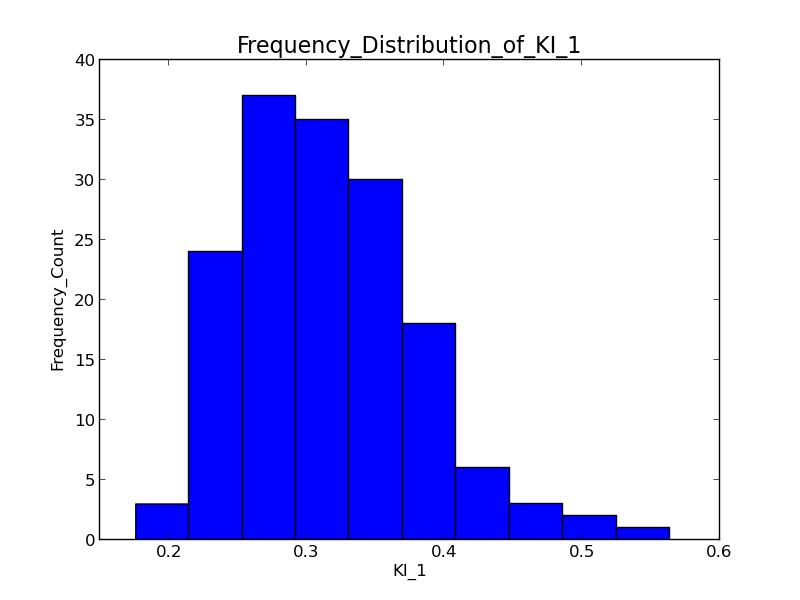} &
\includegraphics[width=3.0in]{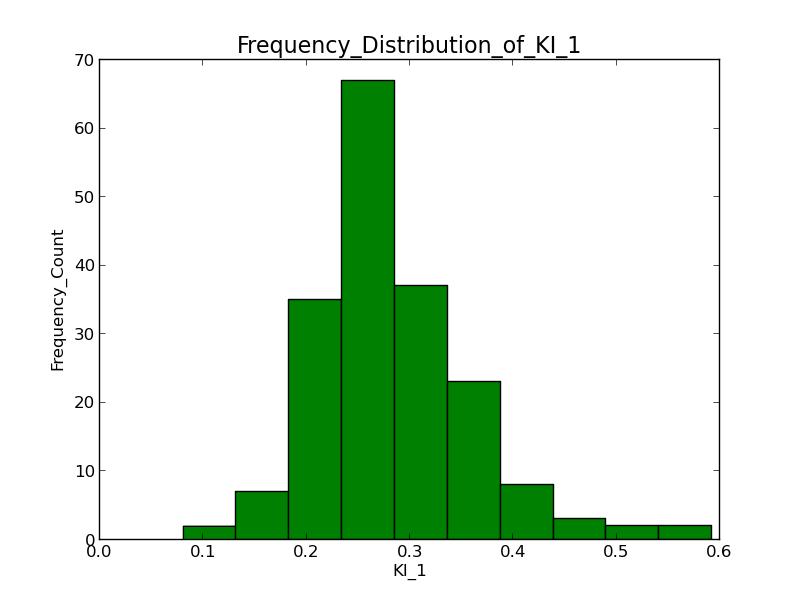} \\
\end{array}$
\end{center}
\caption[The Frequency Distribution of the KI 1 of $H\alpha$ and $H\beta$ emission line profiles]
{Frequency Distribution of the KI 1 of $H\alpha$ and $H\beta$ emission line profiles}
\label{fig:KI_1}
\end{figure*}

\begin{figure*}[!htbp]
\begin{center}$
\begin{array}{cc}
\includegraphics[width=3.0in]{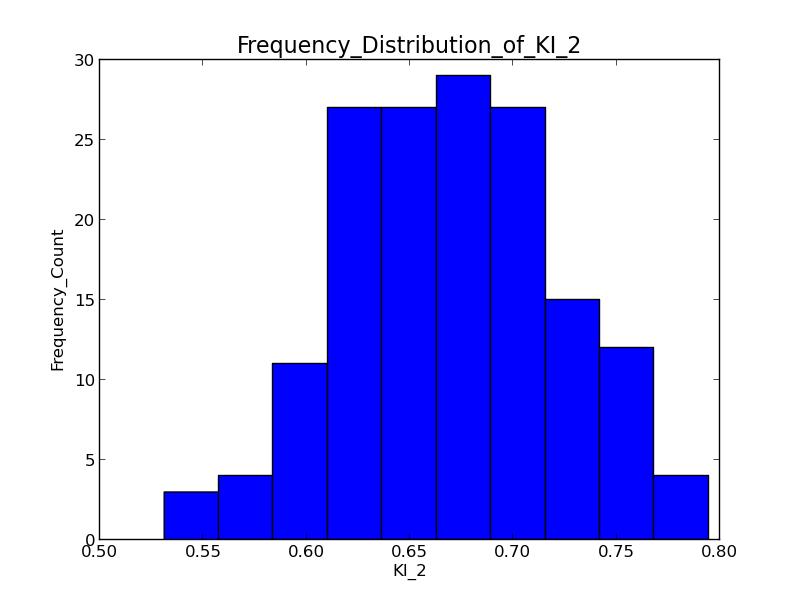} &
\includegraphics[width=3.0in]{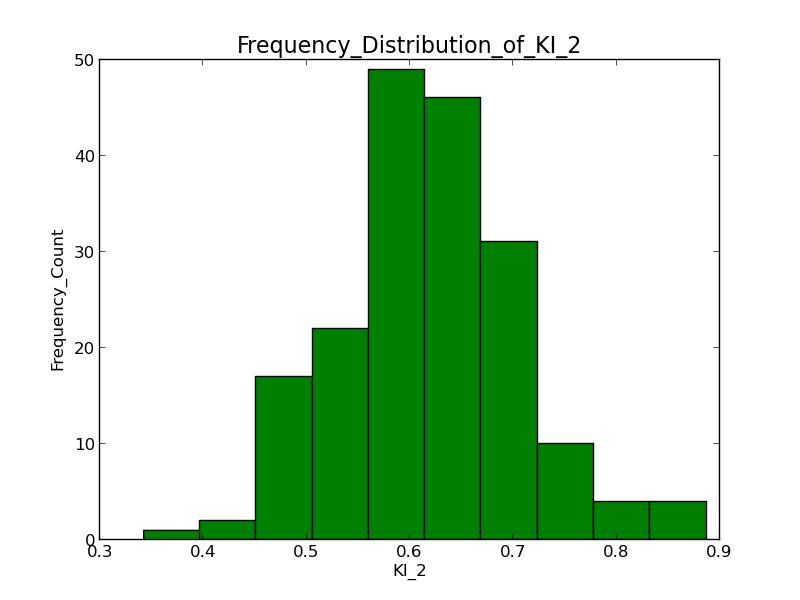} \\
\end{array}$
\end{center}
\caption[The Frequency Distribution of the KI 2 of $H\alpha$ and $H\beta$ emission line profiles]
{Frequency Distribution of the KI 2 of $H\alpha$ and $H\beta$ emission line profiles}
\label{fig:KI_2}
\end{figure*}

\begin{figure*}[!htbp]
\begin{center}$
\begin{array}{cc}
\includegraphics[width=3.0in]{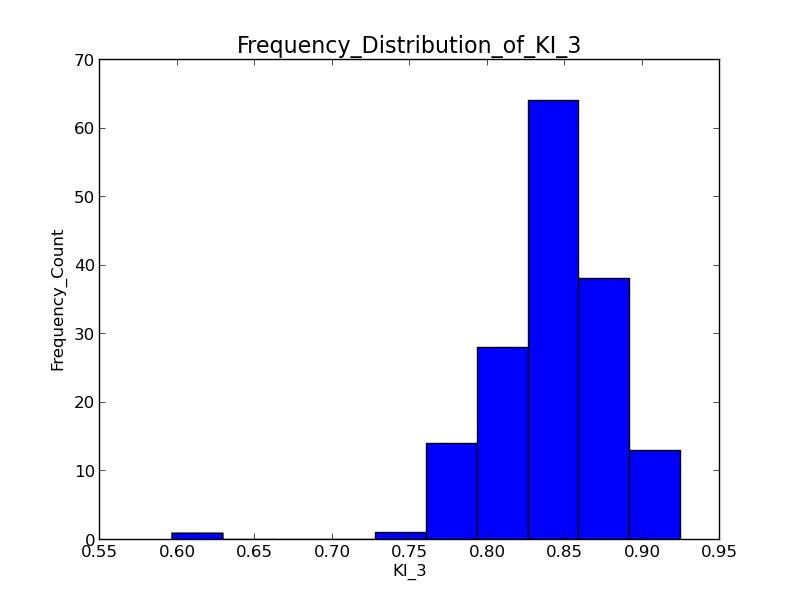} &
\includegraphics[width=3.0in]{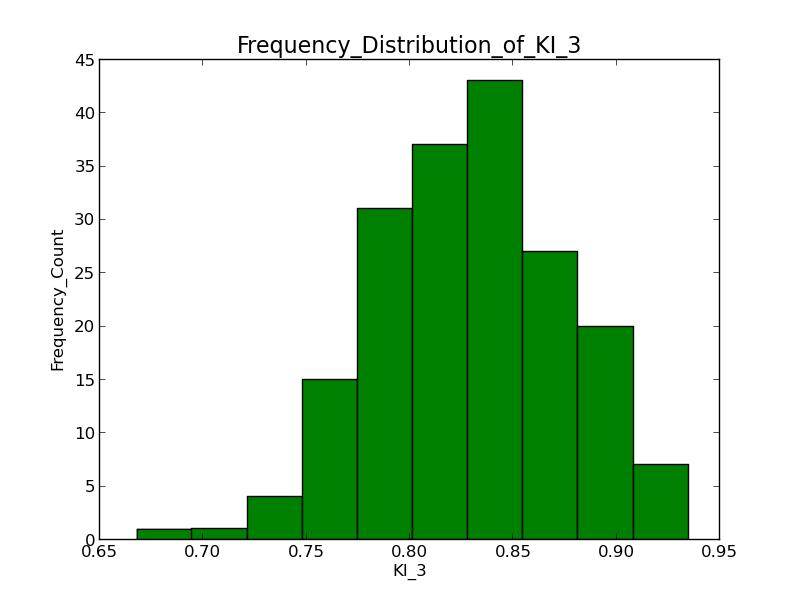} \\
\end{array}$
\end{center}
\caption[The Frequency Distribution of the KI 3 of $H\alpha$ and $H\beta$ emission line profiles]
{Frequency Distribution of the KI 3 of $H\alpha$ and $H\beta$ emission line profiles}
\label{fig:KI_3}
\end{figure*}

\begin{figure*}[!htbp]
\begin{center}$
\begin{array}{cc}
\includegraphics[width=3.0in]{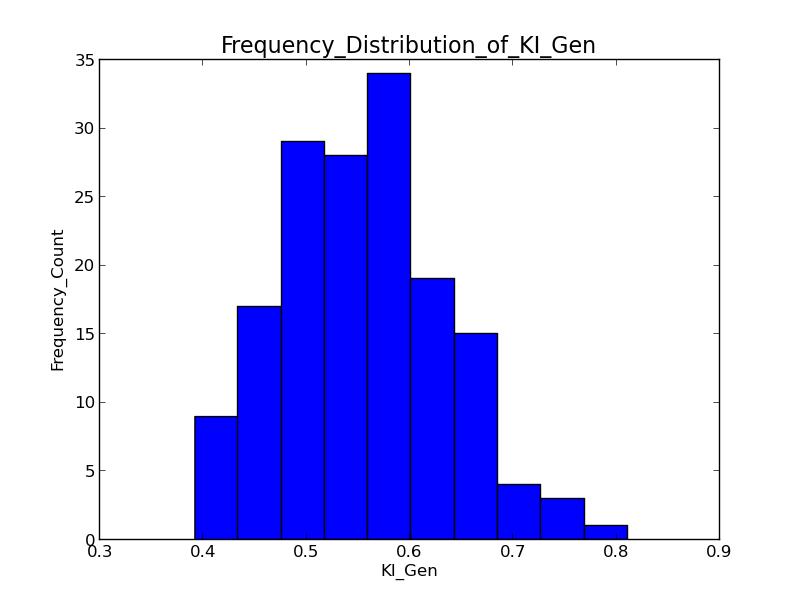} &
\includegraphics[width=3.0in]{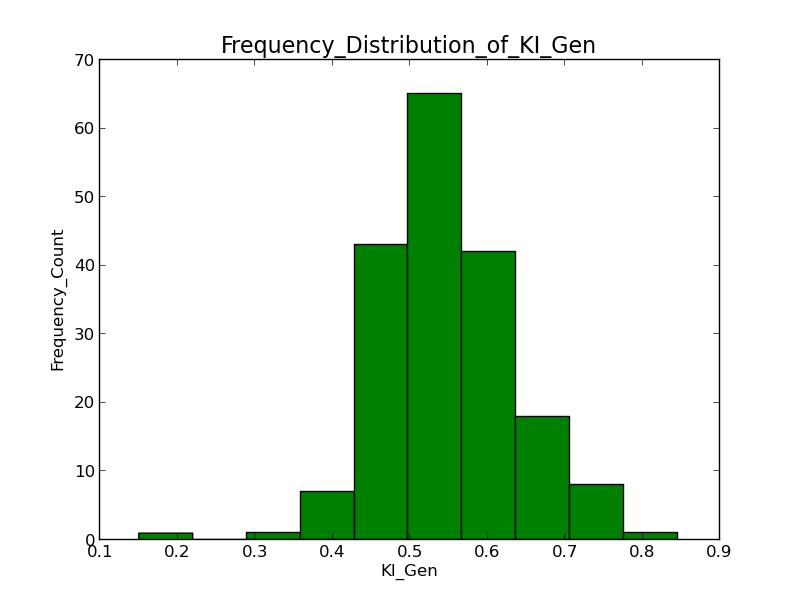} \\
\end{array}$
\end{center}
\caption[The Frequency Distribution of the KI Gen of $H\alpha$ and $H\beta$ emission line profiles]
{Frequency Distribution of the KI Gen of $H\alpha$ and $H\beta$ emission line profiles.}
\label{fig:KI_Gen}
\end{figure*}

Figures \ref{fig:KI_1}, \ref{fig:KI_2}, \ref{fig:KI_3} and \ref{fig:KI_Gen} show the change in profile shape of the $H\alpha$ and $H\beta$ profiles, with the ones to the left (in red) for $H\alpha$
and the ones to the right (in blue) for the $H\beta$ profiles.

It is shown in fig \ref{fig:KI_1} that the change in profile shape for the top parts of the $H\alpha$ and $H\beta$ emission lines is quite high,
peaking at values close to $0.25$.

Fig \ref{fig:KI_2} shows that the change in profile shape for the center parts of the $H\alpha$ and $H\beta$ emission lines is moderate,
peaking at values close to $0.6$.

Fig \ref{fig:KI_3} shows that the change in profile shape for the lower parts of the $H\alpha$ and $H\beta$ emission lines is low,
peaking at values close to $0.8$. \\

\begin{figure*}[!htbp]
\begin{flushleft}$
\begin{array}{cl}
\includegraphics[width=2.8in]{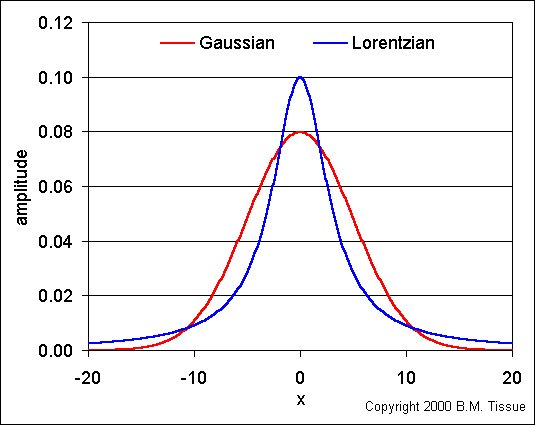} &
\includegraphics[width=4.0in]{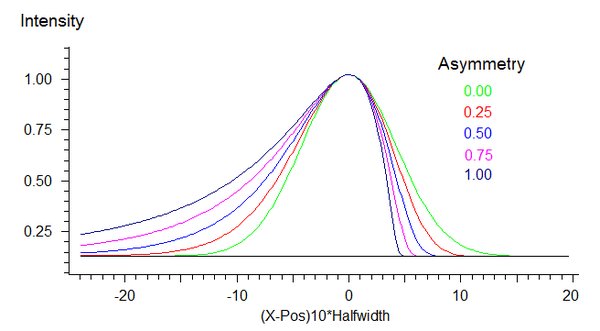} \\
\end{array}$
\end{flushleft}
\caption[Asymmetry and Kurtosis]{Images Showing the percentiles used to calculate values of Kurtosis and Asymmetry.}
\label{fig:ass1}
\end{figure*}

The three measures of steepness, $KI_1$, $KI_2$ and $KI_3$ show a pattern in profile shape change. The change in shape breaks down
as one moves to the base. This can be an indication of a systematic change to Lorentzian from Gaussian, as the change should be smooth
as one moves from around 60\% of the profile downwards. One cannot expect an abrupt change from Gaussian 
to Lorentzian. It could also be due to the selection of the regions of low, center and bottom. Looking at only two regions, 
top and bottom, breaks the smooth transition since the measure of the center of the profile reflects the effects from the same
physical conditions as the lower part of the profile as can be seen in \ref{fig:ass1}.

Fig \ref{fig:KI_Gen} shows that the change in overall profile shape of the $H\alpha$ and $H\beta$ emission lines is moderate,
peaking at values close to $0.55$.

The general trend of profile shape change is expected as not so many sources show high asymmetries, but the mere fact that there are some sources 
in the distribution with high measures in asymmetry and low measures in kurtosis justifies the study. It is now clearly vital to proceed and 
look up other kinematic properties and find out which of them correlate with the high values of asymmetry and low values of Kurtosis.
A previous study on 90 emission line profiles,constituting 30 Balmer lines and 55 forbidden and 5 other permitted lines, from 31 objects 
comprising of S1, S2, S3 classes of Seyfert galaxies, H II regions and QSOs showed that Forbidden lines are found to be narrower and steeper, 
while Balmer lines and other permitted lines are broader and flatter \citep{Basu0}, meaning that the Kurtosis measure will be small as we also 
observed in the $KI_1$.

\begin{figure*}[!htbp]
\begin{center}$
\begin{array}{cc}
\includegraphics[width=3.0in]{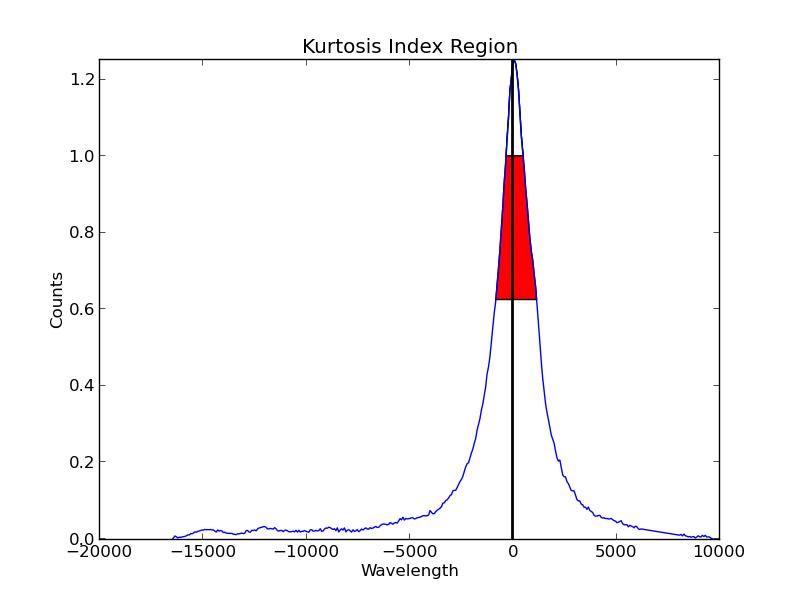} & 
\includegraphics[width=3.0in]{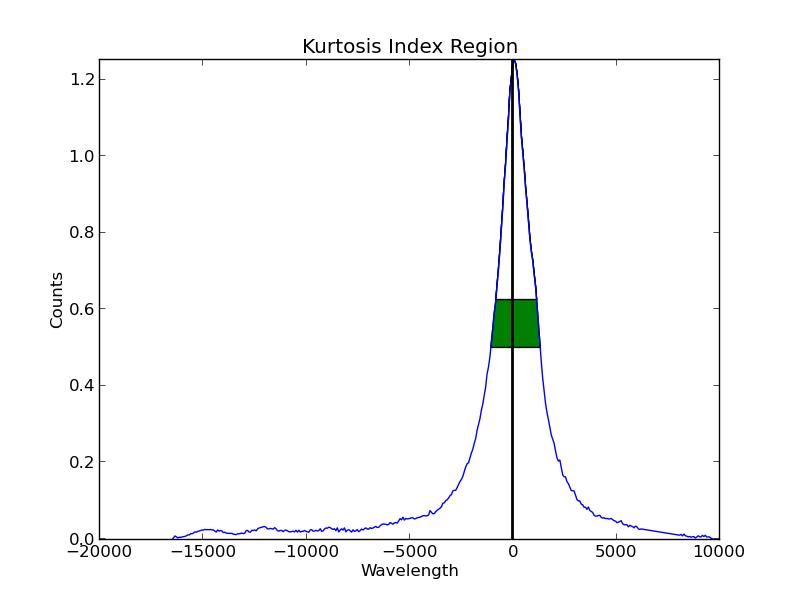} \\
\includegraphics[width=3.0in]{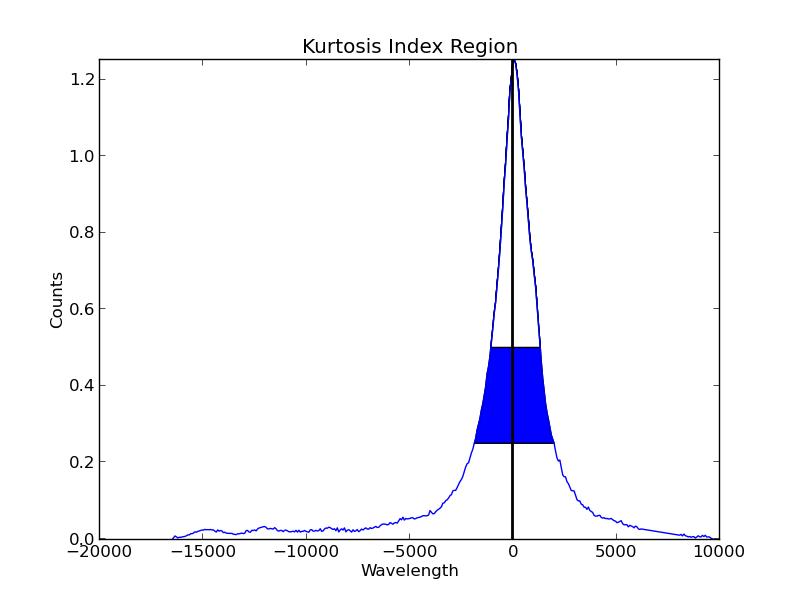} &
\includegraphics[width=3.0in]{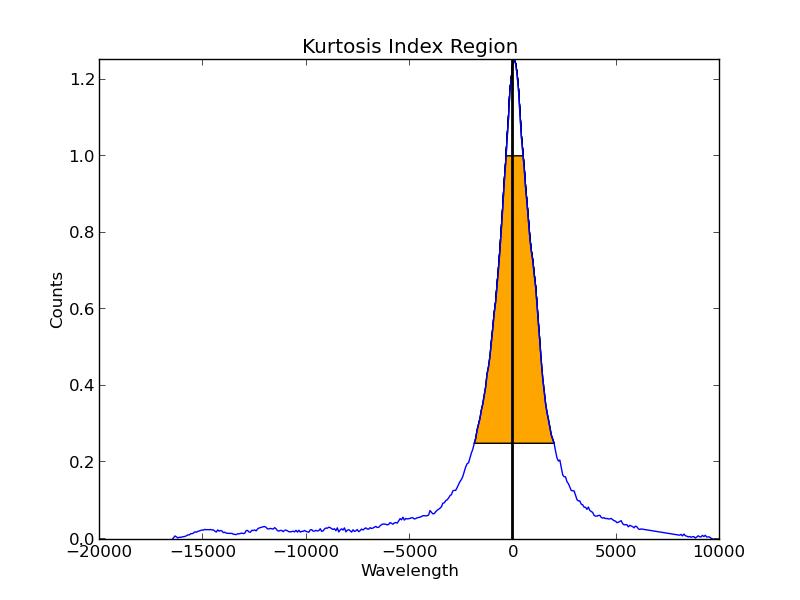} \\
\end{array}$
\end{center}
\caption[Kurtosis Index Regions for the Steep Profiles]
{Kurtosis Index Regions for the Steep Profiles}
\label{fig:KI_narrow}
\end{figure*}

\begin{figure*}[!htbp]
\begin{center}$
\begin{array}{cc}
\includegraphics[width=3.0in]{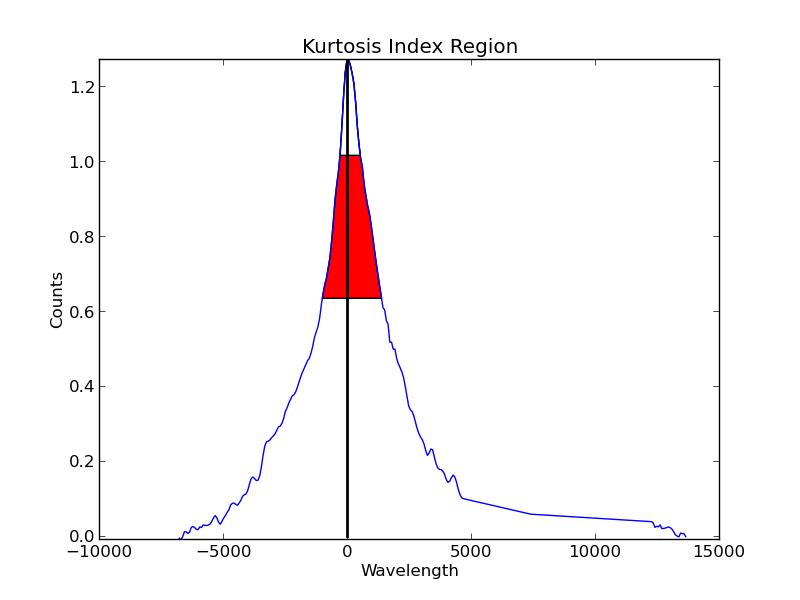} & 
\includegraphics[width=3.0in]{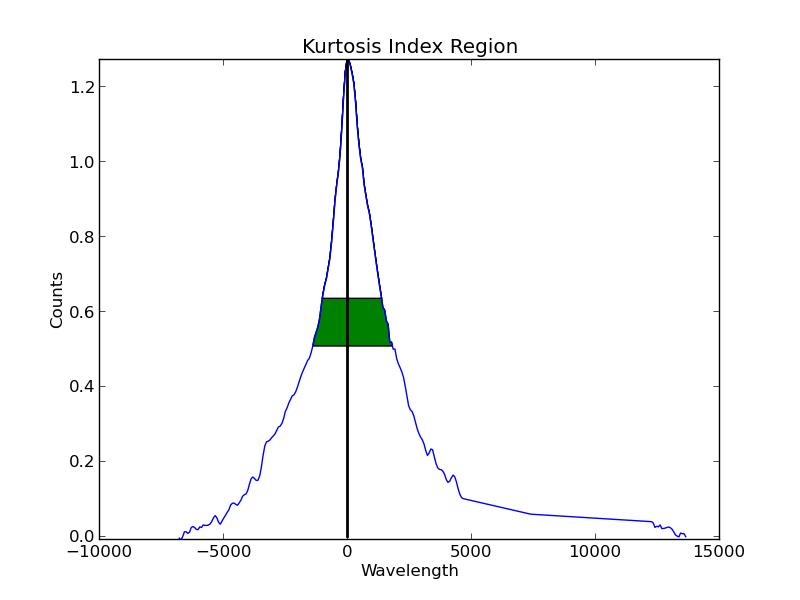} \\
\includegraphics[width=3.0in]{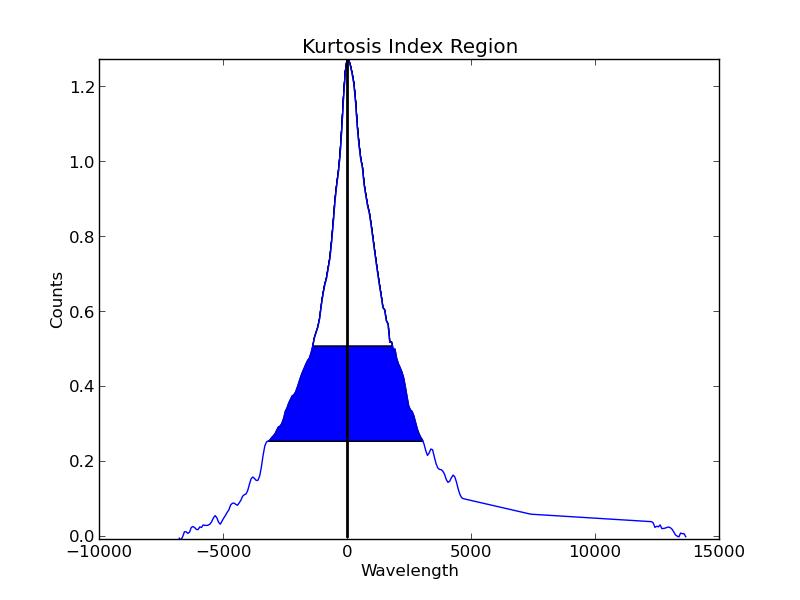} &
\includegraphics[width=3.0in]{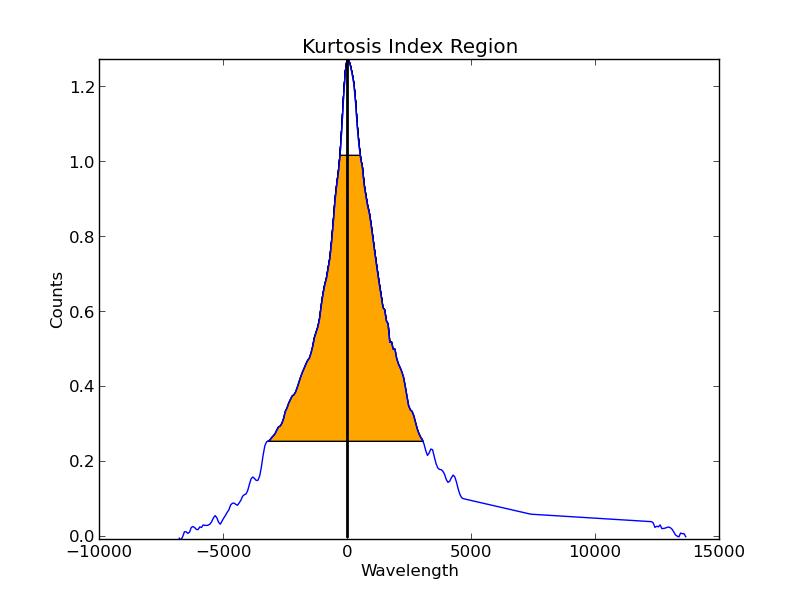} \\
\end{array}$
\end{center}
\caption[Kurtosis Index Regions for the Flatter Profiles]
{Kurtosis Index Regions for the Flatter Profiles}
\label{fig:KI_broad}
\end{figure*}

%----------------------------------------------------------------------------------------

\subsection{Asymmetry Index relation to other kinematic properties}
While we have the Asymmetry Index and Kurtosis Index, its scientifically worthwhile to relate this measure to some other kinematic
properties \citep{Whittle0}. Some of the properties we related Asymmetry Index to are; the FWHM, the luminosity in the V-Band, the Radio Flux, and 
Ionization degree using the Oxygen narrow emission lines.
\subsubsection{Relation of Asymmetry Index to the Line Width(FWHM)}

The Line width is a measure of the strength of the spectral properties in the source which can be used to obtain a number of estimates about the 
size, radius of the region. A relation of this measurement with line asymmetry is important in the quest to obtain the origin of the asymmetry and the nature of the 
mechanism. Thus the asymmetry here will play a great deal in probing the flow of material in and out of the BLR. 
This is in line with the relation here as the base of the profiles shows high asymmetries, thus more effects from the motions of the material
than in the upper portions of the profile.
This is a general trend, which is in agreement with the underlying physics. We also study individual relations with each percentile in the plots
that follow.

The following ten plots are relations of Asymmetry Index and FWHM, those to the left are of $H\alpha$ (in green) while those to the right are 
for $H\beta$ (in blue).

\begin{figure*}[!htbp]
\begin{center}$
\begin{array}{cc}
\includegraphics[width=3.0in]{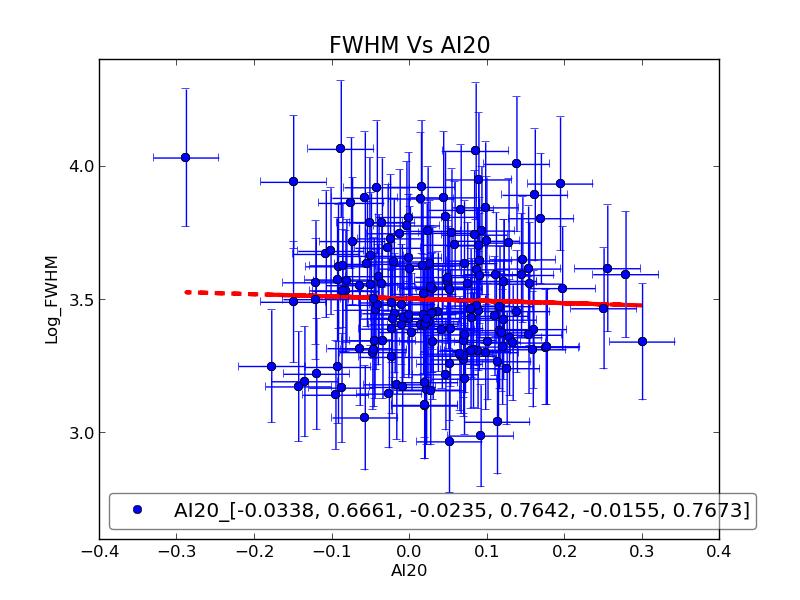} &
\includegraphics[width=3.0in]{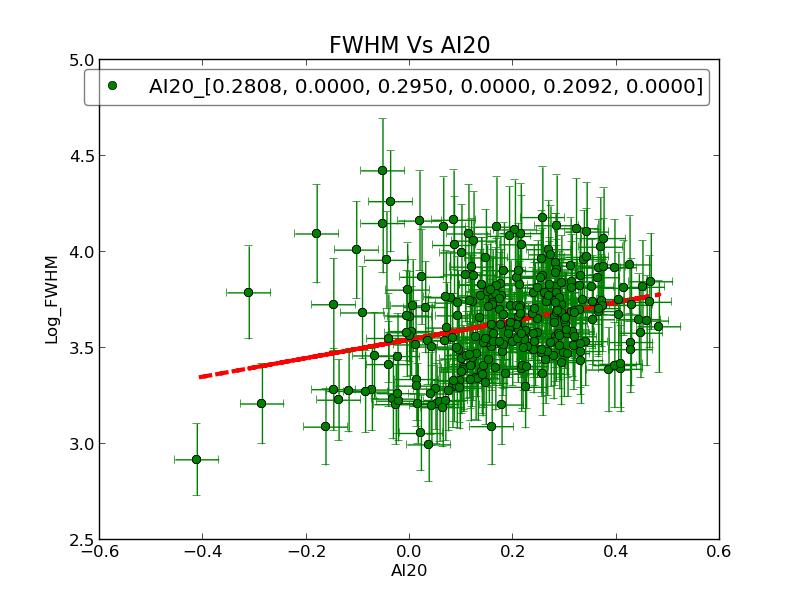} \\
\end{array}$
\end{center}
\caption[The Relation of the AI20 of $H\alpha$ and $H\beta$ emission line profiles with Line Width]
{Relation of the AI20 of $H\alpha$ and $H\beta$ emission line profiles with Line width}
\label{fig:LW_AI20}
\end{figure*}

In Fig\ref{fig:LW_AI20}, the relation between the FWHM and AI20 is shown. The $H\alpha$ does not appear to have a relation with the two variables,
the points are heavily scattered in the plot\! But also the relation observed with the $H\beta$ tends to start breaking down from previous relations.
We observe an increase in asymmetry with line width.

\begin{figure*}[!htbp]
\begin{center}$
\begin{array}{cc}
\includegraphics[width=3.0in]{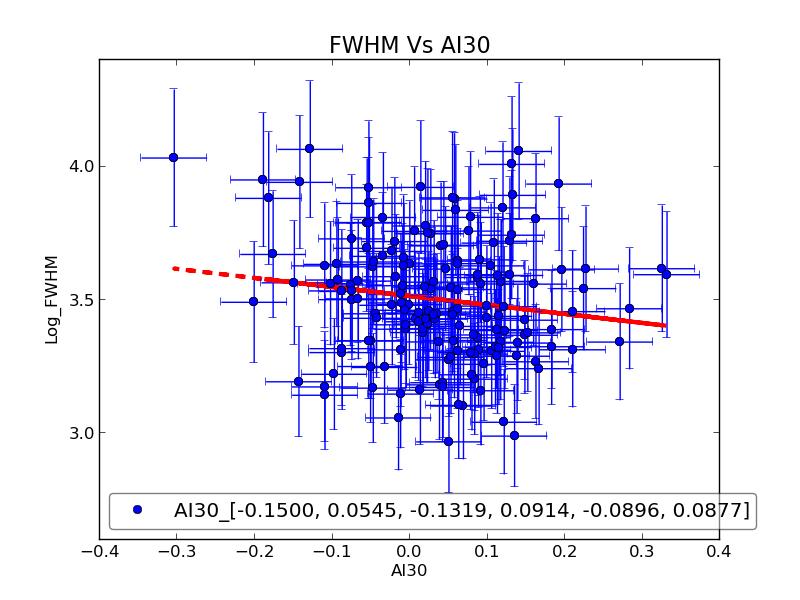} &
\includegraphics[width=3.0in]{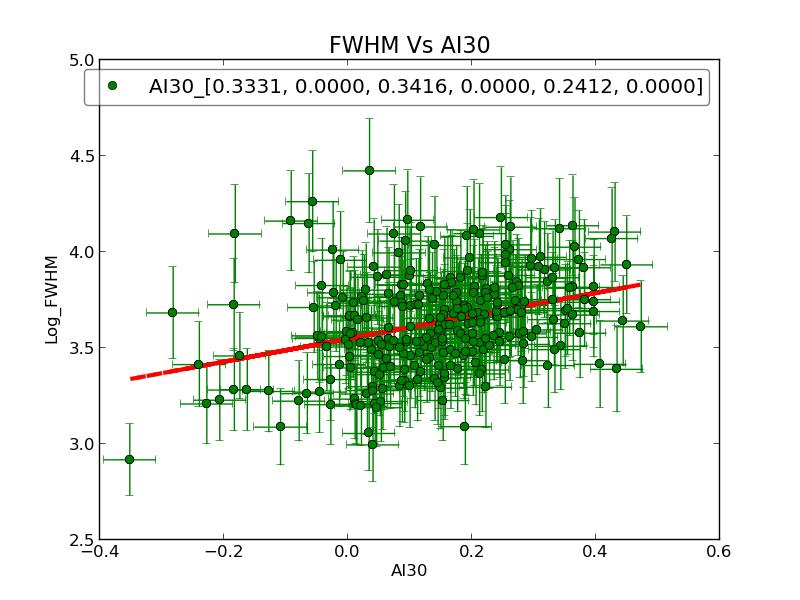} \\
\end{array}$
\end{center}
\caption[The Relation of the AI30 of $H\alpha$ and $H\beta$ emission line profiles with Line Width]
{Relation of the AI30 of $H\alpha$ and $H\beta$ emission line profiles with Line width}
\label{fig:LW_AI30}
\end{figure*}

\begin{figure*}[!htbp]
\begin{center}$
\begin{array}{cc}
\includegraphics[width=3.0in]{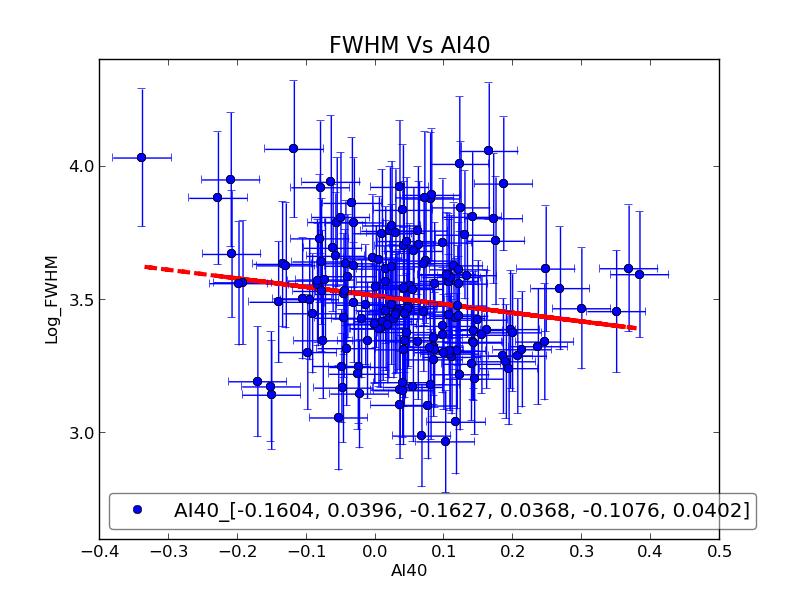} &
\includegraphics[width=3.0in]{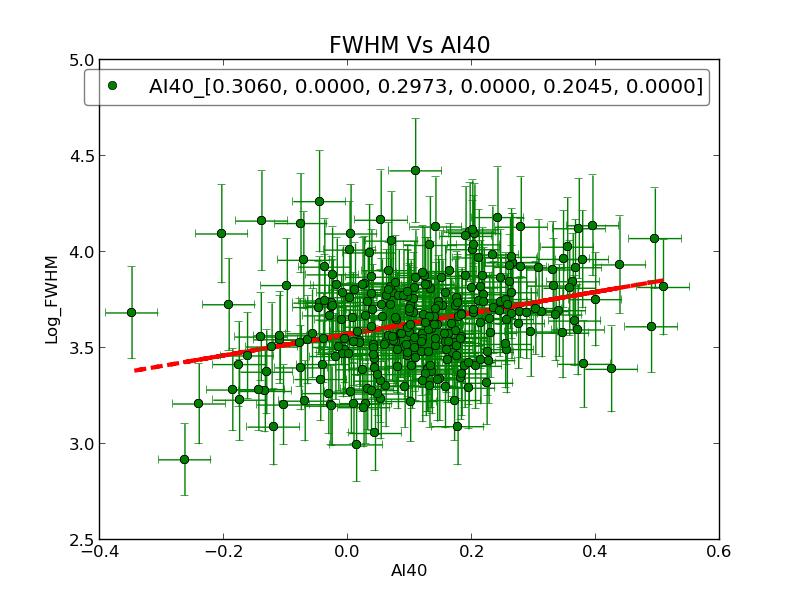} \\
\end{array}$
\end{center}
\caption[The Relation of the AI40 of $H\alpha$ and $H\beta$ emission line profiles with Line Width]
{Relation of the AI40 of $H\alpha$ and $H\beta$ emission line profiles with Line width}
\label{fig:LW_AI40}
\end{figure*}

\begin{figure*}[!htbp]
\begin{center}$
\begin{array}{cc}
\includegraphics[width=3.0in]{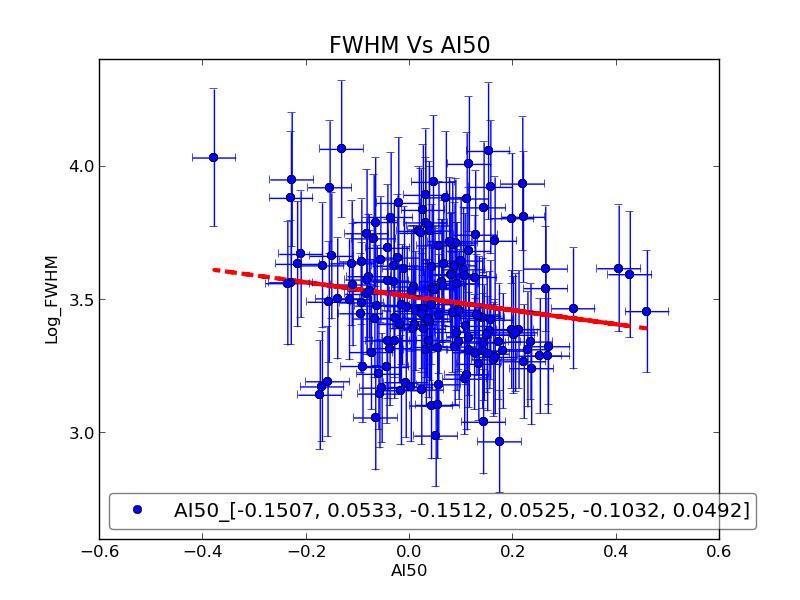} &
\includegraphics[width=3.0in]{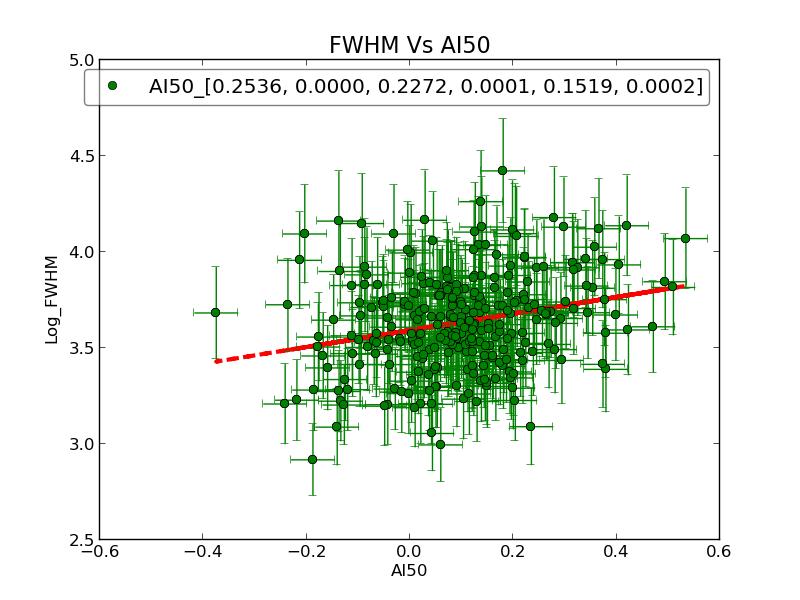} \\
\end{array}$
\end{center}
\caption[The Relation of the AI50 of $H\alpha$ and $H\beta$ emission line profiles with Line Width]
{Relation of the AI50 of $H\alpha$ and $H\beta$ emission line profiles with Line width}
\label{fig:LW_AI50}
\end{figure*}

\begin{figure*}[!htbp]
\begin{center}$
\begin{array}{cc}
\includegraphics[width=3.0in]{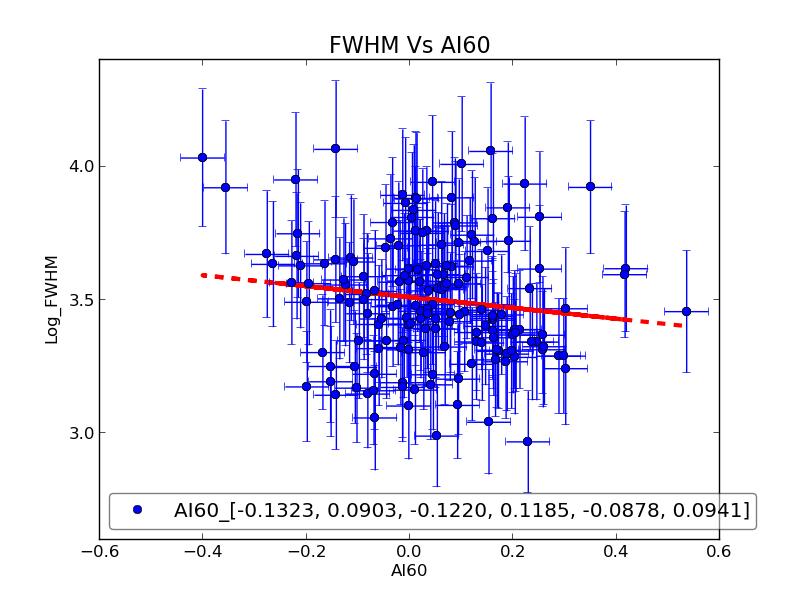} &
\includegraphics[width=3.0in]{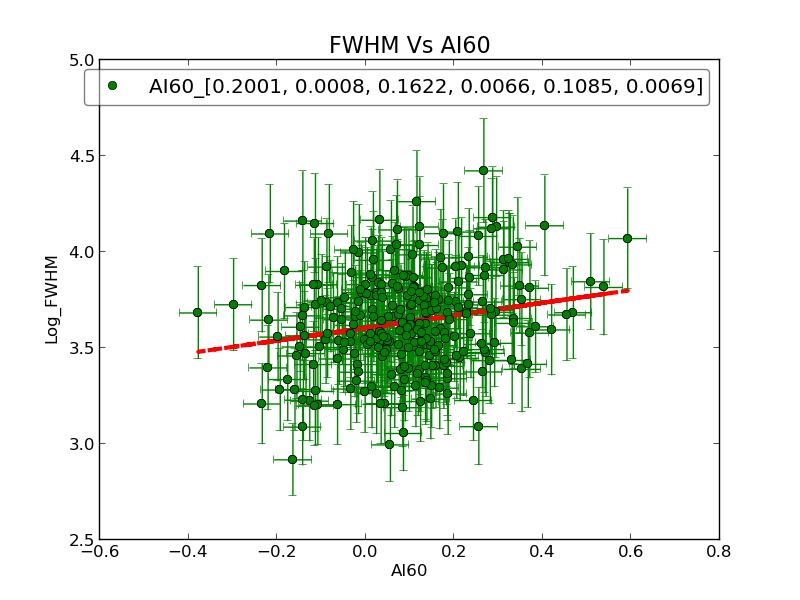} \\
\end{array}$
\end{center}
\caption[The Relation of the AI60 of $H\alpha$ and $H\beta$ emission line profiles with Line Width]
{Relation of the AI60 of $H\alpha$ and $H\beta$ emission line profiles with Line width}
\label{fig:LW_AI60}
\end{figure*}

Figures \ref{fig:LW_AI30}, \ref{fig:LW_AI40}, \ref{fig:LW_AI50} and \ref{fig:LW_AI60} all show a decrease in line width with asymmetry for 
$H\alpha$ and an increase in line width with asymmetry with $H\beta$.

\begin{figure*}[!htbp]
\begin{center}$
\begin{array}{cc}
\includegraphics[width=3.0in]{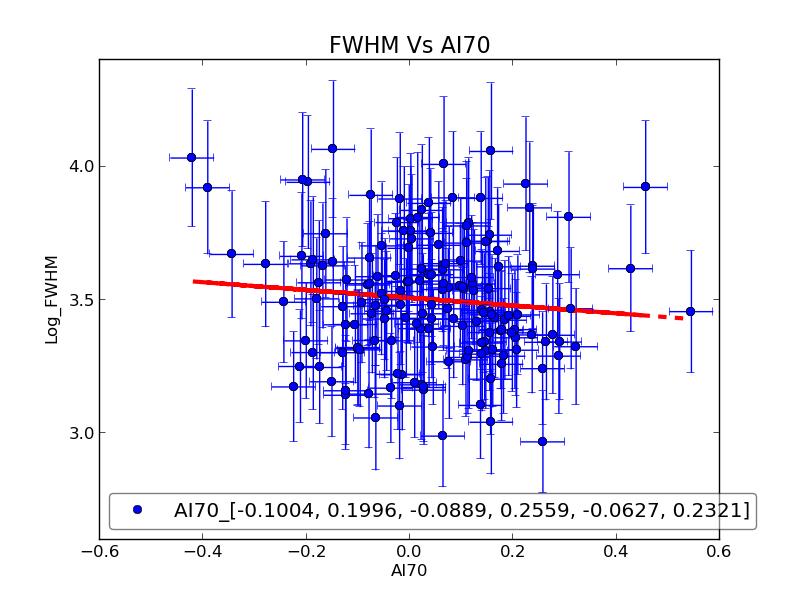} &
\includegraphics[width=3.0in]{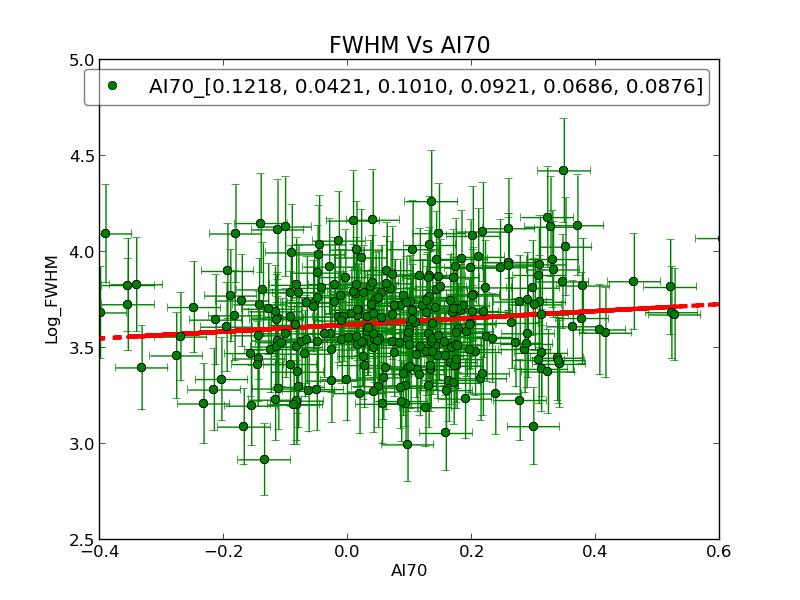} \\
\end{array}$
\end{center}
\caption[The Relation of the AI70 of $H\alpha$ and $H\beta$ emission line profiles with Line Width]
{Relation of the AI70 of $H\alpha$ and $H\beta$ emission line profiles with Line width}
\label{fig:LW_AI70}
\end{figure*}

\begin{table*}[!htbp]
\caption[Correlation Coefficients for Line Width Verses $H\alpha$ Asymmetry]{Correlation Coefficients for Line Width 
Verses $H\alpha$ Asymmetry}
\label{tab:cc_ha_lw}
\begin{center}
\begin{tabular}{ccccccccc}
\hline
Percentile & $\rho_p$ & $P_p{-value}$ & $\rho_s$ & $P_s{-value}$ & $\rho_k$ & $P_k{-value}$ & m & b \\ \hline \hline

20 & -0.0338 & 0.6661 & -0.0235 & 0.7642 & -0.0155 & 0.7673 & -0.0838 & 3.5049 \\ 
30 & -0.1500 & 0.0545 & -0.1319 & 0.0914 & -0.0896 & 0.0877 & -0.3372 & 3.5151 \\ 
40 & -0.1604 & 0.0396 & -0.1627 & 0.0368 & -0.1076 & 0.0402 & -0.3227 & 3.5162 \\ 
50 & -0.1507 & 0.0533 & -0.1512 & 0.0525 & -0.1032 & 0.0492 & -0.2620 & 3.5137 \\ 
60 & -0.1323 & 0.0903 & -0.1220 & 0.1185 & -0.0878 & 0.0941 & -0.2036 & 3.5110 \\ 
70 & -0.1004 & 0.1996 & -0.0889 & 0.2559 & -0.0627 & 0.2321 & -0.1459 & 3.5082 \\ \hline
\end{tabular}
\end{center}
\end{table*}

\begin{table*}[!htbp]
\caption[Correlation Coefficients for Line Width Verses $H\beta$ Asymmetry]{Correlation Coefficients for Line Width 
Verses $H\beta$ Asymmetry}
\label{tab:cc_ha_lw}
\begin{center}
\begin{tabular}{ccccccccc}
\hline
Percentile & $\rho_p$ & $P_p{-value}$ & $\rho_s$ & $P_s{-value}$ & $\rho_k$ & $P_k{-value}$ & m & b \\ \hline \hline

20 & 0.2808 & 0.0000 & 0.2950 & 0.0000 & 0.2092 & 0.0000 & 0.4842 & 3.5448 \\ 
30 & 0.3331 & 0.0000 & 0.3416 & 0.0000 & 0.2412 & 0.0000 & 0.5977 & 3.5481 \\ 
40 & 0.3060 & 0.0000 & 0.2973 & 0.0000 & 0.2045 & 0.0000 & 0.5515 & 3.5719 \\ 
50 & 0.2536 & 0.0000 & 0.2272 & 0.0001 & 0.1519 & 0.0002 & 0.4338 & 3.5919 \\ 
60 & 0.2001 & 0.0008 & 0.1622 & 0.0066 & 0.1085 & 0.0069 & 0.3283 & 3.6045 \\ 
70 & 0.1218 & 0.0421 & 0.1010 & 0.0921 & 0.0686 & 0.0876 & 0.1787 & 3.6210 \\  \hline
\end{tabular}
\end{center}
\end{table*}

In a nutshell, excluding the FWZI, all $H\alpha$ profile percentiles display a negative correlation while $H\beta$ profiles display positive 
correlations excluding their  $00\%$ and $90\%$ percentile. The later correlations are tighter making the $H\beta$ shape parameters better at
analyzing effects causing profile shape in Balmer emission lines. It is also a reflection from the previous statistic on the distribution of 
asymmetry index where we observed that the $H\beta$ profiles displayed more positive asymmetry as opposed to the $H\alpha$ profiles that were
statistically symmetric.

\subsubsection{Relation of Asymmetry Index to the V Band Luminosity}

The relationship of Luminosity with the asymmetry of broad Balmer lines is of great importance because luminosity of one of the properties
that can easily be obtained from a source. Having a clear relationship with this property will help in further scientific relationships
with the AGN Broad Line Region physics. Since the luminosity has already existing relationships with parameters like Black Hole mass,
accretion rate and morphological type, it will help us relate asymmetry of profiles to all there other properties of the galaxies.
One excellent correlation is that between the Black Hole mass and Optical luminosity obtained by \cite{Peterson0}. It was observed that 
Black Hole mass increased with optical luminosity. The correlation between asymmetry and V-Band luminosity therefore will also help us in 
probing this further. \cite{Kaspi0} also found a tight correlation between the luminosity and the radius of the BLR. This will also help
is relate our asymmetry with the radius of the BLR.
The preceding relations are also divided into two, the relation to the left being that of $H\alpha$ profiles and that to the right being
that of the $H\beta$ profiles. But generally, there seems not to be direct relationships between these parameters apart from that of AI00.

\begin{figure*}[!htbp]
\begin{center}$
\begin{array}{cc}
\includegraphics[width=3.0in]{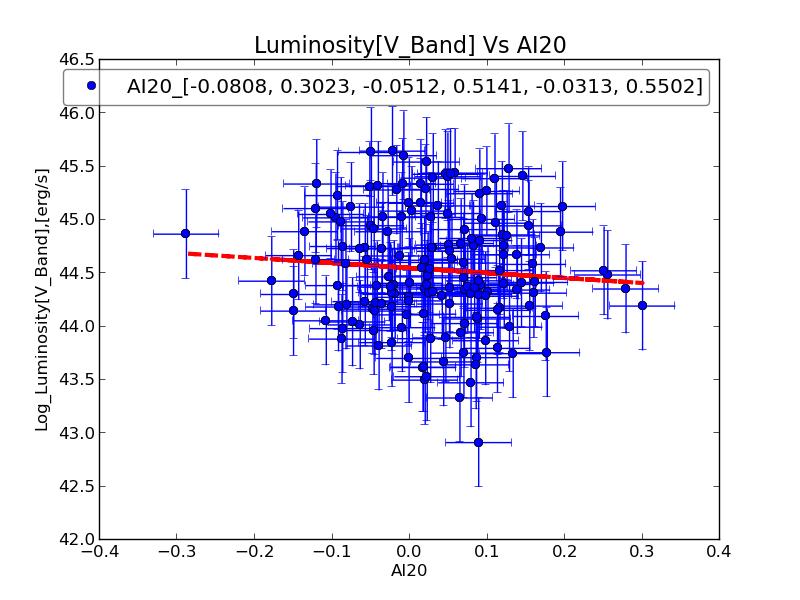} &
\includegraphics[width=3.0in]{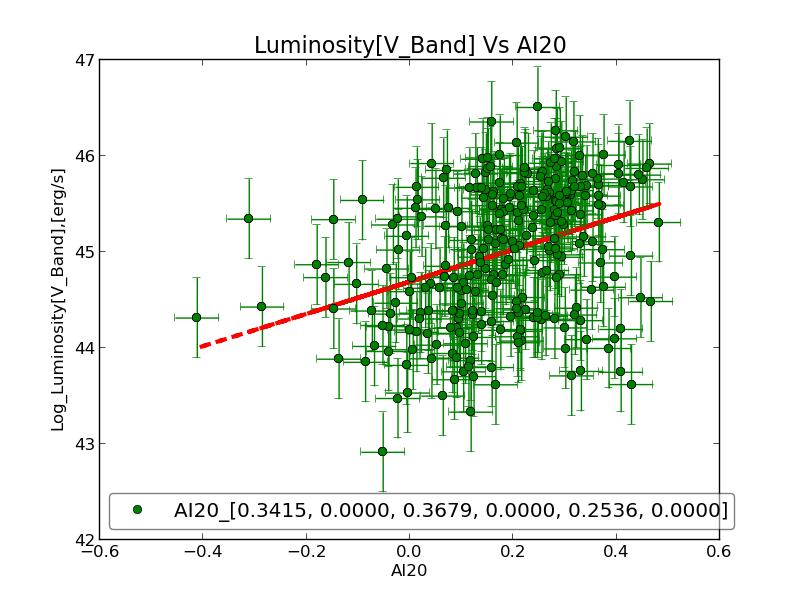} \\
\end{array}$
\end{center}
\caption[The Relation of the AI20 of $H\alpha$ and $H\beta$ emission line profiles with V Band Luminosity]
{Relation of the AI20 of $H\alpha$ and $H\beta$ emission line profiles with V Band Luminosity}
\label{fig:Lum_AI20}
\end{figure*}

\begin{figure*}[!htbp]
\begin{center}$
\begin{array}{cc}
\includegraphics[width=3.0in]{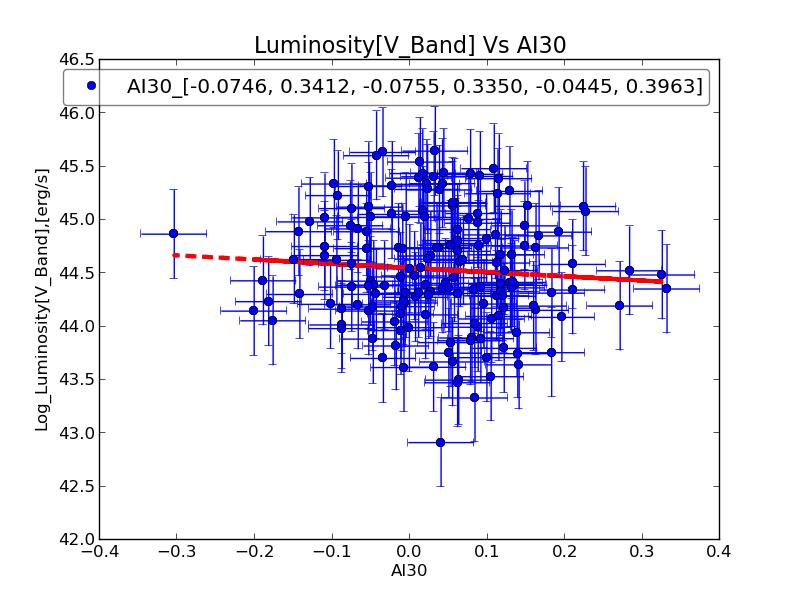} &
\includegraphics[width=3.0in]{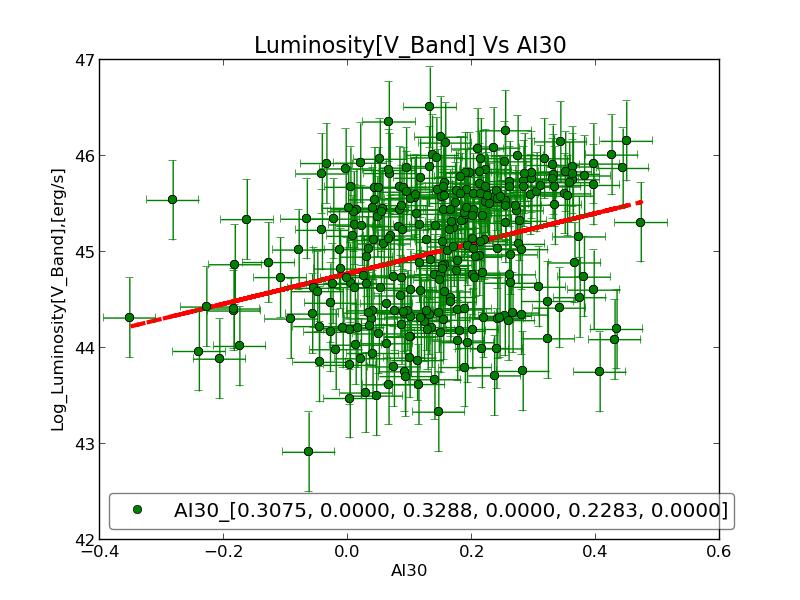} \\
\end{array}$
\end{center}
\caption[The Relation of the AI30 of $H\alpha$ and $H\beta$ emission line profiles with V Band Luminosity]
{Relation of the AI30 of $H\alpha$ and $H\beta$ emission line profiles with V Band Luminosity}
\label{fig:Lum_AI30}
\end{figure*}

\begin{figure*}[!htbp]
\begin{center}$
\begin{array}{cc}
\includegraphics[width=3.0in]{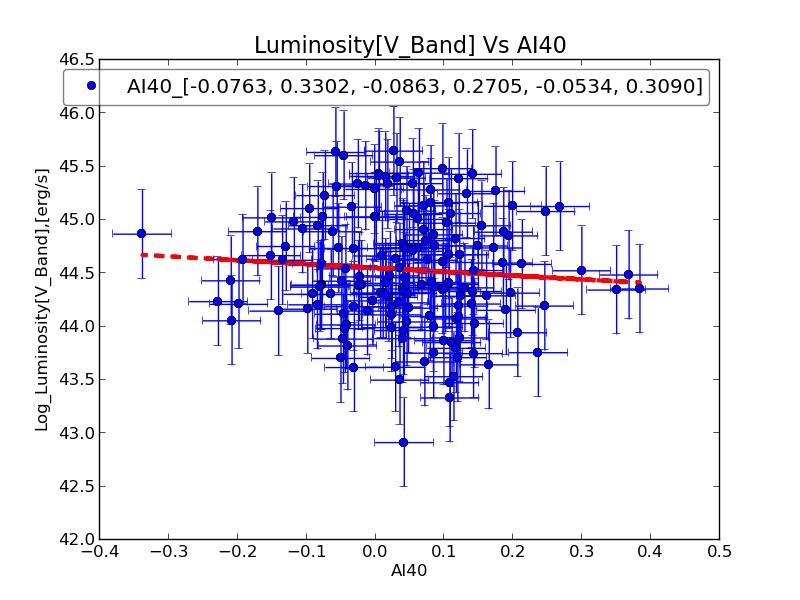} &
\includegraphics[width=3.0in]{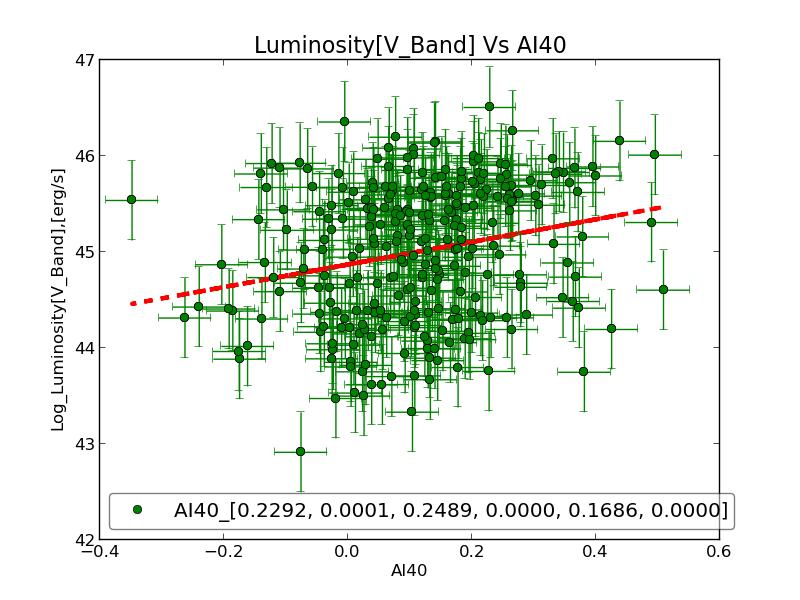} \\
\end{array}$
\end{center}
\caption[The Relation of the AI40 of $H\alpha$ and $H\beta$ emission line profiles with V Band Luminosity]
{Relation of the AI40 of $H\alpha$ and $H\beta$ emission line profiles with V Band Luminosity}
\label{fig:Lum_AI40}
\end{figure*}

\begin{figure*}[!htbp]
\begin{center}$
\begin{array}{cc}
\includegraphics[width=3.0in]{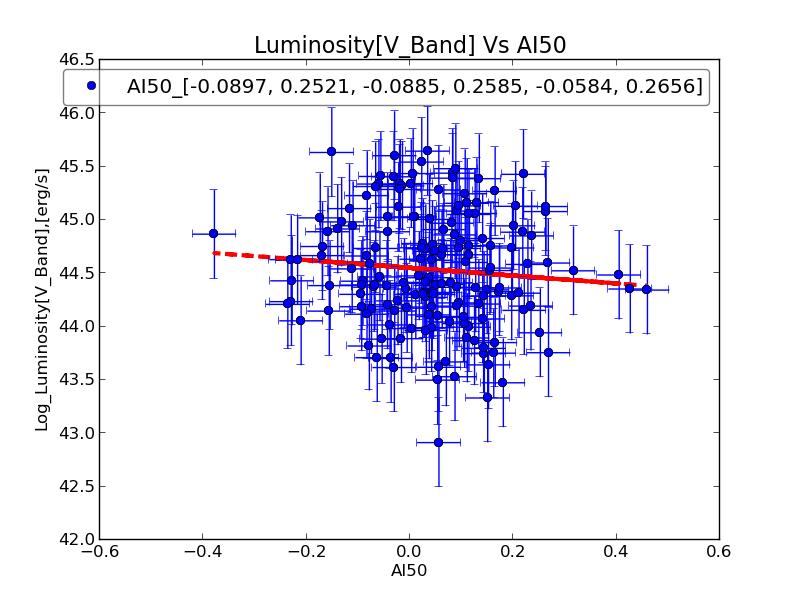} &
\includegraphics[width=3.0in]{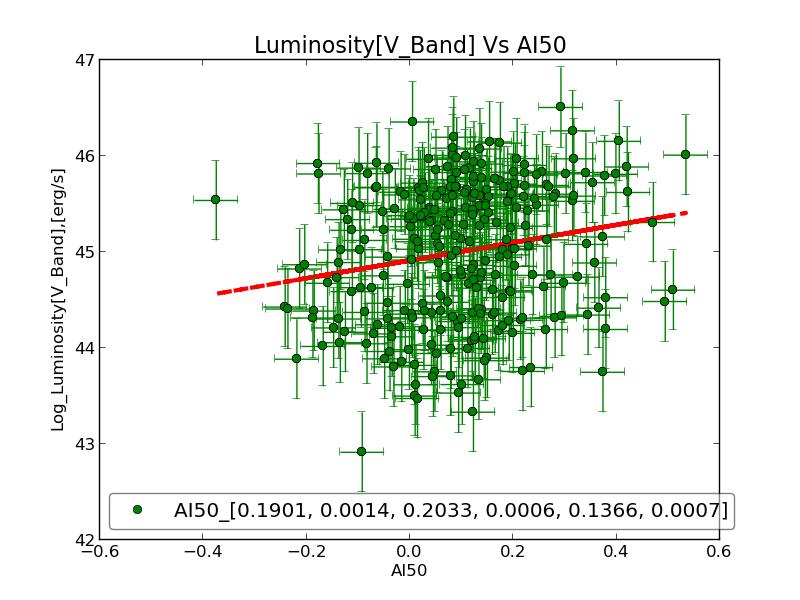} \\
\end{array}$
\end{center}
\caption[The Relation of the AI50 of $H\alpha$ and $H\beta$ emission line profiles with V Band Luminosity]
{Relation of the AI50 of $H\alpha$ and $H\beta$ emission line profiles with V Band Luminosity}
\label{fig:Lum_AI50}
\end{figure*}

\begin{figure*}[!htbp]
\begin{center}$
\begin{array}{cc}
\includegraphics[width=3.0in]{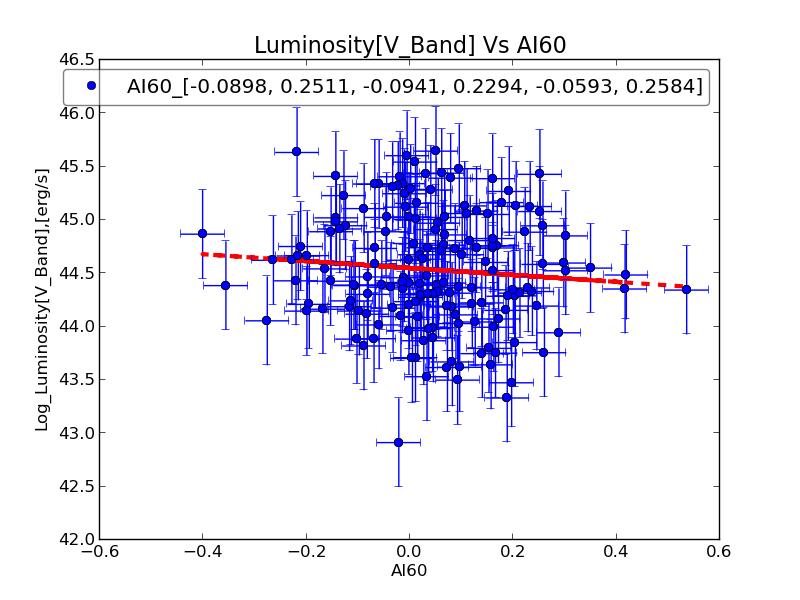} &
\includegraphics[width=3.0in]{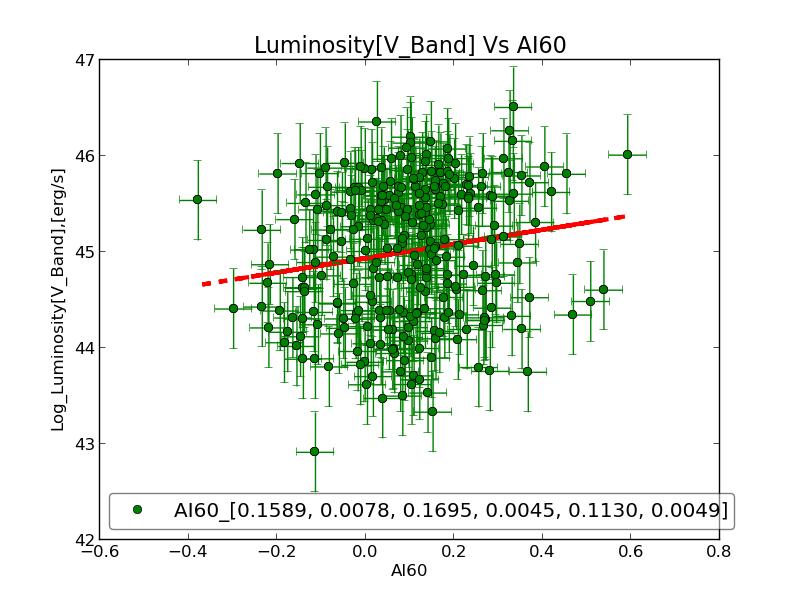} \\
\end{array}$
\end{center}
\caption[The Relation of the AI60 of $H\alpha$ and $H\beta$ emission line profiles with V Band Luminosity]
{Relation of the AI60 of $H\alpha$ and $H\beta$ emission line profiles with V Band Luminosity}
\label{fig:Lum_AI60}
\end{figure*}

\begin{figure*}[!htbp]
\begin{center}$
\begin{array}{cc}
\includegraphics[width=3.0in]{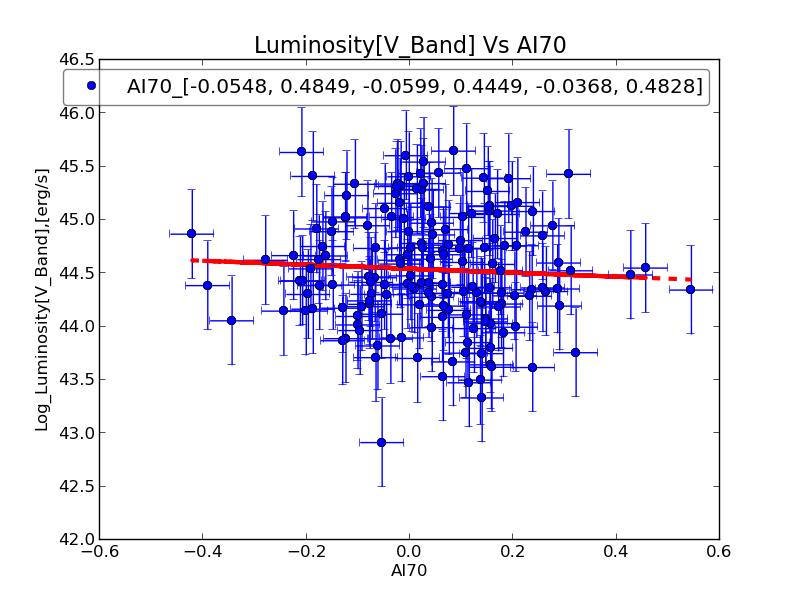} &
\includegraphics[width=3.0in]{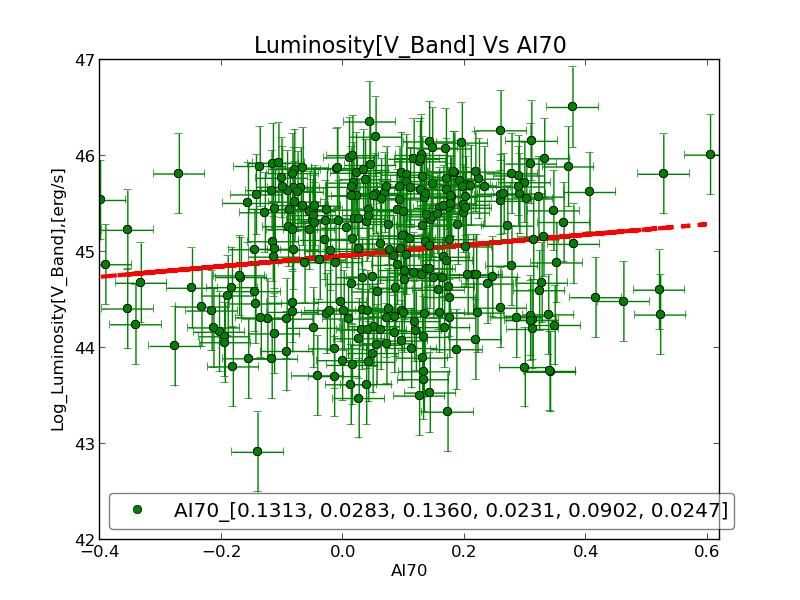} \\
\end{array}$
\end{center}
\caption[The Relation of the AI70 of $H\alpha$ and $H\beta$ emission line profiles with V Band Luminosity]
{Relation of the AI70 of $H\alpha$ and $H\beta$ emission line profiles with V Band Luminosity}
\label{fig:Lum_AI70}
\end{figure*}

\begin{table*}[!htbp]
\caption[Correlation Coefficients for V-Band Luminosity Verses $H\alpha$ Asymmetry]{Correlation Coefficients for V-Band Luminosity 
Verses $H\alpha$ Asymmetry}
\label{tab:cc_ha_lum}
\begin{center}
\begin{tabular}{ccccccccc}
\hline
Percentile & $\rho_p$ & $P_p{-value}$ & $\rho_s$ & $P_s{-value}$ & $\rho_k$ & $P_k{-value}$ & m & b \\ \hline \hline

20 & -0.0808 & 0.3023 & -0.0512 & 0.5141 & -0.0313 & 0.5502 & -0.4690 & 44.5500 \\ 
30 & -0.0746 & 0.3412 & -0.0755 & 0.3350 & -0.0445 & 0.3963 & -0.3932 & 44.5504 \\ 
40 & -0.0763 & 0.3302 & -0.0863 & 0.2705 & -0.0534 & 0.3090 & -0.3599 & 44.5510 \\ 
50 & -0.0897 & 0.2521 & -0.0885 & 0.2585 & -0.0584 & 0.2656 & -0.3655 & 44.5515 \\ 
60 & -0.0898 & 0.2511 & -0.0941 & 0.2294 & -0.0593 & 0.2584 & -0.3243 & 44.5494 \\ 
70 & -0.0548 & 0.4849 & -0.0599 & 0.4449 & -0.0368 & 0.4828 & -0.1867 & 44.5431 \\ \hline
\end{tabular}
\end{center}
\end{table*}

\begin{table*}[!htbp]
\caption[Correlation Coefficients for V-Band Luminosity Verses $H\beta$ Asymmetry]{Correlation Coefficients for V-Band Luminosity 
Verses $H\beta$ Asymmetry}
\label{tab:cc_ha_lum}
\begin{center}
\begin{tabular}{ccccccccc}
\hline
Percentile & $\rho_p$ & $P_p{-value}$ & $\rho_s$ & $P_s{-value}$ & $\rho_k$ & $P_k{-value}$ & m & b \\ \hline \hline

20 & 0.3415 & 0.0000 & 0.3679 & 0.0000 & 0.2536 & 0.0000 & 1.6778 & 44.6916 \\ 
30 & 0.3075 & 0.0000 & 0.3288 & 0.0000 & 0.2283 & 0.0000 & 1.5719 & 44.7758 \\ 
40 & 0.2292 & 0.0001 & 0.2489 & 0.0000 & 0.1686 & 0.0000 & 1.1768 & 44.8697 \\ 
50 & 0.1901 & 0.0014 & 0.2033 & 0.0006 & 0.1366 & 0.0007 & 0.9263 & 44.9124 \\ 
60 & 0.1589 & 0.0078 & 0.1695 & 0.0045 & 0.1130 & 0.0049 & 0.7429 & 44.9354 \\ 
70 & 0.1313 & 0.0283 & 0.1360 & 0.0231 & 0.0902 & 0.0247 & 0.5489 & 44.9612 \\ \hline
\end{tabular}
\end{center}
\end{table*}

In all the relations of Luminosity with asymmetry, it is observed right from figure \ref{fig:Lum_AI20} through to \ref{fig:Lum_AI70}
that the $H\alpha$ asymmetry relations are non-existent or very weak, if found. 
However, the asymmetry in $H\beta$ is observed to be consistent apart from the first and last percentiles.
$H\beta$ positive asymmetry is observed to rise with increasing Luminosity. Increasing luminosity correlates with increasing BLR radius and 
increasing BH mass. This seems to point to a asymmetry arising from any of the factors that positively correlate with luminosity.
This means a tendency of highly luminous sources giving rise to asymmetric $H\beta$ profiles.
The gradient in the relation is observed to fall as one moves from the low percentiles to the high ones, a $10\%$ percentile having a $\sim1.7$
gradient and an $80\%$ percentile having a gradient of $\sim0.3$.

\subsubsection{Relation of Asymmetry Index to the Radio Flux (Core-Radio Flux)}

One of the kinds of AGN are in radio galaxies \citep{Urry0, Peterson1}. A bulk of the emission from such sources
lies in the radio regime. A number of authors \citep{Pearson0, McCarthy0} have studied the behavior of galaxies with radio emission and
all have observed asymmetry in many emission lines from radio galaxies \citep{DeBreuck0}. Even a study on the behavior of jets in
weak radio sources noted asymmetry in spectral lines observed \citep{Laing0}. It is this strong foundation that encourages us to
systematically study the relation of radio Flux to the asymmetry of the emission lines. Our study is important since the spectral line
asymmetry is binned in percentiles, making it easy to notice while part of the emission line responds more to the radio emission.
In this study, the core radio flux at 6.0 arc secs is obtained. This is because radio flux can be extended to several tens of arc secs yet our
interest is the radio flux emanating or very close to the BLR, in which the broad Balmer lines are formed.

The following plots are paired for each percentile, having a relation with the $H\alpha$ emission line on the left and that of the 
$H\beta$ emission line on the right. We analyze each percentile separately starting with the lowest percentile, the FWZI, to the highest
part of te emission line possible, $90\&$ of the broad Balmer line.

\begin{figure*}[!htbp]
\begin{center}$
\begin{array}{cc}
\includegraphics[width=3.0in]{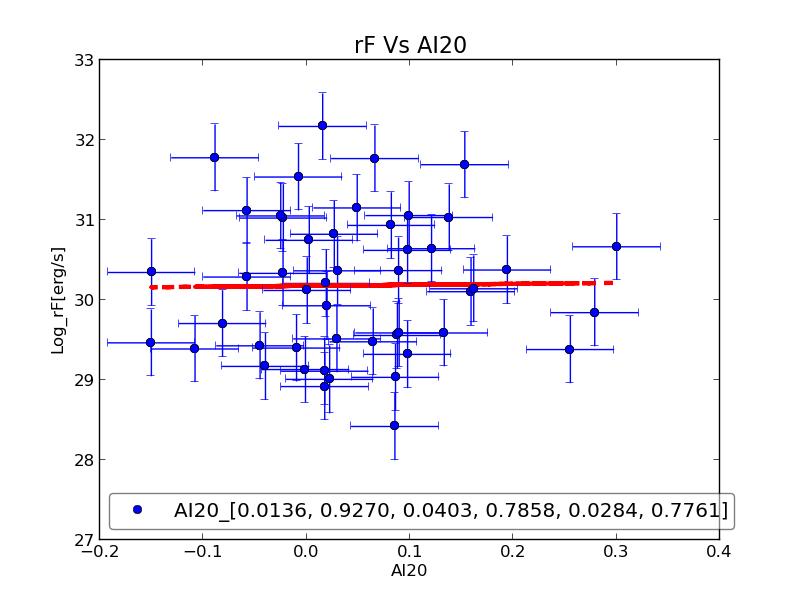} &
\includegraphics[width=3.0in]{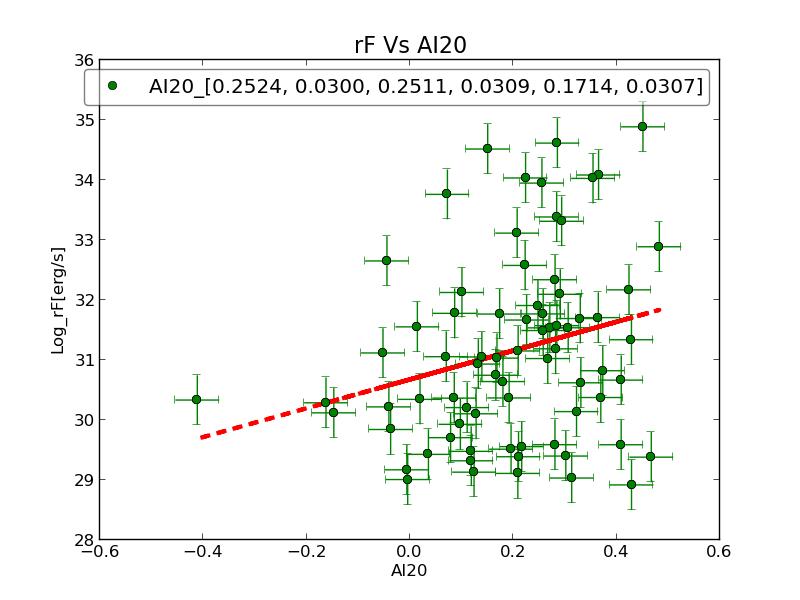} \\
\end{array}$
\end{center}
\caption[The Relation of the AI20 of $H\alpha$ and $H\beta$ emission line profiles with Core Radio Flux]
{Relation of the AI20 of $H\alpha$ and $H\beta$ emission line profiles with Core Radio Flux}
\label{fig:rF_AI20}
\end{figure*}

Figure \ref{fig:rF_AI20} is displaying the correlation between radio flux and asymmetry at $20\%$ of the line profiles. The same behavior of
$H\beta$ profile asymmetry being positively correlated with radio flux. However too in this case, the $H\alpha$ profile asymmetry correlation
is flat.

\begin{figure*}[!htbp]
\begin{center}$
\begin{array}{cc}
\includegraphics[width=3.0in]{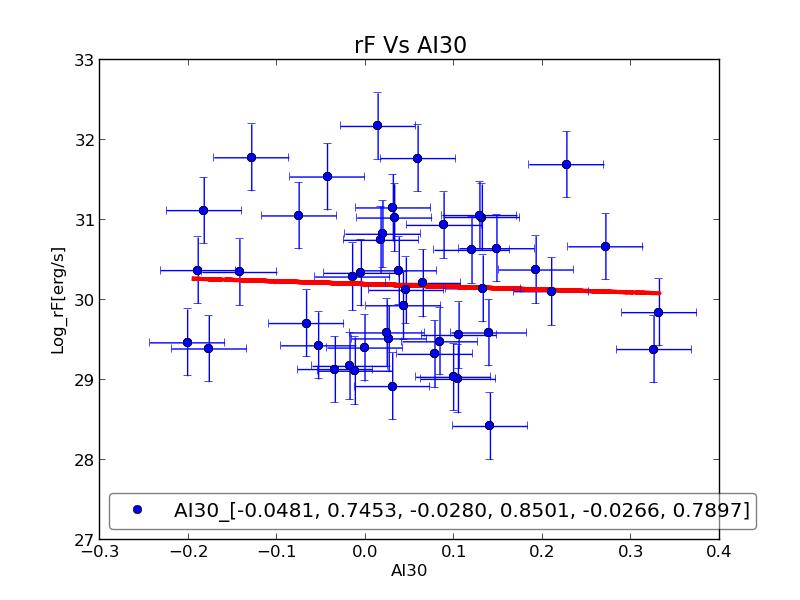} &
\includegraphics[width=3.0in]{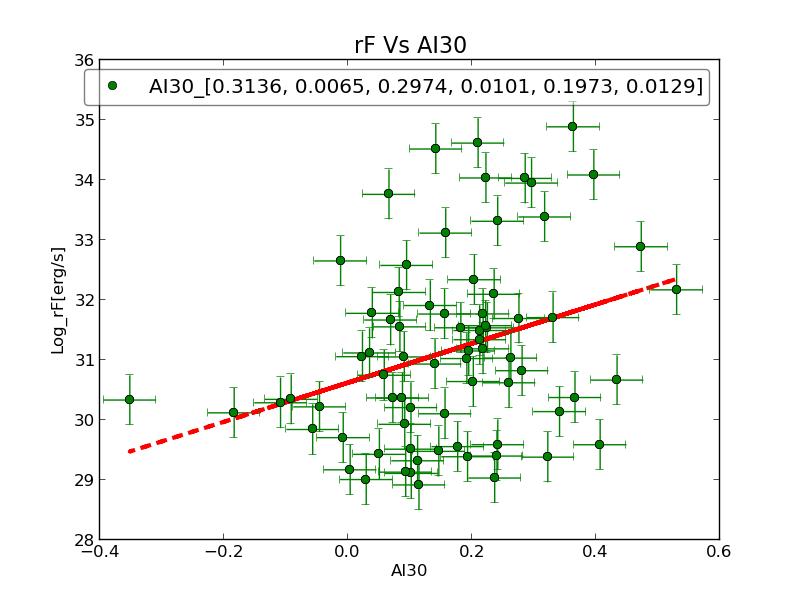} \\
\end{array}$
\end{center}
\caption[The Relation of the AI30 of $H\alpha$ and $H\beta$ emission line profiles with Core Radio Flux]
{Relation of the AI30 of $H\alpha$ and $H\beta$ emission line profiles with Core Radio Flux}
\label{fig:rF_AI30}
\end{figure*}

\begin{figure*}[!htbp]
\begin{center}$
\begin{array}{cc}
\includegraphics[width=3.0in]{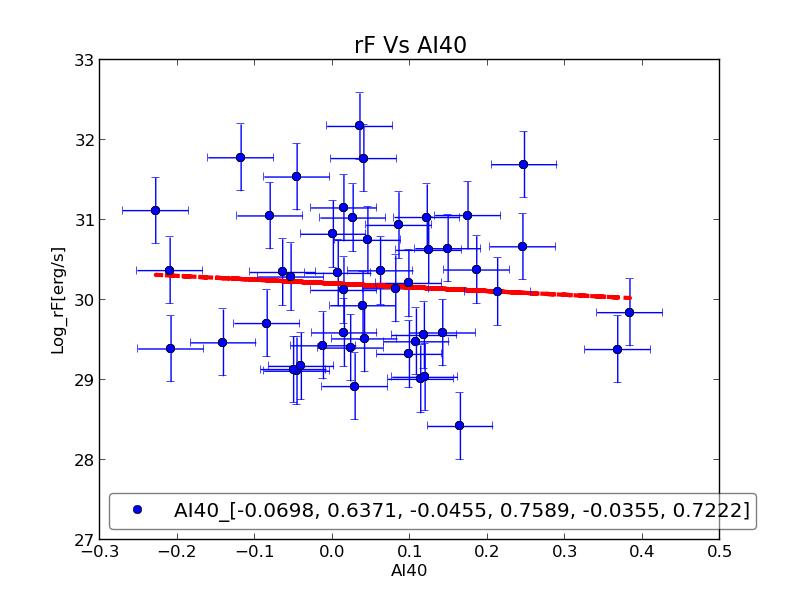} &
\includegraphics[width=3.0in]{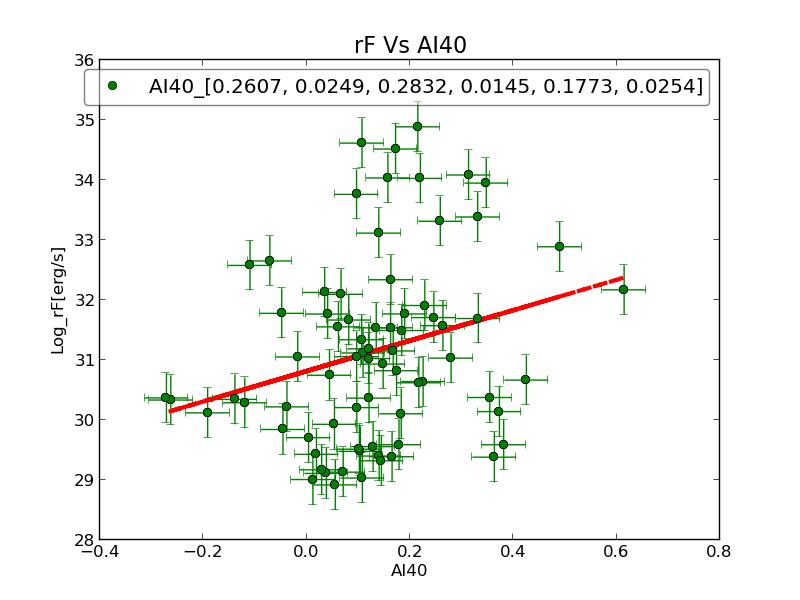} \\
\end{array}$
\end{center}
\caption[The Relation of the AI40 of $H\alpha$ and $H\beta$ emission line profiles with Core Radio Flux]
{Relation of the AI40 of $H\alpha$ and $H\beta$ emission line profiles with Core Radio Flux}
\label{fig:rF_AI40}
\end{figure*}

\begin{figure*}[!htbp]
\begin{center}$
\begin{array}{cc}
\includegraphics[width=3.0in]{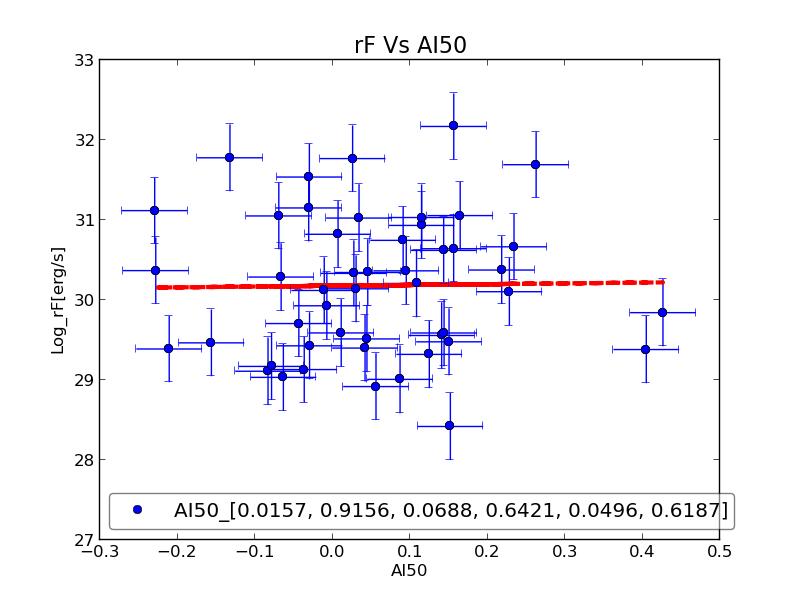} &
\includegraphics[width=3.0in]{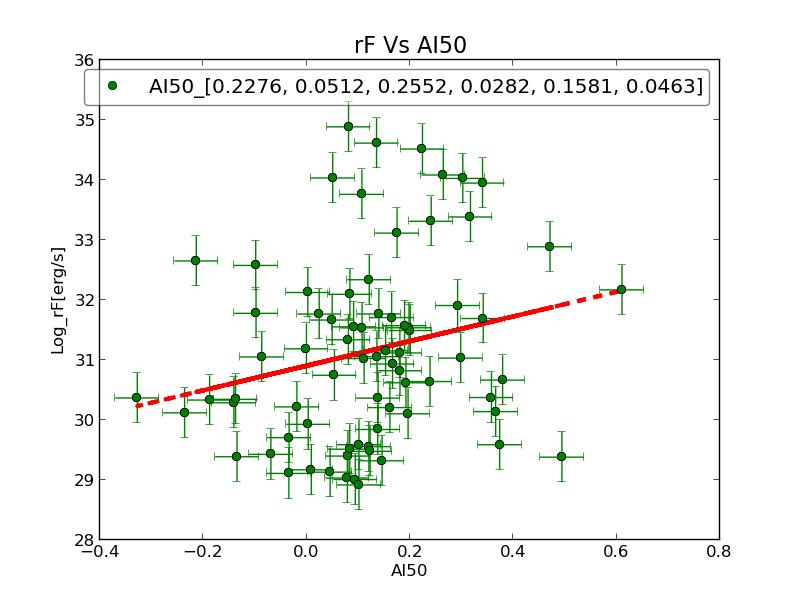} \\
\end{array}$
\end{center}
\caption[The Relation of the AI50 of $H\alpha$ and $H\beta$ emission line profiles with Core Radio Flux]
{Relation of the AI50 of $H\alpha$ and $H\beta$ emission line profiles with Core Radio Flux}
\label{fig:rF_AI50}
\end{figure*}

\begin{figure*}[!htbp]
\begin{center}$
\begin{array}{cc}
\includegraphics[width=3.0in]{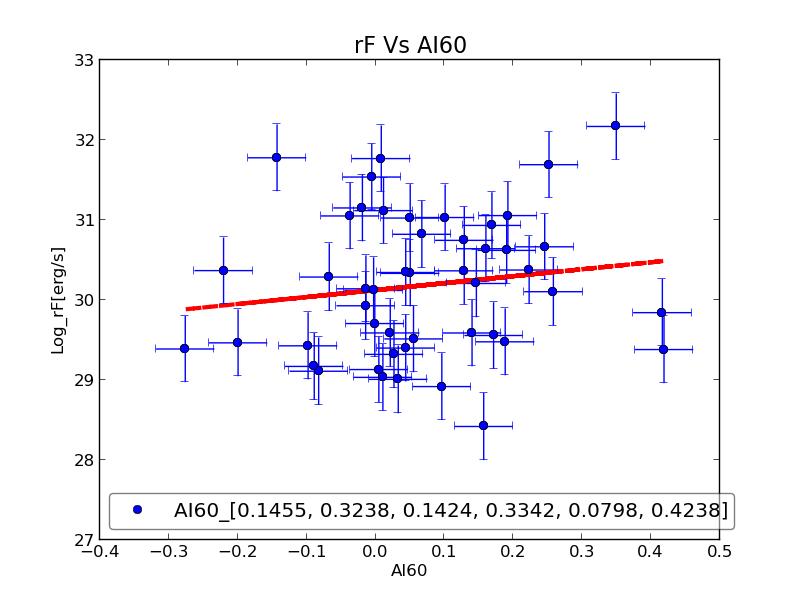} &
\includegraphics[width=3.0in]{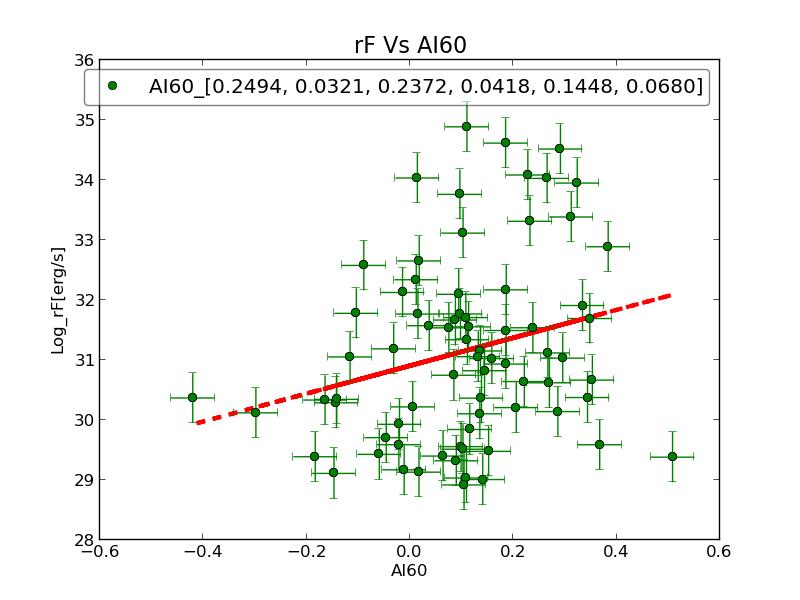} \\
\end{array}$
\end{center}
\caption[The Relation of the AI60 of $H\alpha$ and $H\beta$ emission line profiles with Core Radio Flux]
{Relation of the AI60 of $H\alpha$ and $H\beta$ emission line profiles with Core Radio Flux}
\label{fig:rF_AI60}
\end{figure*}

\begin{figure*}[!htbp]
\begin{center}$
\begin{array}{cc}
\includegraphics[width=3.0in]{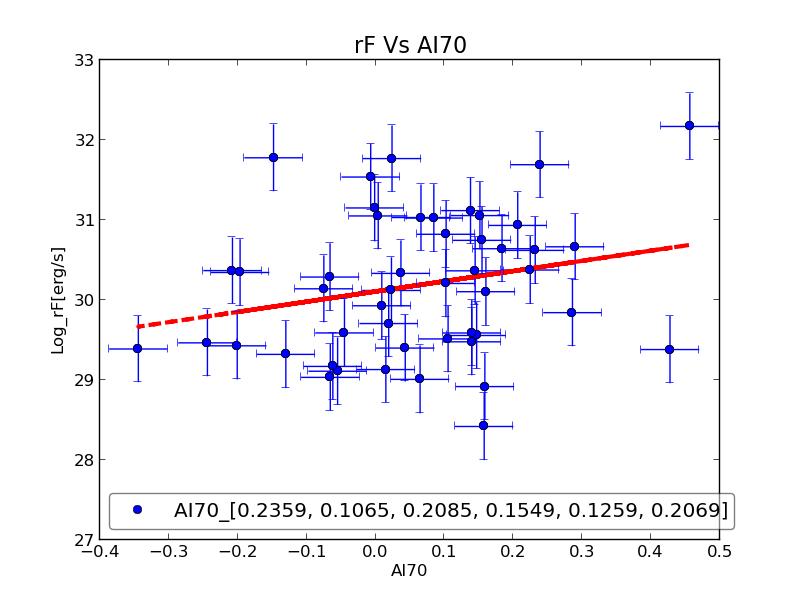} &
\includegraphics[width=3.0in]{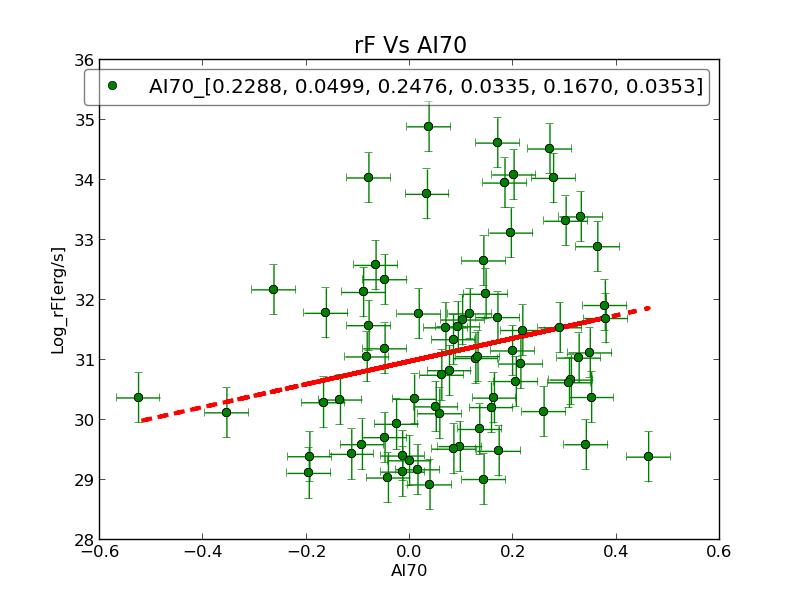} \\
\end{array}$
\end{center}
\caption[The Relation of the AI70 of $H\alpha$ and $H\beta$ emission line profiles with Core Radio Flux]
{Relation of the AI70 of $H\alpha$ and $H\beta$ emission line profiles with Core Radio Flux}
\label{fig:rF_AI70}
\end{figure*}

Figures \ref{fig:rF_AI30}, \ref{fig:rF_AI40}, \ref{fig:rF_AI50}, \ref{fig:rF_AI60} and \ref{fig:rF_AI70} all show the pattern of
the most asymmetric $H\alpha$ and $H\beta$ profiles positively correlating with core radio flux. 

\begin{table*}[!htbp]
\caption[Correlation Coefficients for $H\alpha$ relations]{Correlation Coefficients for $H\alpha$ relations}
\label{tab:cc_ha_rf}
\begin{center}
\begin{tabular}{ccccccccc}
\hline
Percentile & $\rho_p$ & $P_p{-value}$ & $\rho_s$ & $P_s{-value}$ & $\rho_k$ & $P_k{-value}$ & m & b \\ \hline \hline

20 & 0.0136 & 0.9270 & 0.0403 & 0.7858 & 0.0284 & 0.7761 & 0.1207 & 22.2540 \\ 
30 & -0.0481 & 0.7453 & -0.0280 & 0.8501 & -0.0266 & 0.7897 & -0.3450 & 22.2752 \\ 
40 & -0.0698 & 0.6371 & -0.0455 & 0.7589 & -0.0355 & 0.7222 & -0.4732 & 22.2829 \\ 
50 & 0.0157 & 0.9156 & 0.0688 & 0.6421 & 0.0496 & 0.6187 & 0.0988 & 22.2544 \\ 
60 & 0.1455 & 0.3238 & 0.1424 & 0.3342 & 0.0798 & 0.4238 & 0.8695 & 22.2000 \\ 
70 & 0.2359 & 0.1065 & 0.2085 & 0.1549 & 0.1259 & 0.2069 & 1.2763 & 22.1829 \\ \hline
\end{tabular}
\end{center}
\end{table*}

\begin{table*}[!htbp]
\caption[Correlation Coefficients for $H\beta$ relations]{Correlation Coefficients for $H\beta$ relations}
\label{tab:cc_hb_rf}
\begin{center}
\begin{tabular}{ccccccccc}
\hline
Percentile & $\rho_p$ & $P_p{-value}$ & $\rho_s$ & $P_s{-value}$ & $\rho_k$ & $P_k{-value}$ & m & b \\ \hline \hline

20 & 0.2524 & 0.0300 & 0.2511 & 0.0309 & 0.1714 & 0.0307 & 2.4011 & 22.7527 \\ 
30 & 0.3136 & 0.0065 & 0.2974 & 0.0101 & 0.1973 & 0.0129 & 3.2619 & 22.6989 \\ 
40 & 0.2607 & 0.0249 & 0.2832 & 0.0145 & 0.1773 & 0.0254 & 2.5361 & 21.6919 \\ 
50 & 0.2276 & 0.0512 & 0.2552 & 0.0282 & 0.1581 & 0.0463 & 2.0519 & 22.9814 \\ 
60 & 0.2494 & 0.0321 & 0.2372 & 0.0418 & 0.1448 & 0.0680 & 2.3258 & 22.9817 \\ 
70 & 0.2288 & 0.0499 & 0.2476 & 0.0335 & 0.1670 & 0.0353 & 1.9114 & 23.0574 \\ \hline
\end{tabular}
\end{center}
\end{table*}

From the relations of asymmetry with radio flux, it has been noted that binning the profile is important in finding out which section
of the profile responds more to the radio emission. The top part of the profile does not respond to changes in radio emission as the centroid of the
profile.
It is also noted that the $H\alpha$ and $H\beta$ profile shape is consistent with its response to radio flux throughout all the percentiles, although
it the relation is weaker in the  $H\alpha$ asymmetry.

\subsubsection{Relation of Asymmetry Index to the Ionization degree}

The Ionization degree is a measure of the excitation degree of the region from where the lines used to measure the ratio arise from.
It is vital for us, since it will give us a clue to how asymmetric profiles relate to this ionization potential of the regions. A 
direct relation will give more insight to a wind scenario from the broad line region to the narrow line region in which these emission lines
are obtained.

In the plots to follow, we have analyzed two measures of excitation, [OIII]/[OI] in red, and [OIII]/[OII] in blue. We go ahead to analyze
the relations separately for each percentile. As in previous plots, $H\alpha$ profile analysis is on the left and $H\beta$ profile
analysis on the right of each figure.

\begin{figure*}[!htbp]
\begin{center}$
\begin{array}{cc}
\includegraphics[width=3.0in]{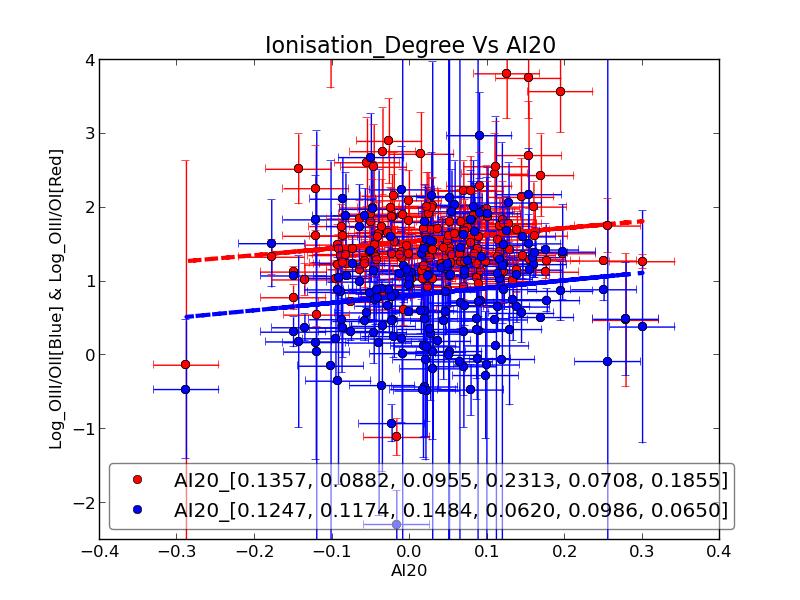} &
\includegraphics[width=3.0in]{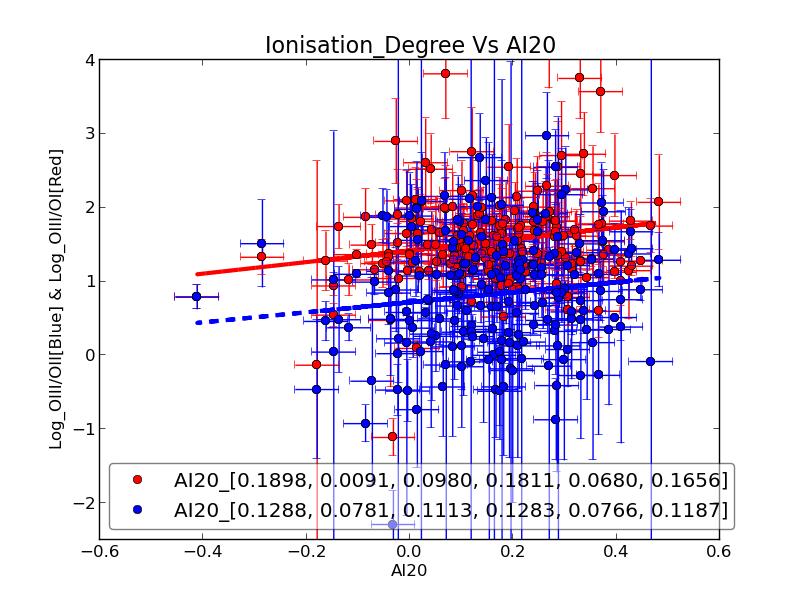} \\
\end{array}$
\end{center}
\caption[The Relation of the AI20 of $H\alpha$ and $H\beta$ emission line profiles with Ionization Degree]
{Relation of the AI20 of $H\alpha$ and $H\beta$ emission line profiles with Ionization Degree}
\label{fig:ID_AI20}
\end{figure*}

\begin{figure*}[!htbp]
\begin{center}$
\begin{array}{cc}
\includegraphics[width=3.0in]{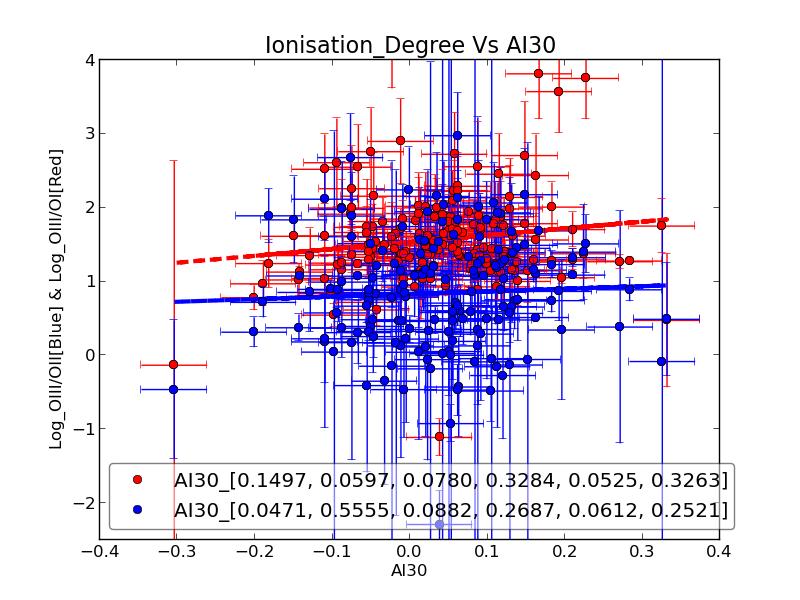} &
\includegraphics[width=3.0in]{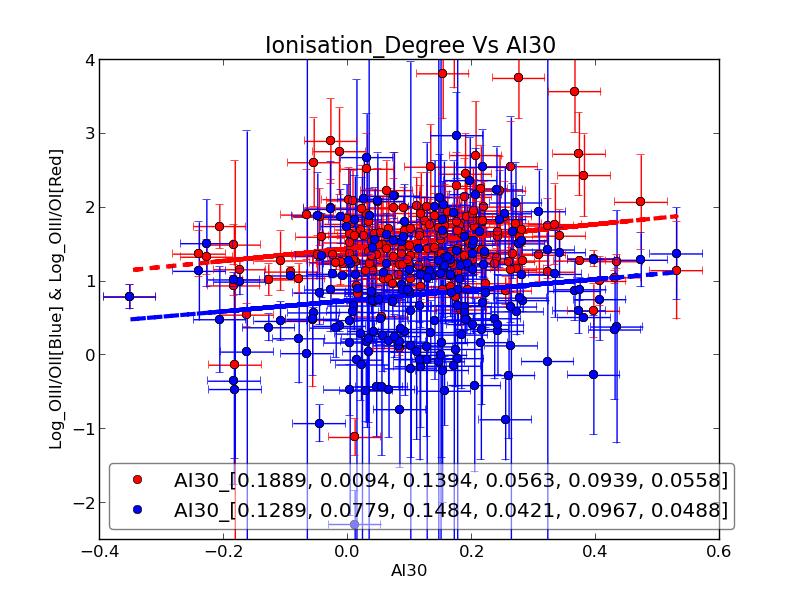} \\
\end{array}$
\end{center}
\caption[The Relation of the AI30 of $H\alpha$ and $H\beta$ emission line profiles with Ionization Degree]
{Relation of the AI30 of $H\alpha$ and $H\beta$ emission line profiles with Ionization Degree}
\label{fig:ID_AI30}
\end{figure*}

\begin{figure*}[!htbp]
\begin{center}$
\begin{array}{cc}
\includegraphics[width=3.0in]{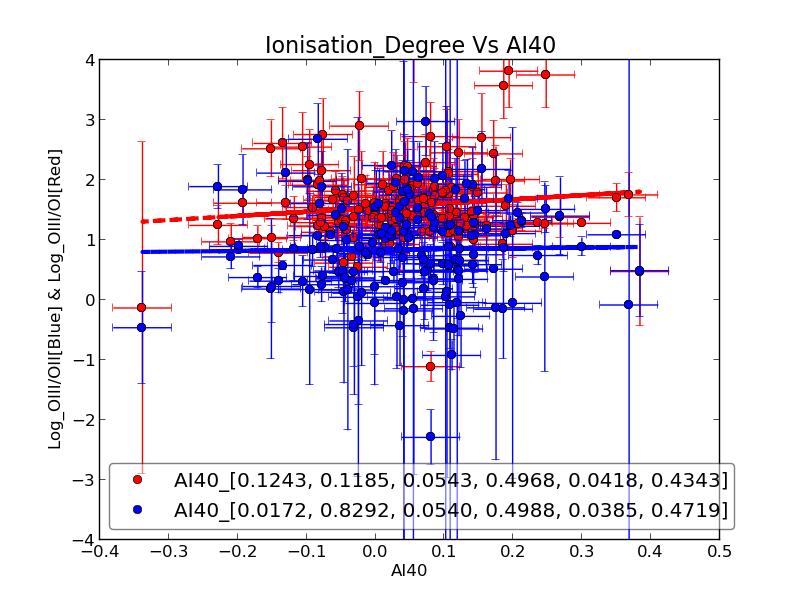} &
\includegraphics[width=3.0in]{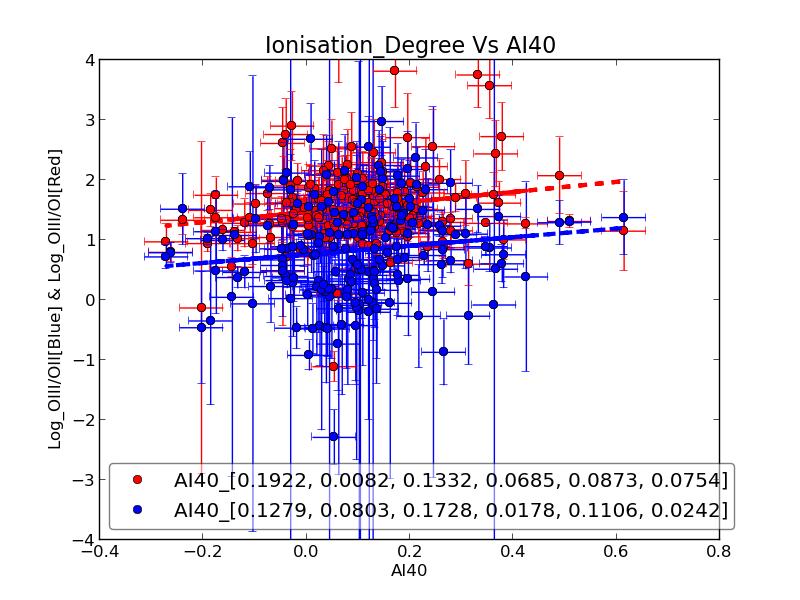} \\
\end{array}$
\end{center}
\caption[The Relation of the AI40 of $H\alpha$ and $H\beta$ emission line profiles with Ionization Degree]
{Relation of the AI40 of $H\alpha$ and $H\beta$ emission line profiles with Ionization Degree}
\label{fig:ID_AI40}
\end{figure*}

\begin{figure*}[!htbp]
\begin{center}$
\begin{array}{cc}
\includegraphics[width=3.0in]{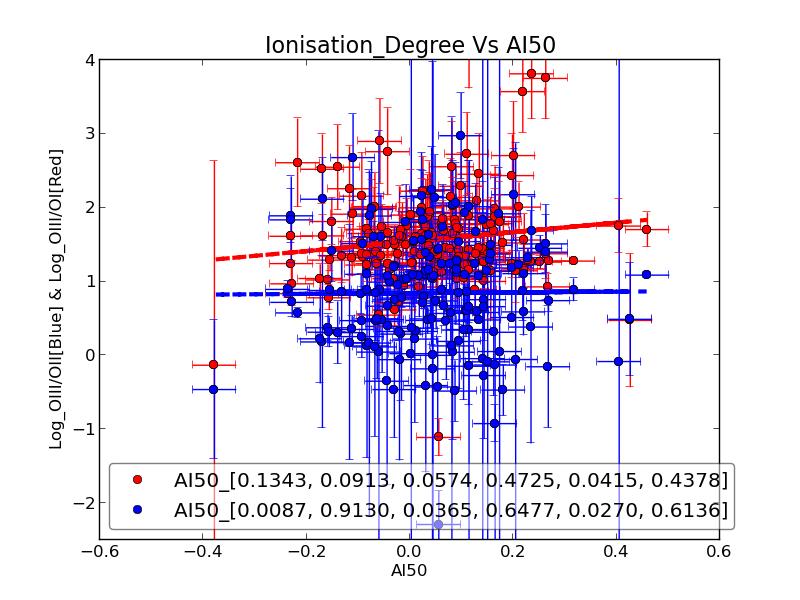} &
\includegraphics[width=3.0in]{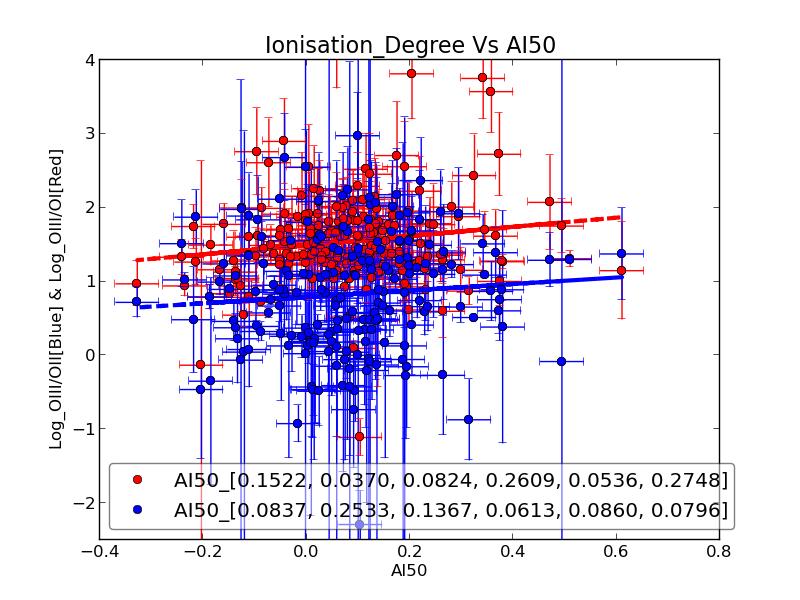} \\
\end{array}$
\end{center}
\caption[The Relation of the AI50 of $H\alpha$ and $H\beta$ emission line profiles with Ionization Degree]
{Relation of the AI50 of $H\alpha$ and $H\beta$ emission line profiles with Ionization Degree}
\label{fig:ID_AI50}
\end{figure*}

\begin{figure*}[!htbp]
\begin{center}$
\begin{array}{cc}
\includegraphics[width=3.0in]{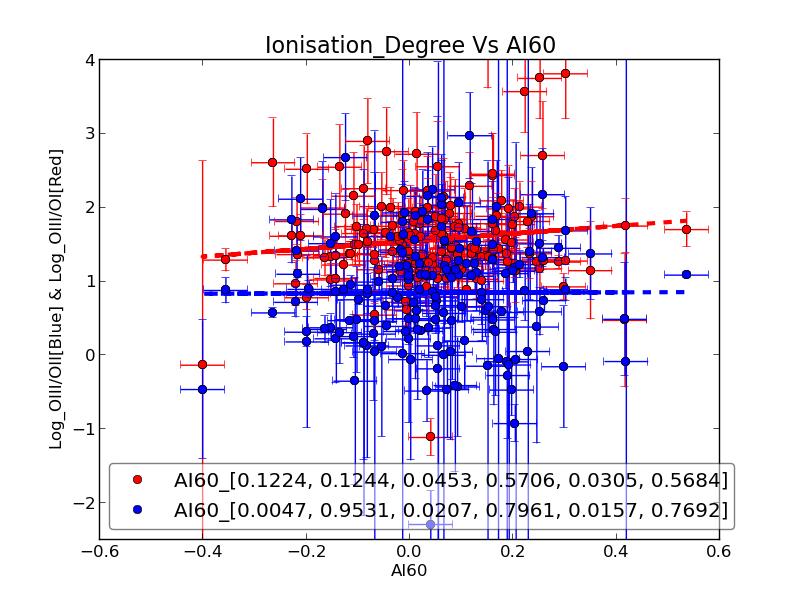} &
\includegraphics[width=3.0in]{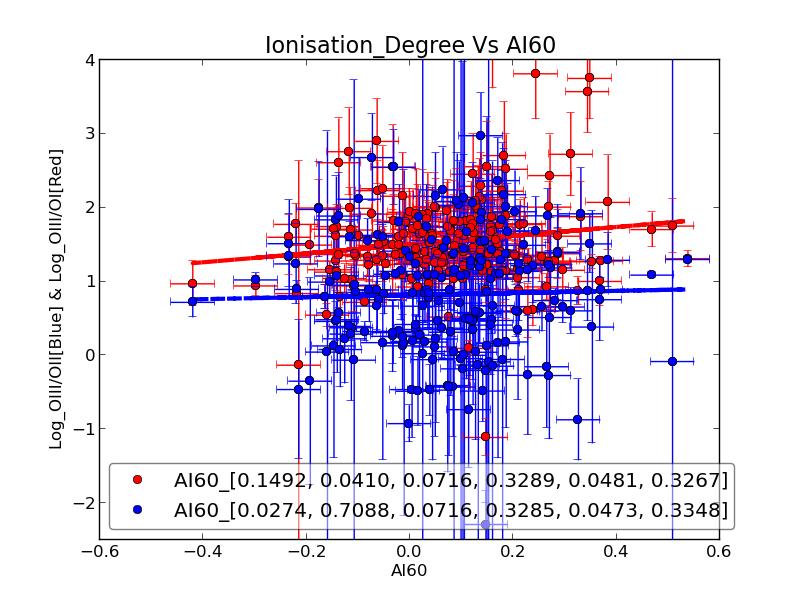} \\
\end{array}$
\end{center}
\caption[The Relation of the AI60 of $H\alpha$ and $H\beta$ emission line profiles with Ionization Degree]
{Relation of the AI60 of $H\alpha$ and $H\beta$ emission line profiles with Ionization Degree}
\label{fig:ID_AI60}
\end{figure*}

\begin{figure*}[!htbp]
\begin{center}$
\begin{array}{cc}
\includegraphics[width=3.0in]{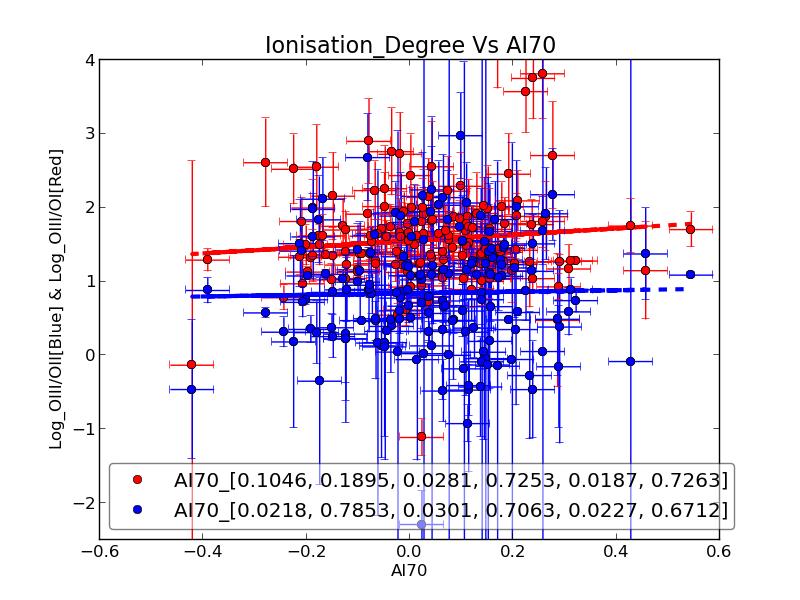} &
\includegraphics[width=3.0in]{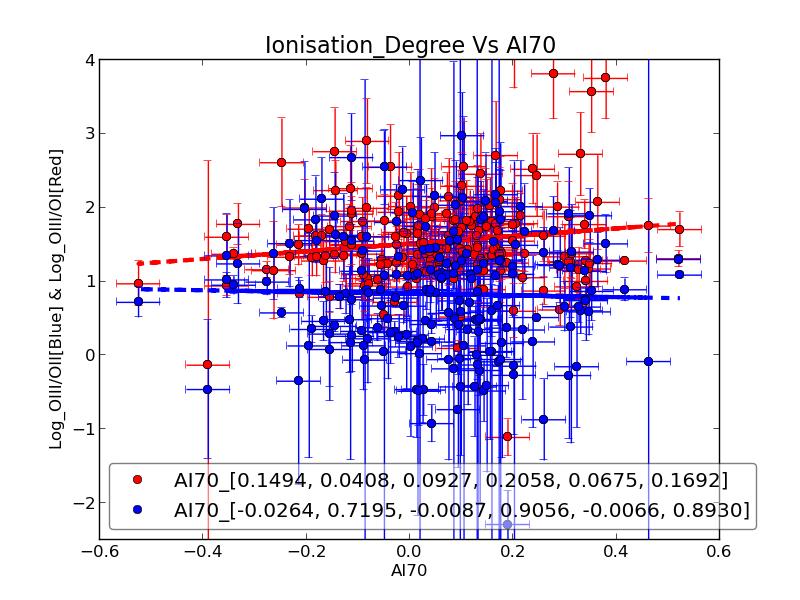} \\
\end{array}$
\end{center}
\caption[The Relation of the AI70 of $H\alpha$ and $H\beta$ emission line profiles with Ionization Degree]
{Relation of the AI70 of $H\alpha$ and $H\beta$ emission line profiles with Ionization Degree}
\label{fig:ID_AI70}
\end{figure*}

For all the relations of ionization degree with asymmetry, it is maintained from figure \ref{fig:ID_AI20} all 
through to \ref{fig:ID_AI70} that it is very difficult to obtain a precise correlation of ionization degree with Balmer line asymmetry. 

Statistical analyses show that the $H\beta$ relations are more reliable than $H\alpha$ measurements as seen from the $P-values$ and correlation
coefficients.

%----------------------------------------------------------------------------------------

\subsection{Kurtosis Index relation to other kinematic properties}

The way a profile changes, or how steep it may be, the Kurtosis Index, is a useful tool to use in studying other kinematic
properties of the source because its a visual property. Once one knows how this measure is related to other indirect kinematic
properties, it becomes easier to infer such properties by a simple analysis of the profile shape.

\subsubsection{Relation of Kurtosis Index with Line Width (FWHM)}
Since the Line width is a measure of the strength of the spectral features in the emission line, relating the Kurtosis with line width is simply 
relating the strength of the spectral features to the steepness (shape) of a profile.

\begin{figure*}[!htbp]
\begin{center}$
\begin{array}{cc}
\includegraphics[width=3.0in]{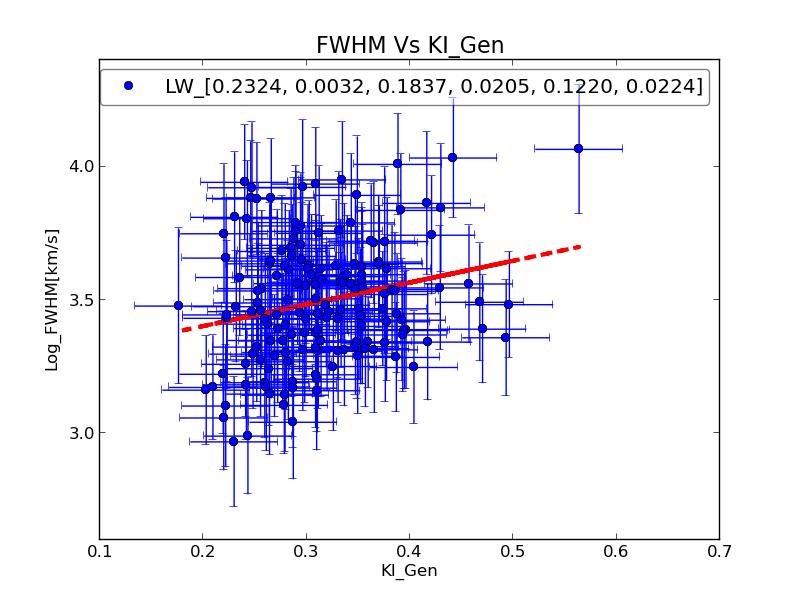} &
\includegraphics[width=3.0in]{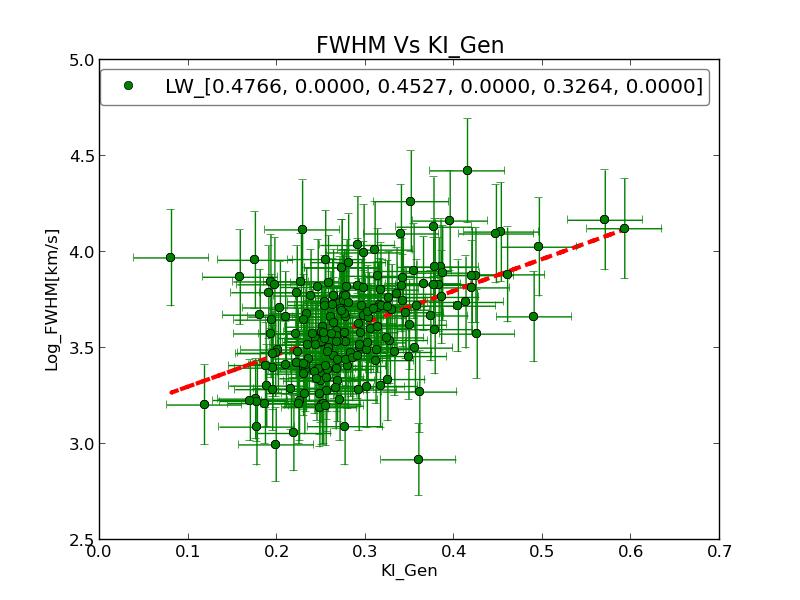} \\
\end{array}$
\end{center}
\caption[The Relation of the Line Width of $H\alpha$ and $H\beta$ emission line profiles with Kurtosis Index]
{Relation of the Line Width of $H\alpha$ and $H\beta$ emission line profiles with Kurtosis Index}
\label{fig:KI_Gen_FWHM}
\end{figure*}

Fig \ref{fig:KI_Gen_FWHM} shows how the Kurtosis varies with both Balmer emission lines. The $H\beta$ providing a tighter
correlation than the $H\alpha$. But still one thing in evident, higher values of Kurtosis, translate to higher line widths generally.
This suggests that broader profiles are flatter as previous studies show (\cite{Basu0}).
\cite{Basu0} found a good correlation between FWHM and kurtosis, in which he noted that Balmer lines and 
other permitted lines are broader and flatter. This means that very broad lines experience less change in their shape parameters than less
broadened lines.
As seen in tables \ref{tab:cc_ha_KI} and \ref{tab:cc_hb_KI}, the correlations are reliable with the $H\beta$ profile kurtosis index being
tighter and steeper.

\subsubsection{Relation of Kurtosis Index with Luminosity (V Band)}

\begin{figure*}[!htbp]
\begin{center}$
\begin{array}{cc}
\includegraphics[width=3.0in]{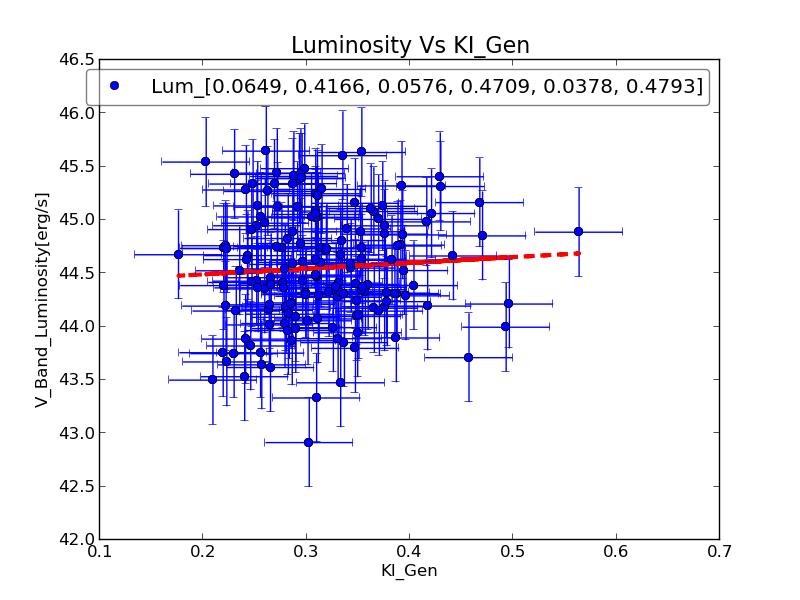} &
\includegraphics[width=3.0in]{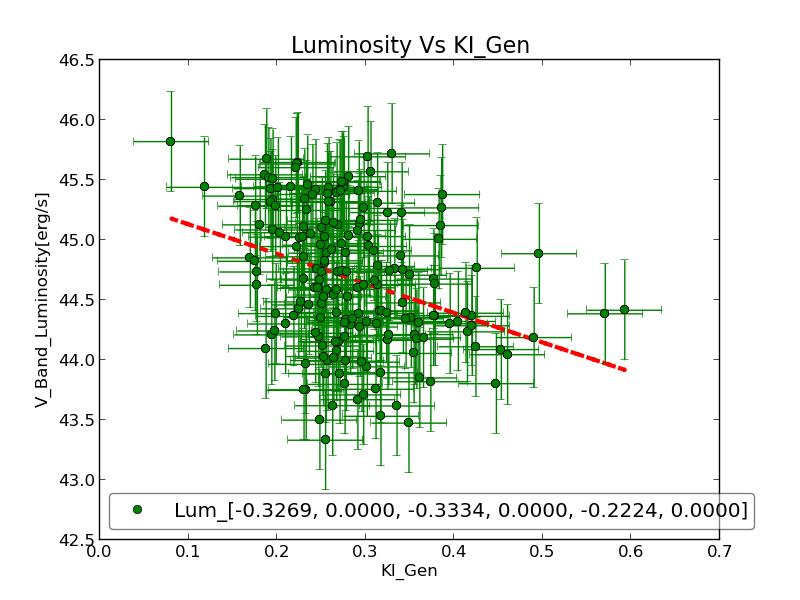} \\
\end{array}$
\end{center}
\caption[The Relation of the V Band Luminosity with Kurtosis Index]
{Relation of the V Band Luminosity of with Kurtosis Index}
\label{fig:KI_Gen_lum}
\end{figure*}

The relation of Kurtosis with Luminosity is also clearer with the $H\beta$ profile kurtosis. $H\alpha$ profile kurtosis relation is flat . 
This trend suggests that the most luminous sources will most likely have $H\beta$ profiles with extended wings. 
Although they both show Kurtosis Index decreasing with increasing Luminosity, the $P-values$ of the correlations are too high for us to rely
on this trend only. However, this points to the direction that luminosity most likely is one tool to look at when studying accretion disk winds.

\subsubsection{Relation of Kurtosis Index with Radio Flux (Core-Radio Flux)}

\begin{figure*}[!htbp]
\begin{center}$
\begin{array}{cc}
\includegraphics[width=3.0in]{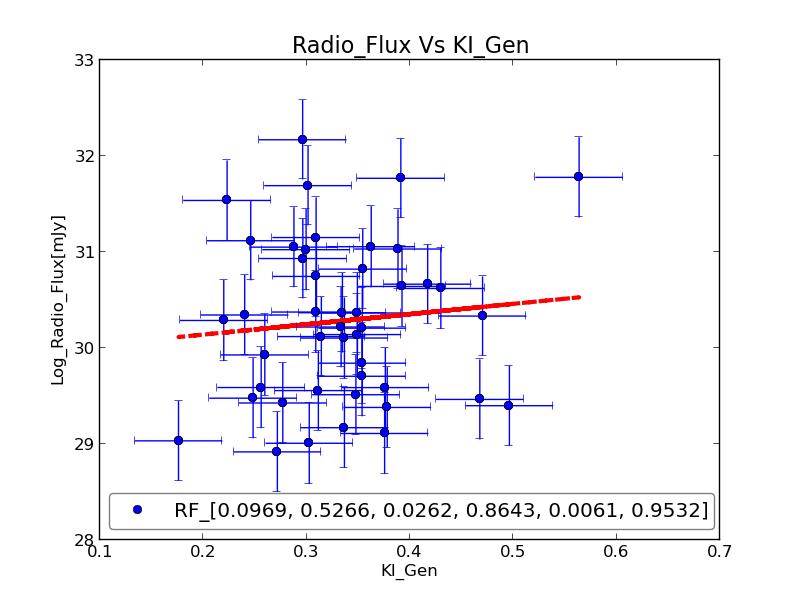} &
\includegraphics[width=3.0in]{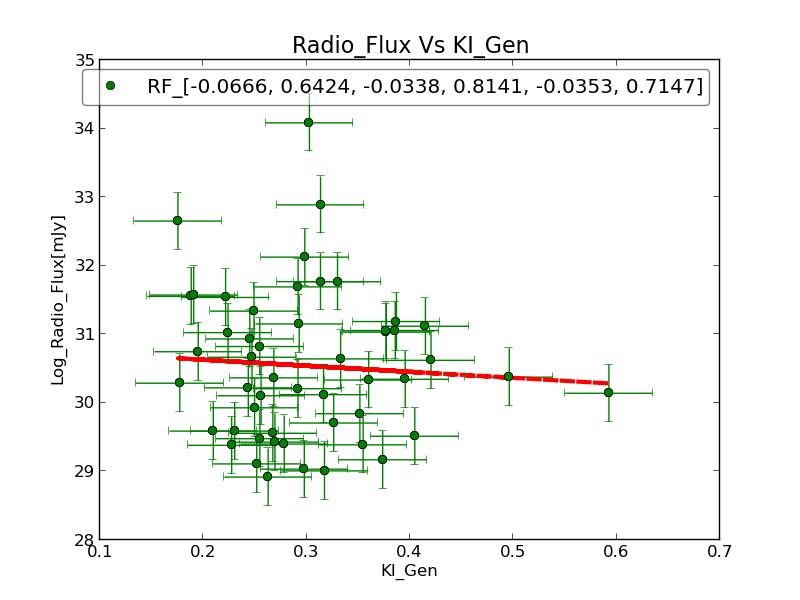} \\
\end{array}$
\end{center}
\caption[The Relation of the core Radio Flux with Kurtosis Index]
{Relation of the core Radio Flux with Kurtosis Index}
\label{fig:KI_Gen_rf}
\end{figure*}

$H\alpha$ kurtosis displays a positive correlation while $H\beta$ kurtosis shows an almost
flat correlation. These relations of radio flux to kurtosis make is challenging to extract a meaningful trend of radio flux with profile kurtosis. 
Previous studies by \cite{Whittle} showed a that radio flux varied for different kinds of AGN. In our data, we did not separate the kind of AGN, 
thus making it impossible to see any relation. He found out that there was a significant difference between the profile kurtosis of linear sources
and non-linear sources, with linear sources having steeper sided profiles. He went ahead to say that such observations could be naturally 
explained if the NLR gas at each end of the jet were radiating with opposite Doppler shifts relative to the synthetic(line-center) velocity. 
This means that the jets associated with the linear sources would perturb the outer parts of the velocity field, broadening the core to produce 
a high-kurtosis value, although the perturbation is not strong enough to influence the profile asymmetry.

\begin{table*}[!htbp]
\caption[Correlation Coefficients for $\frac{[OIII]}{[OI]}$ Ionization Degree Verses $H\alpha$ Asymmetry]{Correlation Coefficients for $\frac{[OIII]}{[OI]}$ Ionization 
Degree Verses $H\alpha$ Asymmetry}
\label{tab:cc_ha_id1}
\begin{center}
\begin{tabular}{ccccccccc}
\hline
Percentile & $\rho_p$ & $P_p{-value}$ & $\rho_s$ & $P_s{-value}$ & $\rho_k$ & $P_k{-value}$ & m & b \\ \hline \hline

20 & 0.1357 & 0.0882 & 0.0955 & 0.2313 & 0.0708 & 0.1855 & 0.9175 & 1.5411 \\ 
30 & 0.1497 & 0.0597 & 0.0780 & 0.3284 & 0.0525 & 0.3263 & 0.9266 & 1.5335 \\ 
40 & 0.1243 & 0.1185 & 0.0543 & 0.4968 & 0.0418 & 0.4343 & 0.6905 & 1.5390 \\ 
50 & 0.1343 & 0.0913 & 0.0574 & 0.4725 & 0.0415 & 0.4378 & 0.6414 & 1.5407 \\ 
60 & 0.1224 & 0.1244 & 0.0453 & 0.5706 & 0.0305 & 0.5684 & 0.5182 & 1.5459 \\ 
70 & 0.1046 & 0.1895 & 0.0281 & 0.7253 & 0.0187 & 0.7263 & 0.4225 & 1.5504 \\ \hline
\end{tabular}
\end{center}
\end{table*}

\begin{table*}[!htbp]
\caption[Correlation Coefficients for $\frac{[OIII]}{[OII]}$Ionization Degree Verses $H\alpha$ Asymmetry]{Correlation Coefficients for $\frac{[OIII]}{[OII]}$ Ionization
Degree Verses $H\alpha$ Asymmetry}
\label{tab:cc_ha_id2}
\begin{center}
\begin{tabular}{ccccccccc}
\hline
Percentile & $\rho_p$ & $P_p{-value}$ & $\rho_s$ & $P_s{-value}$ & $\rho_k$ & $P_k{-value}$ & m & b \\ \hline \hline

20 & 0.1247 & 0.1174 & 0.1484 & 0.0620 & 0.0986 & 0.0650 & 1.0134 & 0.8136 \\ 
30 & 0.0471 & 0.5555 & 0.0882 & 0.2687 & 0.0612 & 0.2521 & 0.3505 & 0.8314 \\ 
40 & 0.0172 & 0.8292 & 0.0540 & 0.4988 & 0.0385 & 0.4719 & 0.1151 & 0.8399 \\ 
50 & 0.0087 & 0.9130 & 0.0365 & 0.6477 & 0.0270 & 0.6136 & 0.0501 & 0.8428 \\ 
60 & 0.0047 & 0.9531 & 0.0207 & 0.7961 & 0.0157 & 0.7692 & 0.0239 & 0.8439 \\ 
70 & 0.0218 & 0.7853 & 0.0301 & 0.7063 & 0.0227 & 0.6712 & 0.1057 & 0.8402 \\ \hline
\end{tabular}
\end{center}
\end{table*}

\begin{table*}[!htbp]
\caption[Correlation Coefficients for $\frac{[OIII]}{[OI]}$ Ionization Degree Verses $H\beta$ Asymmetry]{Correlation Coefficients for $\frac{[OIII]}{[OI]}$ Ionization 
Degree Verses $H\beta$ Asymmetry}
\label{tab:cc_hb_id1}
\begin{center}
\begin{tabular}{ccccccccc}
\hline
Percentile & $\rho_p$ & $P_p{-value}$ & $\rho_s$ & $P_s{-value}$ & $\rho_k$ & $P_k{-value}$ & m & b \\ \hline \hline

20 & 0.1898 & 0.0091 & 0.0980 & 0.1811 & 0.0680 & 0.1656 & 0.7812 & 1.4197 \\ 
30 & 0.1889 & 0.0094 & 0.1394 & 0.0563 & 0.0939 & 0.0558 & 0.8271 & 1.4452 \\ 
40 & 0.1922 & 0.0082 & 0.1332 & 0.0685 & 0.0873 & 0.0754 & 0.8391 & 1.4654 \\ 
50 & 0.1522 & 0.0370 & 0.0824 & 0.2609 & 0.0536 & 0.2748 & 0.6199 & 1.4930 \\ 
60 & 0.1492 & 0.0410 & 0.0716 & 0.3289 & 0.0481 & 0.3267 & 0.5974 & 1.4997 \\ 
70 & 0.1494 & 0.0408 & 0.0927 & 0.2058 & 0.0675 & 0.1692 & 0.5168 & 1.5130 \\ \hline
\end{tabular}
\end{center}
\end{table*}

\begin{table*}[!htbp]
\caption[Correlation Coefficients for $\frac{[OIII]}{[OII]}$ Ionization Degree Verses $H\beta$ Asymmetry]{Correlation Coefficients for $\frac{[OIII]}{[OII]}$ Ionization
Degree Verses $H\beta$ Asymmetry}
\label{tab:cc_hb_id2}
\begin{center}
\begin{tabular}{ccccccccc}
\hline
Percentile & $\rho_p$ & $P_p{-value}$ & $\rho_s$ & $P_s{-value}$ & $\rho_k$ & $P_k{-value}$ & m & b \\ \hline \hline

20 & 0.1288 & 0.0781 & 0.1113 & 0.1283 & 0.0766 & 0.1187 & 0.6798 & 0.7197 \\ 
30 & 0.1289 & 0.0779 & 0.1484 & 0.0421 & 0.0967 & 0.0488 & 0.7236 & 0.7414 \\ 
40 & 0.1279 & 0.0803 & 0.1728 & 0.0178 & 0.1106 & 0.0242 & 0.7158 & 0.7608 \\ 
50 & 0.0837 & 0.2533 & 0.1367 & 0.0613 & 0.0860 & 0.0796 & 0.4372 & 0.7920 \\ 
60 & 0.0274 & 0.7088 & 0.0716 & 0.3285 & 0.0473 & 0.3348 & 0.1407 & 0.8178 \\ 
70 & -0.0264 & 0.7195 & -0.0087 & 0.9056 & -0.0066 & 0.8930 & -0.1170 & 0.8355 \\  \hline
\end{tabular}
\end{center}
\end{table*}

\subsubsection{Relation of Kurtosis Index with Ionization degree}

\begin{figure*}[!htbp]
\begin{center}$
\begin{array}{cc}
\includegraphics[width=3.0in]{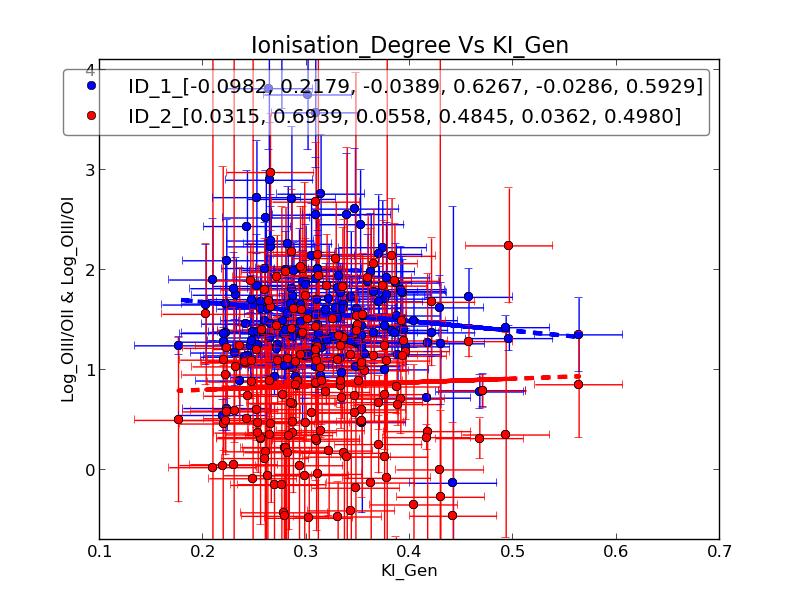} &
\includegraphics[width=3.0in]{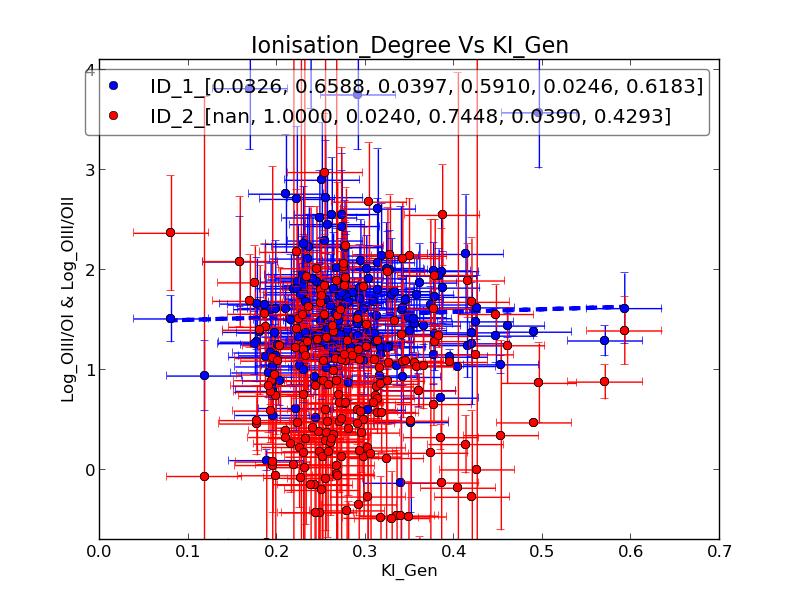} \\
\end{array}$
\end{center}
\caption[The Relation of the Ionization Degree with Kurtosis Index]
{Relation of the Ionization Degree with Kurtosis Index}
\label{fig:KI_Gen_id}
\end{figure*}

The ionization degree here will be related to the excitation state of the NLR. The correlations between profile shape and the NLR excitation
state are potentially very important because they reveal the presence of dynamical interactions between moving BLR moving gas(winds) and their
environment.
We have looked at two parameters of excitation degree, [OIII]/[OI] in red, and [OIII]/[OII] in blue. These values were obtained from the
spectra of the data downloaded from SDSS.
In general, the dispersion in excitation degree reduces with flatter profiles as observed in both $H\alpha$, on the left, and in $H\beta$ to the right
of figure \ref{fig:KI_Gen_id}.
Since the flatter profiles are generally the more asymmetric ones, then there is a view that more asymmetric profiles have less excitation degrees.
\cite{Whittle0} found out that there was a correlation between dust and asymmetry, where it was noted that an increase in extinction matched
higher asymmetric profiles. High asymmetry correlates with low values of kurtosis. This meant that dust not only causes profile asymmetry but also 
softens the ionizing radiation field which consequently produces lower excitation NLR.
Thus in the presence of dust, it is not easy to obtain a direct correlation between Kurtosis and the BLR winds or moving clouds.

\begin{table*}[!htbp]
\caption[Correlation Coefficients for the Kinematic Properties Verses $H\alpha$ Kurtosis Index]{Correlation Coefficients
for the Kinematic Properties Verses $H\alpha$ Kurtosis Index}
\label{tab:cc_ha_KI}
\begin{center}
\begin{tabular}{ccccccccc}
\hline
Relation & $\rho_p$ & $P_p{-value}$ & $\rho_s$ & $P_s{-value}$ & $\rho_k$ & $P_k{-value}$ & m & b \\ \hline \hline
LW & 0.2324 & 0.0032 & 0.1837 & 0.0205 & 0.1220 & 0.0224 & 0.8158 & 3.2392 \\ 
RF & 0.0969 & 0.5266 & 0.0262 & 0.8643 & 0.0061 & 0.9532 & 1.0674 & 29.9262 \\
L & 0.0649 & 0.4166 & 0.0576 & 0.4709 & 0.0378 & 0.4793 & 0.5398 & 44.3827 \\ 
$ID_1$ & -0.0982 & 0.2179 & -0.0389 & 0.6267 & -0.0286 & 0.5929 & -0.9613 & 1.8726 \\ 
$ID_2$ & 0.0315 & 0.6939 & 0.0558 & 0.4845 & 0.0362 & 0.4980 & 0.3699 & 0.7284 \\
\hline
\end{tabular}
\end{center}
\end{table*}

\begin{table*}[!htbp]
\caption[Correlation Coefficients for the Kinematic Properties Verses $H\beta$ Kurtosis Index]{Correlation Coefficients
for the Kinematic Properties Verses $H\beta$ Kurtosis Index}
\label{tab:cc_hb_KI}
\begin{center}
\begin{tabular}{ccccccccc}
\hline
Relation & $\rho_p$ & $P_p{-value}$ & $\rho_s$ & $P_s{-value}$ & $\rho_k$ & $P_k{-value}$ & m & b \\ \hline \hline
LW & 0.4766 & 0.0000 & 0.4527 & 0.0000 & 0.3264 & 0.0000 & 1.6614 & 3.1338 \\
RF & -0.0666 & 0.6424 & -0.0338 & 0.8141 & -0.0353 & 0.7147 & -0.8794 & 30.8045 \\
L & -0.3269 & 0.0000 & -0.3334 & 0.0000 & -0.2224 & 0.0000 & -2.4625 & 45.3767 \\ 
$ID_1$ & 0.0326 & 0.6588 & 0.0397 & 0.5910 & 0.0246 & 0.6183 & 0.2657 & 1.4750 \\ 
$ID_2$ & nan & 1.0000 & 0.0240 & 0.7448 & 0.0390 & 0.4293 & nan & nan \\
\hline
\end{tabular}
\end{center}
\end{table*}

\section{Discussion}
In the last section, we displayed the results obtained from the analysis of the Asymmetry Index and Kurtosis Index. We also looked at how the Asymmetry Index 
and Kurtosis Index relates to some other kinematic properties such as line width, luminosity,radio flux and ionization degree. It this chapter
we shall discuss some of the most striking relations that were observed and, if possible, study the most likely implications of the relation to 
understanding the structure and properties of the accretion disk winds in AGN.

\subsection{Asymmetry Index}

Astrophysical line profiles are observed to display a variety of shapes. This is because the shape of a line profile can depend on
a number of parameters which include; shocks, inflows and outflows of a wind component, obscuration by dust and rotation. If we only take the effect 
of the velocity field, that is shocks, Doppler motions, turbulence, inflow and outflow wind components and rotation, we can separate the region 
of the line profile that suffers from each effect. Since the line width is an excellent measure of strength of the properties of the line profile, 
we shall keep its relation to asymmetry in mind while discussing the rest of the kinematic properties.\\
We shall not include relations obtained at $00\%$ percentiles since errors in that measurement were higher than for other parts of the profile due to 
the difficulty in estimating the continuum levels at both ends of the broad Balmer lines. 
\citep{Balcells0} studied asymmetric line profiles in merger remnants, the study in his models suggesting that the asymmetry is not due to a disk 
of accreted secondary material; it is intrinsic to the distribution of primary particles, and has been added during the merger. 
It is argued that the asymmetries are a consequence of the transfer of the orbital energy and angular momentum to the primary particles during 
the merger; for the mergers studied in his models, profile asymmetries are a relic of the formation dynamics rather than the signature of 
superimposed components. Relating this to our case of the BLR, the asymmetries may be due to the transfer of the orbital energy and angular momentum
of the accreting material and it is a relic of the material falling in or leaving the accretion disk. This material flows in or out in form of a wind.
According to \citet{Gaskell0}, a line shifts, thus asymmetry, in the base of a profile is a reflection of high accretion rates of the AGN. They
demonstrated that high accretion rate AGNs will show line blue shifts, both in their models and observational data.
For our case, most of the Balmer emission line asymmetries were red-shifted, indicating outflow of material (wind) rather than inflow (accretion).

\subsubsection{Line Width}
Line width are a good measure in investigating the strength of the spectral features in a system. Some authors use it to differentiate
Seyfert galaxies from other emission line galaxies \citep{Feldman0}. Others, use it to separate quasars in two populations, Population A and Population B
in which the later have extremely broad Balmer emission lines exceeding a width of 3000 km/s. The line width of the emission line profiles
might depend on the the velocity field, geometry of the line emiting gas, obscuring effects, the superposition of line emission from different regions
and on isotropy/anisotropy of the line emission \citep{Kollat0}. In the relations with asymmetry, it was observed that
$H\alpha$ profiles showed a negative correlation, while the $H\beta$ profiles displayed a positive correlation.
This would mean that the line width as a measure of the spectral features strength of the line emitting region will provide better results when we
use the $H\beta$ line asymmetry relation to line width.
High values in line width were observed to correlate with higher values in asymmetry for the $H\beta$ profiles. This may suggest that systems
with high values in $H\beta$ profile asymmetry have within them lots of information that we can obtain on the velocity field and geometry of the 
line emitting gas, whereas those with less $H\beta$ asymmetry carried less information that can be extracted about the geometry of the line emitting
gas.

In this case it is the $20\%, 30\%, 40\%, 50\%~and~60\%$ $H\beta$ emission line percentiles displaying good correlations to support this. 

\begin{figure*}[!htbp]
\begin{center}$
\begin{array}{cc}
\includegraphics[width=3.0in]{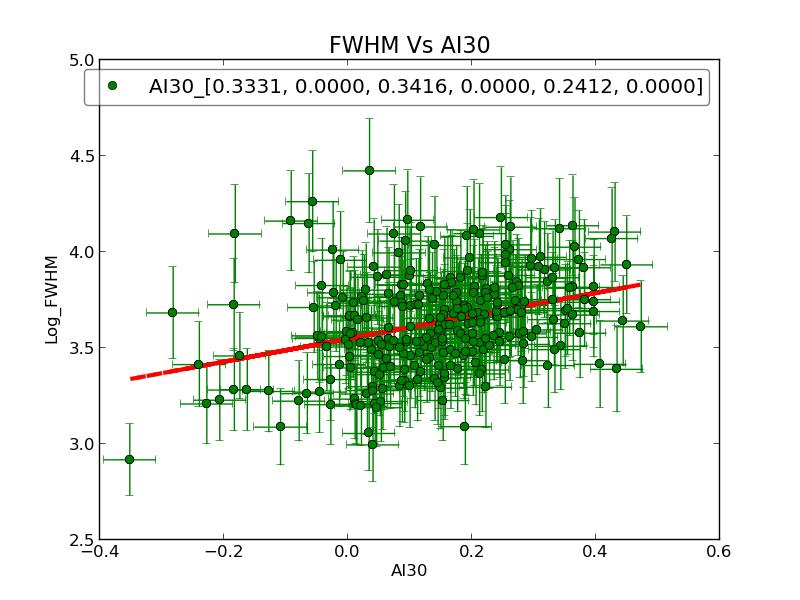} & 
\includegraphics[width=3.0in]{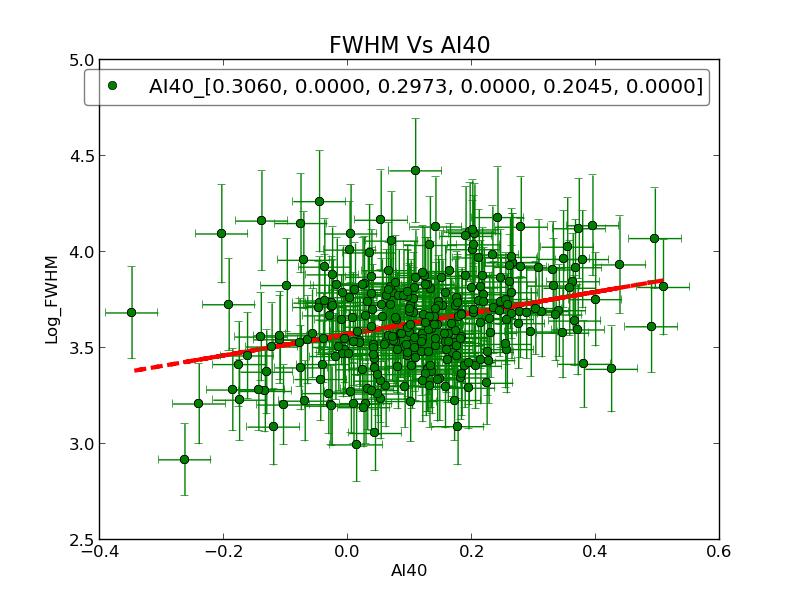} \\
\includegraphics[width=3.0in]{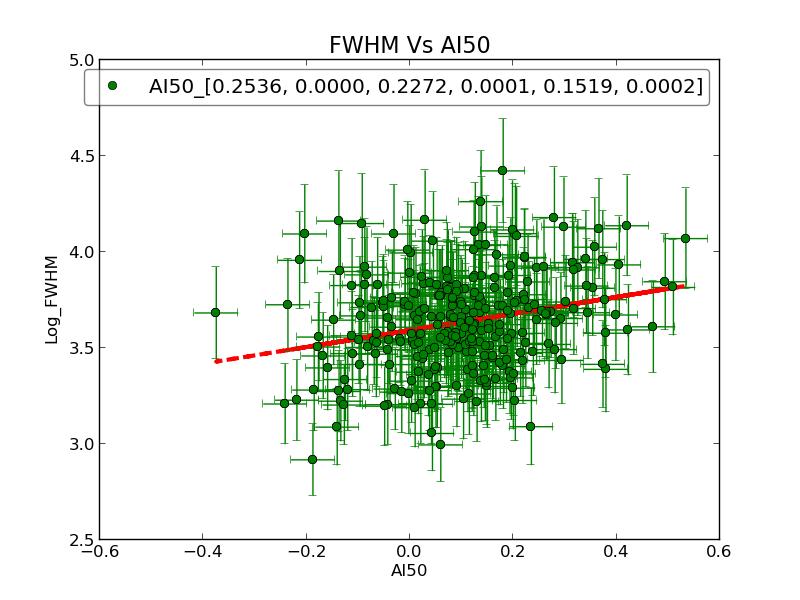} &
\includegraphics[width=3.0in]{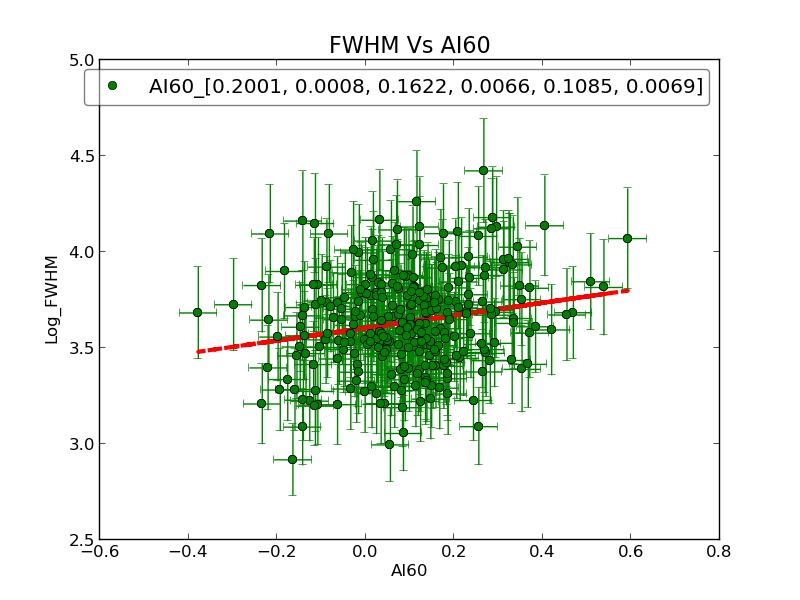} \\
\end{array}$
\end{center}
\caption[The Relation of the AI00 of $H\beta$ and AI10 $H\beta$ emission line profiles with Line Width]
{Relation of the AI00 of $H\alpha$ and AI10 $H\beta$ emission line profiles with Line width}
\label{fig:LW_Disc}
\end{figure*}

\subsubsection{Luminosity}

Luminosity is another property that is widely studied, having many relations \citep{Rafanelli0} with other properties of host galaxies 
\citep{Lamura0} like among other properties, the BLR size and BH mass.
Although the scatter was high, some plots appeared to suggest that highly asymmetric profiles correlated with high luminosities.
This means these luminosities are being driven by the same mechanism giving rise to asymmetry. This means winds will create asymmetry and also
rise the luminosity of the source as seen in such relations.
The best relations were obtained in $H\beta$ profile asymmetry for the $10\%, 20\%, 30\%,~and~40\%$ percentiles.
The moderately tight positive correlations point to a direction in which we notice that the most luminous sources have high asymmetries in their
$H\beta$ emission line profiles.

\begin{figure*}[!htbp]
\begin{center}$
\begin{array}{cc}
\includegraphics[width=3.0in]{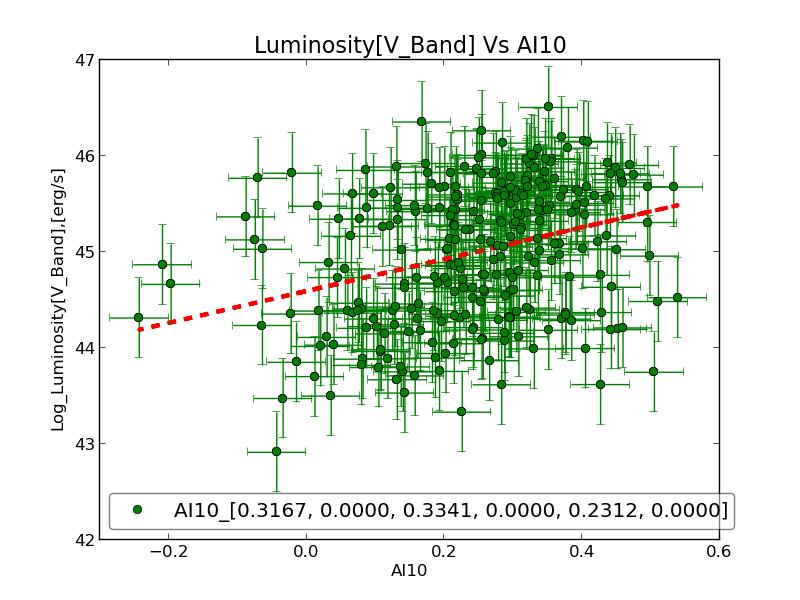} & 
\includegraphics[width=3.0in]{Hb_AI20_Luminosity} \\
\includegraphics[width=3.0in]{Hb_AI30_Luminosity} &
\includegraphics[width=3.0in]{Hb_AI40_Luminosity} \\
\end{array}$
\end{center}
\caption[The Relation of the AI00 of $H\beta$ and AI10 $H\beta$ emission line profiles with Line Width]
{Relation of the AI00 of $H\alpha$ and AI10 $H\beta$ emission line profiles with Line width}
\label{fig:Lum_Disc}
\end{figure*}

\subsubsection{Radio Flux}

\begin{figure*}[!htbp]
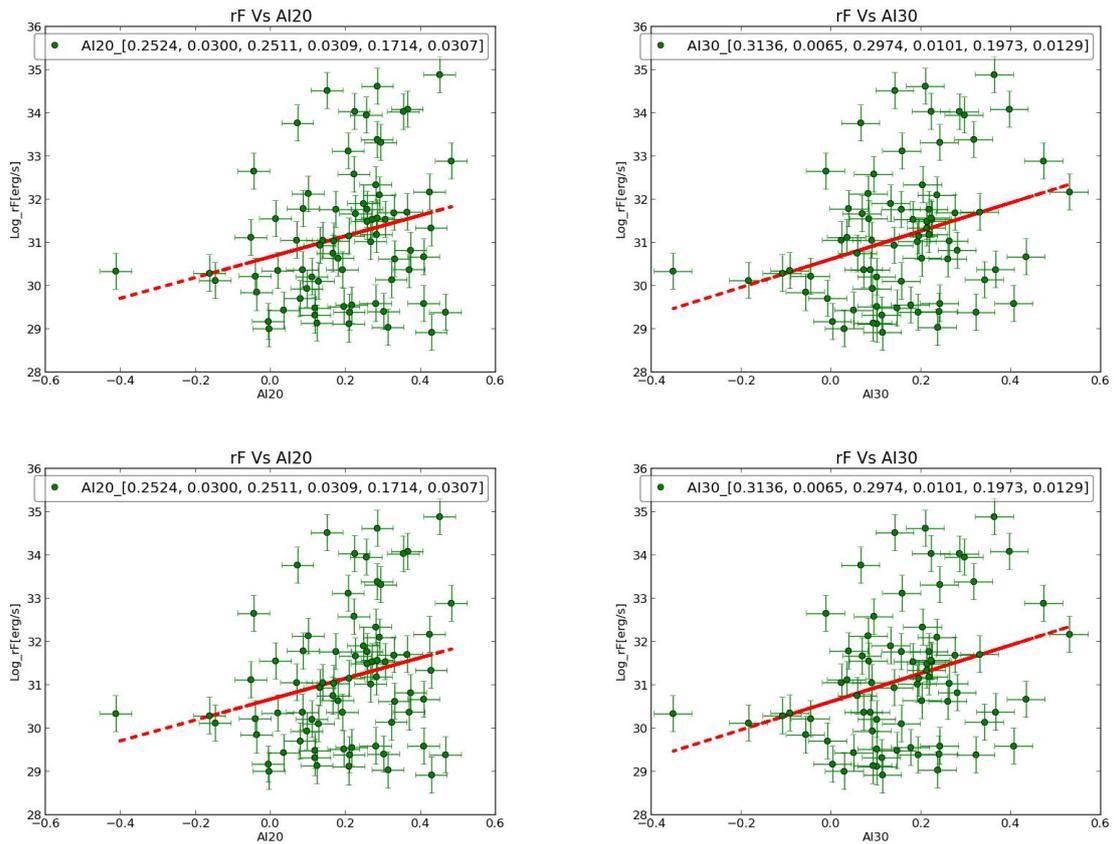

\begin{center}$
\begin{array}{cc}
\includegraphics[width=3.0in]{Hb_AI20_rf} &
\includegraphics[width=3.0in]{Hb_AI30_rf} \\
\includegraphics[width=3.0in]{Hb_AI20_rf} &
\includegraphics[width=3.0in]{Hb_AI30_rf} \\
\end{array}$
\end{center}
\caption[The Relation of the AI20 of $H\beta$ and AI30 of $H\beta$ emission line profiles with Core Radio Flux]
{The Relation of the AI20 of $H\beta$ and AI30 of $H\beta$ emission line profiles with Core Radio Flux}
\label{fig:hb_rf_20_30}
\end{figure*}

The radio flux scaled in such a way that high $H\beta$ emission line asymmetries positively correlated with radio flux. 
The clearest relations are in $20\%, 30\%, 40\%, 50\%~and~60\%$ $H\beta$ emission line percentiles as seen in fig \ref{fig:hb_rf_20_30}.
\citep{Corbin0}, in his study of ''The Emission-Line Properties of Low-Redshift Quasi-stellar Objects. II. The Relation to Radio Type'', found 
out that FRS quasars have significantly wider and more red-ward asymmetric $H\beta$ profiles.
Studies on the $H\alpha$ Balmer emission line in Solar Physics by \citep{Ichimoto} noted red asymmetry of H-alpha flare line profiles. 
They believed that the Red-shifted emission streaks of H-alpha line are found at the initial phase of almost all flares which occur near 
the disk center, and are considered to be substantial features of the asymmetry.  It is found that a downward motion in the flare chromospheric
region is the cause of the red-shifted emission streak. The downward motion abruptly increases at the onset of a flare, attains its maximum 
velocity of about 40 to 100 km/s shortly before the impulsive peak of the microwave burst, and rapidly decreases before the intensity of 
H-alpha line reaches its maximum. This proves that high radio emission can cause asymmetry in emission line shapes and the asymmetry is a consequence
of motion material that is giving rise to the emission line.

\subsubsection{Ionization Degree}
The ionization degree in the narrow line region has also been noted to respond with effects taking place in the broad line region.
For an outflow,a wind from an accretion disk would drive gas and dust out of the broad line region \citep{Kollat6}. This would cause both cooling
and heating consequences in the narrow line region and this has been observed in the high variance in ionization degree for higher asymmetries.
An inflow would also rid the narrow line region of material that would contribute to either heating or cooling, thus having still a 
dispersion in ionization degree measurements.\\

In all the properties studied, it seems plausible to confirm that accretion disk winds are part of the reason to the asymmetry in 
emission line profiles, both directly and indirectly. It is also noted that whether its inflow or outflow, the results would be similar
although the most studied relations favor outflows.

\begin{figure*}[!htbp]
\begin{center}$
\begin{array}{cc}
\includegraphics[width=3.0in]{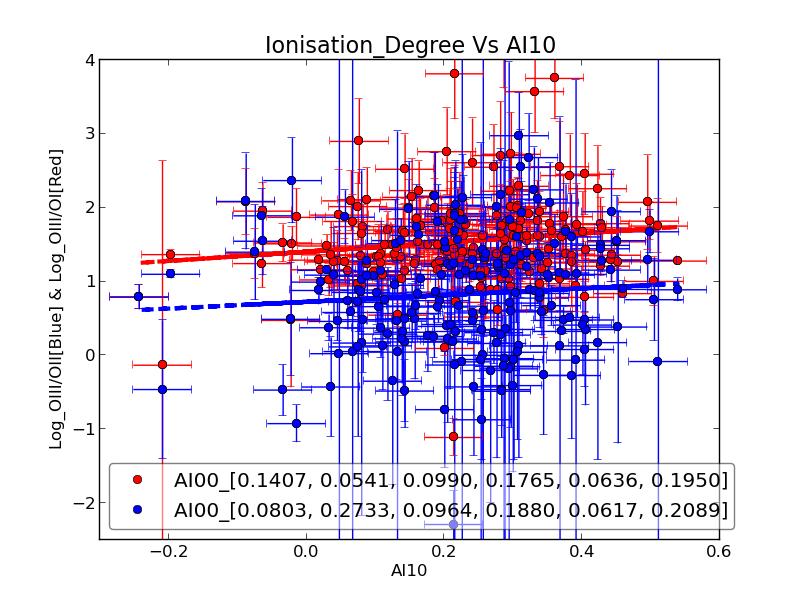} &
\includegraphics[width=3.0in]{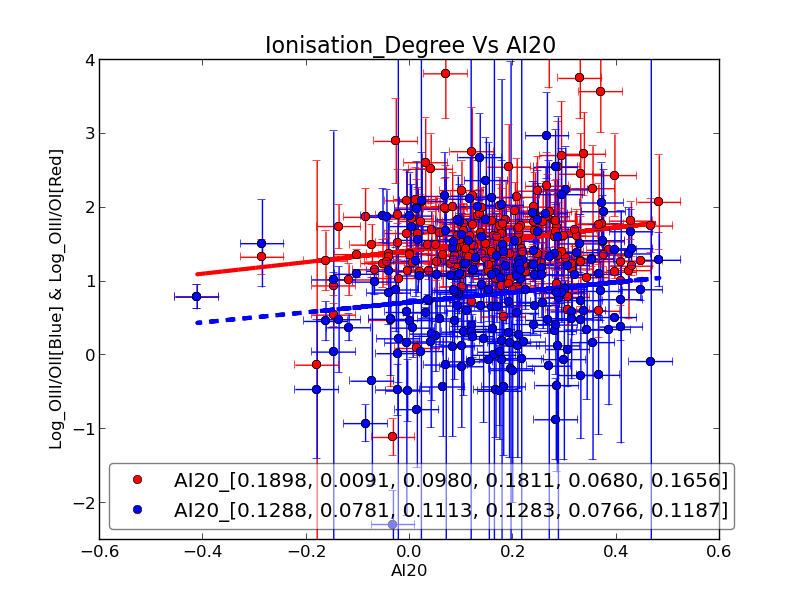} \\
\includegraphics[width=3.0in]{Hb_AI30_Ionisation_Degree} &
\includegraphics[width=3.0in]{Hb_AI40_Ionisation_Degree} \\
\end{array}$
\end{center}
\caption[The Relation of the AI10, AI20, AI30, AI40 of $H\beta$ emission line profiles with Ionization Degree]
{The Relation of the AI10, AI20, AI30, AI40 of $H\beta$ emission line profiles with Ionization Degree}
\label{fig:hb_10-40_id}
\end{figure*}

The clear relations in this case can be seen in figure \ref{fig:hb_10-40_id} which shows those of the 10th, 20th, 30th and 40th
percentile from top left to right.

\section{Conclusion}
In this study we investigated the asymmetry in the first two broad Balmer emission lines, ($H\alpha$ \& $H\beta$), with the aim of probing 
accretion disk winds in Active Galactic Nuclei (AGN). 
Our sample was chosen from the SDSS DR7 with slightly over 300 spectra. 
We extracted the individual broad Balmer emission lines from each spectrum, normalized them to have the emission line peak have a value of unity,
and measured the line shifts at each chosen percentile after centering the profiles at the a wavelength value corresponding to the emission line peak.
Line shift values were then used to calculate the Asymmetry Index (A.I) and Kurtosis Index (K.I) for each respective percentile, and
statistical distribution of these measurements was analyzed, after which, both the AI and KI were plotted
against some kinematic properties like line width (FWHM), Luminosity, Radio Flux, and Ionization degree ($\frac{[OIII]\lambda~5007}{[OI]\lambda~6302}$
\& $\frac{[OIII]\lambda~5007}{[OII]\lambda~3727}$).

According to our results, we come to the following conclusions:
\begin{itemize}
 \item Asymmetry in $H\beta$ profiles positively correlated with Line width, V-band Luminosity, Radio Flux and Ionization degree.
 \item Asymmetry in $H\beta$ profiles shows a stronger correlation with Line width, V-band Luminosity, Radio Flux and Ionization degree than
 that of $H\alpha$ emission line profiles. Its correlations were tighter, with steeper gradients.
 \item From the statistical distribution of the AI and KI, the asymmetry of the profiles is predominantly red-shifted.
 \item Broader lines showed more asymmetry than narrower ones. 
 \item Overplots of $H\alpha$ profiles with $H\beta$ profiles from the same spectrum showed that the $H\beta$ profiles had an extended red asymmetry
 from their $H\alpha$ counterparts.%, as seen in Fig.\ref{fig:ha_asoverplots}.
 \item The flux ratio $\frac{[OIII]\lambda~5007}{[OI]\lambda~6302}$, as a measure of ionization degree is more reliable than $\frac{[OIII]\lambda~5007}{[OII]\lambda~3727}$
 since $[OII]\lambda~3727$ values contained more errors in their values.
 \item The observed distribution of the asymmetry is consistent with previous studies on many other emission and absorption lines, most showing
 positive asymmetries.% \citep{Heckman0, Verhamme0, Zamfir, Brotherton1}.
 \item Percentiles of $20\%, 30\%, 40\%, 50\%, 60\%$ and $80\%$ contain the most reliable information about profile shape since they are least affected
 by errors due to continuum estimation and line peak estimation.
 \item The line profile shape parameters carry much detailed information about kinematics of the BLR, which can be better understood by means
  of techniques exploiting the profile by region, rather than measuring general shape parameters.
\end{itemize}

Although this analysis may be an advancement in probing accretion disk winds, a study with more sources and incorporating many more other emission
and absorption lines would be much better in drawing conclusive relations with the kinematic properties and thus key to understanding the structure
of accretion disk winds in AGN.

%\section*{References}
\bibliographystyle{aa} % style aa.bst
\bibliography{paper} % your references Yourfile.bib 

\end{document}